%% file: ChainGraphMainV3.tex
\documentclass[12pt, onecolumn]{IEEEtran}

\usepackage{setspace}
\usepackage{amsmath}
\usepackage{amssymb}
\usepackage{amsthm}
\usepackage{subfigure}
\usepackage[dvips]{color}
\usepackage{graphicx}
\usepackage{mathcomSTEP}
\usepackage{anyfontsize}
\usepackage{enumerate}

\allowdisplaybreaks
\theoremstyle{plain}    

\theoremstyle{definition} \newtheorem{definition}{Definition}   

\theoremstyle{remark} \newtheorem*{remark}{Remark}  

\theoremstyle{remark}

\newtheorem{assumption}{Assumption}

\theoremstyle{empty}
\newtheorem{cond}{Condition}

\begin{document}

\title{
A unified graphical approach to \\
random coding for multi-terminal networks
%
}

\author{
  \IEEEauthorblockN{Stefano Rini\IEEEauthorrefmark{1}\IEEEauthorrefmark{2}   and Andrea Goldsmith\IEEEauthorrefmark{1}\\}
  \IEEEauthorblockA{
  \IEEEauthorrefmark{1}
Department of Electrical Engineering, Stanford University,  USA\\
Email: \texttt{andrea@wsl.stanford.edu}}\\
 \IEEEauthorblockA{%
\IEEEauthorrefmark{2}
Department of Communication Engineering, National Chiao-Tung University, Taiwan\\
E-mail: \texttt{stefano@nctu.edu.tw}}
 }

\maketitle
%

\input{intro.tex}

\input{BackgroundAndContributions.tex}

\input{NetworkModel.tex}
\input{UserVrtualization}

\input{ChainGraphRepresentation}

\input{GMMassociated}

\input{CodebookConstruction}
\input{AchievableRateRegion}
\input{ConcludingRemark}
\input{Conclusion}
%
%
\newpage
\bibliographystyle{IEEEtran}
\bibliography{steBib2}

\newpage
\appendices

\input{GraphicalMarkovModels}
\input{Appendix}
\end{document}

%% file: intro.tex
\begin{abstract}
A unified approach to the derivation of rate regions for single-hop memoryless networks is presented.
A general transmission scheme  for any memoryless, single-hop, $k$-user channel with or without common information, is defined through two steps.
The first step is user virtualization: each user is divided into multiple virtual sub-users according to a chosen rate-splitting strategy which preserves the rates of the original messages.
This results in an enhanced channel with a possibly larger number of users
for which more coding possibilities are available.
Moreover, user virtualization provides a simple mechanism to encode common messages to any subset of users.
Following user virtualization, the message of each user in the enhanced model is coded using a chosen combination of coded time-sharing, superposition coding and joint binning.
A graph is used to represent the chosen coding strategies: nodes in the graph represent codewords while edges represent coding operations.
This graph is used to construct a graphical Markov model which illustrates the statistical dependency among codewords that can be introduced by the superposition coding or joint binning.
%
Using this statistical representation of the overall codebook distribution, the error probability of the code is shown to vanish via a unified analysis.
The rate bounds that define the achievable rate region are obtained by linking the error analysis to the properties of the graphical Markov model.
This proposed framework makes it possible to numerically obtain an achievable rate region by specifying a user virtualization strategy and describing a set of coding operations.
The largest achievable rate region can be obtained by considering all the  possible rate-splitting strategies and taking the union over all the possible ways to superimpose or bin codewords.
The achievable rates obtained based on this unified graphical approach to random coding encompass the best random coding achievable rates for all memoryless single-hop networks known to date, including broadcast, multiple access, interference, and cognitive radio channels, as well as new results for topologies not previously studied.
We demonstrate the technique for several single-hop network topologies to illustrate the steps by which achievable regions can be efficiently computed.
\end{abstract}


{\IEEEkeywords
Wireless network, Random coding, Achievable rate region, User virtualization, Chain Graph, Graphical Markov model, Coded time-sharing, Rate-splitting, Superposition coding, Binning, Gelfand-Pinsker coding.
}

\thanks{This paper was presented in part at the 2013 IEEE Information Theory and Applications (ITA) Workshop, San Diego, USA.
}

\section{Introduction}

Random coding was originally developed by Shannon as the capacity-achieving strategy for point-to-point channels  \cite{shannon48}.
Shannon's notion of random codebook generation and jointly-typical set decoding was later extended to single-hop multi-user channels by introducing new techniques such as superposition coding, rate-splitting, coded time-sharing, and joint binning.
The main contribution of this paper is a unified graphical approach to random coding that produces an achievable rate region for any single-hop memoryless network based on random coding schemes involving rate-splitting, coded  time-sharing, superposition coding and joint binning for both common and private information.
%
%
We show that any such scheme can be described by a matrix which details the splitting of the messages and by a graph which represents the coding operations.
The rate-splitting matrix defines the user virtualization, that is, how the original users can be split into multiple virtual sub-users.
The coding operations after user virtualization are expressed using a graph in which nodes represents codewords, one set of edges represents superposition
coding while another set binning.

Once this representation is established, we construct a Graphical Markov Model (GMM) \cite{pearl1988probabilistic} of the coding operation and use it to describe the factorization of the distribution of the codewords.
This GMM is obtained by associating a distribution to a graph by letting nodes represent random variables
while edges specify conditional dependence among the variables.
%
%
Surprisingly, this simple approach in specifying local dependencies among variables allows GMMs to compactly capture complex dependence structures among a large set of random variables.
%
%
%
By linking the code construction to the codebook distribution through GMMs, we are able to provide a unified error analysis based on the packing and covering lemmas for any scheme that can be described through this formalism.
Consequently, we obtain a description of the achievable rate region in terms of the properties of the graph which details the construction of the code.
This expression is particularly compact and can be easily evaluated numerically for channel models with a large number of users,  as we describe in more detail in Sec. \ref{sec:ConcludingRemark}.

%
%
%
%
%
%
\bigskip
The derivation of achievable rate regions based on random coding is a widely-studied topic in network information theory.
The contribution of this work is to generalize the derivation of achievable rate regions via random coding by establishing
a systematic framework for user virtualization and
%
a representation of the coding operations which links the encoding and decoding operations to the error events analysis.
%
%
The resulting graphical approach to random coding unifies the derivation of achievable rate regions for any single-hop discrete memoryless k-user channel, with the most general message sets, via all known random coding techniques, including superposition, binning, rate splitting and coded time sharing.
This work thus subsumes all of the best known achievable rate regions for the Broadcast Channel (BC), Multiple Access Channel (MAC), 2-user InterFerence Channel (IFC), and 2-user Cognitive IFC (CIFC) within one unified framework, while also providing the framework to extend these known results to any number of users and/or more general message sets. In addition to extending these previous results, our framework can be used to characterize achievable rate regions under all combinations of the above coding schemes for single-hop discrete memoryless multi-terminal topologies not previously studied.
The application of our framework to several such topologies are discussed later in the manuscript.
\subsection{Prior work}

Due to the complexity associated with the possible coding strategies, achievable rate regions for single-hop networks have generally been limited to two users and two private messages, with some
treatments of common information.
There has been some prior work towards a unified theory simplifying the derivation of achievable rates for certain multi-terminal networks.
%
A first approach in this direction can be found in \cite{gunduz2010capacity} where the capacity of the Multiple Access Channel (MAC) with common
messages is studied.
The capacity of this channel was first derived by Han \cite{han1979capacity} and can be achieved using independent codewords and joint decoding.
The authors of \cite{gunduz2010capacity} identify a special hierarchy of common messages for which the capacity region
 is characterized by fewer inequalities.
This compact characterization is obtained by superimposing the common codewords over the private ones.
Although the capacity of this channel had already been established, \cite{gunduz2010capacity} is the first instance in which a coding scheme for a general
channel model is studied.
More specifically, an acyclic digraph is used in \cite{gunduz2010capacity} to describe the coding scheme for any given channel: nodes in the graph represent codewords while edges specify superposition coding among codewords.
%
%

The prior work in \cite{gunduz2010capacity} presents a unified approach to determining achievable rates in MACs with common information
using superposition coding for a specific hierarchy of the common messages.
A systematic approach to the analysis of general achievable schemes employing superposition coding is also alluded to in \cite{el2011network}, where tables are utilized to derive the error events for such transmission schemes.
Even if a general procedure is not explicitly detailed, \cite{el2011network} suggests a systematic derivation of the achievable rate regions.
An attempt to generalize the derivation of achievable regions using binning is provided in \cite{khosravi2011interference}, but no
closed-form characterization of the achievable rate is provided.

A different approach to the study of general achievable regions for multi-terminal channels is represented by the concept of ``multi-cast regions'' in \cite{grokop2008fundamental} and of ``latent capacity'' in \cite{tian2011latent}.
In a multi-cast channel, an achievable region can be obtained from another by shifting information from common rates to private rates and vice versa.
Accordingly, an attainable region can be enlarged by taking the union over all possible such manipulations, which corresponds to linear transformations of the original region.
%
%
The resulting region has a natural polyhedral description and an interesting question is whether there exists
a simpler characterization of the capacity region in terms of those achievable points that cannot be obtained as linear combinations of other points, a set termed ``latent capacity''.
This question has been partially answered in the positive only for a few channels: in \cite{tian2011latent} the latent capacity region for the 3-users symmetric broadcast channel is characterized while, in  \cite{salimi2014polyhedral}, this result is extended to a general $k$ user symmetrical broadcast channel.

%
%
%

The numerical computation of an achievable rate region based on the coding strategies represented by the chain graph entails the derivation of a large number of linear rate bounds involving mutual information terms.
 This computation for specific topologies has been developed in our prior works \cite{rini2012combining,rini2013interference} to improve upon the best-known achievable rate regions for the 2-user Gaussian CIFC and for the 2-user IFC with common messages, respectively, in the latter case achieving the capacity region for that channel.
 In addition, our work \cite{rini2013rate} provides the computation of achievable rate regions based on our framework for a topology not previously studied, that of a broadcast transmitter assisted by any number of (wired) relays sending information to an arbitrary number of users.
 These earlier works demonstrate numerically that the proposed unified graphical approach can both improve upon existing achievable rate regions and derive new results for complex topologies whose achievable rate regions would be otherwise computationally-prohibitive to obtain.
 More details on the numerical computation of achievable rate regions, using the channels from these prior works as specific examples, are provided in Sec. \ref{sec:ConcludingRemark}.

\subsection{Paper organization:}
The remainder of the paper will provide the necessary background material, describe our system model, summarize our contributions, and develop the unified graphical approach to random coding, as follows:
Section \ref{sec:schemes} introduces the coding strategies for single-hop networks and our contributions.
Section~\ref{sec:Network Model} presents the network model. 
Section~\ref{sec:User Virtualization} introduces the user-virtualization procedure.
In Section~\ref{sec:The chain graph representation of an achievable region} we introduce a general achievable scheme which utilizes graphs to represent coding operations; in Section~\ref{eq:GMM associated with the CGRAS} these graphs are associated with a graphical Makov model to represent the distribution of the codewords in the codebook.
%
Section~\ref{sec:Codebook construction, encoding and decoding operations} details the construction of the codebook, the encoding and the decoding associated with any given strategy.
Section~\ref{sec:The Achievable Rate Region of the CGRAS} derives the rate bounds that define the achievable rate region based on the proposed graph representation.
%
Section \ref{sec:ConcludingRemark} illustrates the application of the proposed framework to improve on existing achievable regions for canonical channels and to derive achievable regions for topologies not previously studied.
Finally Section~\ref{sec:Conclusion} concludes the paper.

%% file: BackgroundAndContributions.tex
\section{Random Coding Strategies and Summary of Approach}
\label{sec:schemes}

In this section, we first review the random coding strategies widely used in studying the capacity of single-hop networks that will be part of our unified approach.
Following this, we summarize the steps in our general approach to the derivation of achievable rate regions based on random coding for  one hop multi-terminal networks.

\subsection{Random coding strategies for single-hop networks}
\label{sec:Coding strategies for single-hop networks}
Combinations  of the following random coding techniques have been widely used in the literature on capacity and achievable rates for single-hop networks:
rate-splitting, superposition coding,
joint binning,  and coded time-sharing.
%
%
\begin{itemize}

\item
{\bf Rate-splitting} was originally introduced by Han and Kobayashi in deriving an achievable region for the IFC \cite{Han_Kobayashi81}:
%
it consists of dividing the network message into multiple sub-messages which are associated with different virtual sub-users.
%
%
The rate of the original message is preserved when each sub-messages is encoded  by a (possibly) smaller set of encoders than the original message and decoded by a (possibly) larger set of receivers.
%
%
%
In the classical achievable scheme of \cite{Han_Kobayashi81}, the message of each user is divided into a private and a common part: the private part
is decoded only at the intended receiver while the common part is decoded by both receivers.
%
%

\item
{\bf Superposition coding} was first introduced by Cover in~\cite{CoverCapacityDetBC} for the degraded BC and intuitively consists of
``stacking'' the codebook of one user over the codebook of another.
Destinations in the channel decode (some of the) codewords starting from the bottom of the stack, while treating the remaining codewords as noise.
%
%
This strategy achieves capacity in a number of channels, such as the degraded BC \cite{cover1972broadcast},
 the MAC with common messages \cite{slepian1973coding} and the IFC in the
 ``very strong interference'' regime \cite{sato1981capacity,carleial}.
%
%

\item
{\bf Gel'fand-Pinsker binning}, often simply referred to as binning~\cite{GelfandPinskerClassic}, allows a transmitter to pre-code (portions of) the message against the interference experienced at the destination when this interference is known at the transmitter itself.
%
%
It  was originally  devised by Slepian and Wolf \cite{slepian1973noiseless} for distributed lossless compression, and it also achieves capacity in the Gelf'and-Pinsker (GP) problem \cite{GelfandPinskerClassic}.
%
%
Binning is used by Marton \cite{MartonBroadcastChannel} to derive the largest known achievable region for the BC and is a crucial transmission strategy in many other models, usually with some form of ``broadcast'' element, including the CIFC~\cite{RTDjournal1}.

%

\item
{\bf Coded time-sharing} was also proposed by Han and Kobayashi \cite{Han_Kobayashi81} in their derivation of an achievable region for
the IFC.
In (simple) time-sharing the transmitters uses one codebook for some fraction of the time and another codebook for the remaining fraction of the time.
Coded time-sharing extends (simple) time-sharing and consists of choosing a specific transmission codebook according to a random sequence.
%
Coded time-sharing generalizes TDM/FDM strategies and potentially improves upon the convex hull of the achievable rates attained by each strategy \cite{chong2008han}.
\end{itemize}

\smallskip

Although many other encoding strategies have been proposed in the literature, the relatively simple strategies described above are sufficient to achieve capacity for a large number of memoryless, single-hop channels with no feedback or cooperation.
For this reason we focus on these basic ingredients and consider a general achievable scheme which can be obtained with any combination of them.

\bigskip

Capacity-approaching transmission strategies which are not considered in our framework are mainly strategies for multi-hop channels and structured codes such as lattice codes.
In particular, strategies such as decode-and-forward \cite{cover1979capacity},  partial-decode-and-forward \cite{kramer2006cooperative}, and compute-and-forward \cite{nazer2011compute}
are useful in a multi-hop scenario, in which the intermediate nodes need to code in a causal fashion.
These strategies are also relevant for channels with causal transmitter or receiver cooperation, that is, channels in which transmitters or receivers can communicate
directly with each other.
%
%
These transmission strategies are based on random coding, as binning and superposition coding are, but the decoding error analysis is fundamentally different from these single-hop strategies. Therefore an extension in this direction would likely not lead to elegant and compact expressions as we obtain with single-hop strategies.
Similarly for channels with feedback, achievable regions must efficiently introduce dependency between channel inputs and past channel outputs and the analysis of such schemes is far from straightforward.
%

Another class of strategies which we do not consider are lattice codes \cite{zamir2002nested}, which can be stacked and nested to form
structured transmission strategies.
These schemes are especially useful in additive, symmetric channels since the sum of two codewords is still a codeword. This makes it possible to decode
the sum of multiple interference signals as if they were produced by a sole interferer \cite{etkin2009degrees}.
Also in this case, although extensions of our framework to include these transmission strategies are possible, such generalizations are not pursued here.

\subsection{Summary of approach}
\label{sec:Contributions}

\begin{figure}
\centering
\includegraphics[width= 1 \textwidth ]{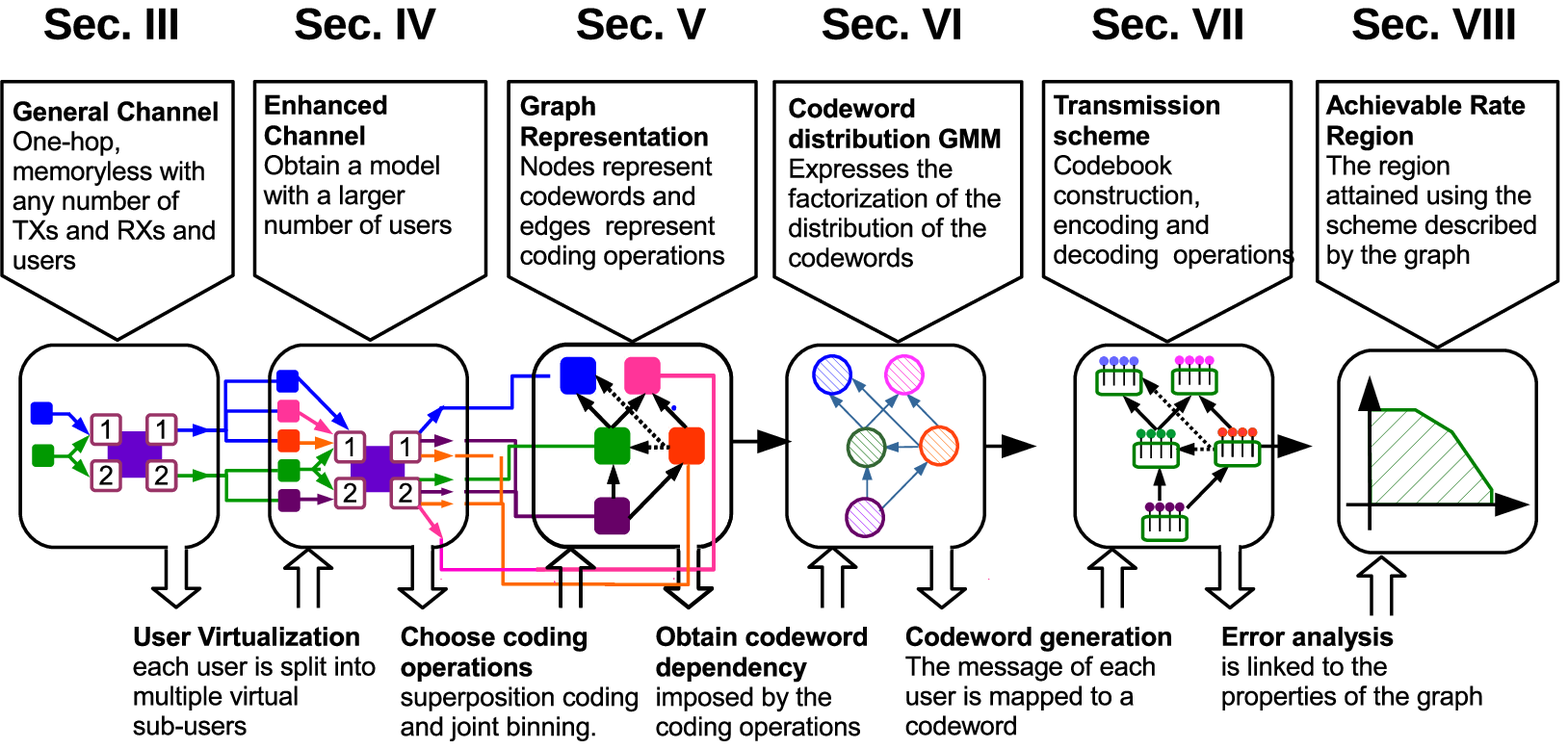}
\vspace{- 5 cm}
\caption{A conceptual representation of our approach.}
\label{fig:contributionRecap}
\vspace{-.5 cm}
\end{figure}

We summarize our approach to unified random coding in Fig. \ref{fig:contributionRecap} and as follows:

\begin{itemize}

\item
{\bf Step 1, Sec. \ref{sec:Network Model}:  Network model. \\}
We introduce a general formalism to describe one-hop memoryless network with any number of transmitters, receivers and any number of private and common messages.

\item
{\bf Step 2, Sec. \ref{sec:User Virtualization}:  User virtualization. \\}
User virtualization, which consists of splitting users into multiple virtual sub-users, was first introduced by Han and Kobayashi in \cite{han1979capacity} when studying the
capacity of the InterFerence Channel (IFC).
%
%
In our approach user virtualization generalizes the approach in \cite{han1979capacity} by allowing for a broader mapping of messages between original users and virtual users
and can be systematically employed to produce a channel model with a larger number of users.
An achievable region for this enhanced channel can then be projected back to the original channel through a rate-splitting strategy that preserves
the rates of the users.

\item
{\bf Step 3, Sec. \ref{sec:The chain graph representation of an achievable region}: Graph representation of the achievable scheme.\\}
This new formalism provides a simple unified framework to represent achievable  schemes based on  coded time-sharing, rate-splitting, superposition coding and  joint binning.
It also offers a compact description of the codebook generation, as well as  encoding and decoding procedures.

\item
{\bf Step 4, Sec.  \ref{eq:GMM associated with the CGRAS}: Express the codeword joint distribution  through a GMM.\\}
The proposed graph representation also describes the factorization of the distribution of the codewords
in the codebook. The GMM embeds the conditions upon which dependency can be established through graph properties such as cycles and connected sets.

\item {\bf Step 5, Sec. \ref{sec:Codebook construction, encoding and decoding operations}: Describe the codebook construction, encoding and decoding operations. \\}
%
The GMM can also be used to describe how the codebook to transmit a message can be generated using random, iid draws.  A codebook to transmit each message is generated using the superposition coding steps in the graph. After the codebook has been generated, binning is used to select the codewords for transmission.

\item {\bf Step 6, Sec. \ref{sec:The Achievable Rate Region of the CGRAS}: Show vanishing error probability and obtain the achievable region.\\}
The probability distribution expressed by the GMM describes the joint distribution among codewords: an error is committed when the incorrect codeword appears to have the correct joint distribution with the remaining transmitted codewords.
As the rate of a codeword increases, this event is increasingly likely and the covering and packing lemma \cite{el2011network} can be used to derive the highest rate for which the probability of incorrectly decoding a codeword is vanishing with the block-length.
The set of conditions that grants correct decoding correspond to the achievable region.
\end{itemize}

For the last step, we  shall consider three classes of coding schemes with increasing complexity and, in each scenario, derive the achievable rate region in terms of the structure of the graph representing the coding operations.
In particular, we first consider schemes with only superposition coding. Next we include binning and finally we consider the most general case which includes superposition coding, binning and  joint binning.

\begin{figure}
\centering
\includegraphics[width= 0.75 \textwidth ]{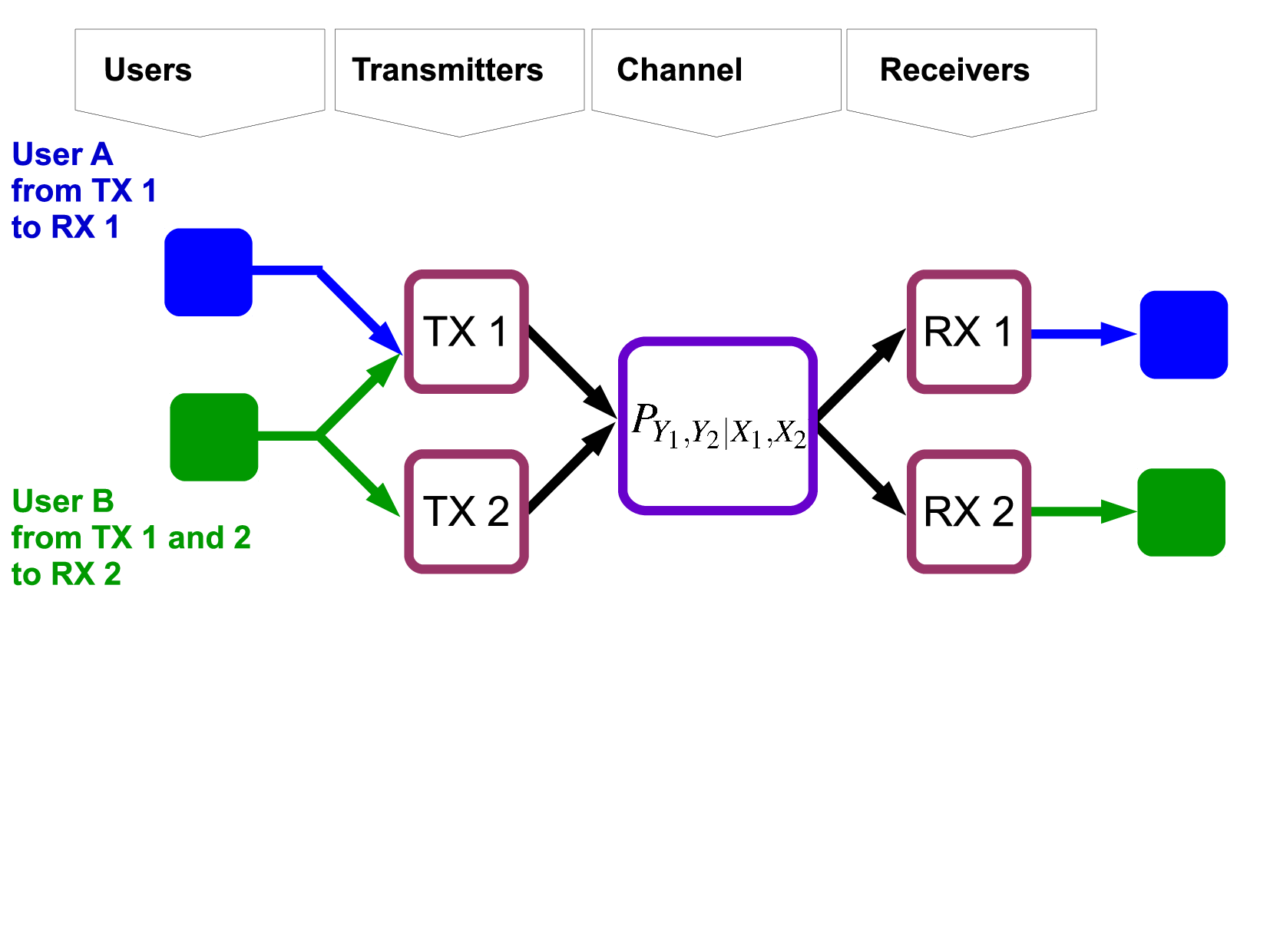}
\vspace{- 3 cm}
\caption{A conceptual representation of the communication system under consideration.}
\label{fig:channelModelIntro}
\end{figure}

A conceptual representation of the communication system under consideration and of our approach is provided in Fig. \ref{fig:channelModelIntro}:
we consider any memoryless, one-hop channel with any number of transmitters and receivers and without feedback or cooperation. %
We additionally allow a message to be provided to multiple transmitters and decoded at multiple receivers.
%
We refer to the set of transmitters encoding a message together with the set of receivers decoding the message as a ``user''.
%

%

%% file: NetworkModel.tex
\section{Network model}
\label{sec:Network Model}
We consider a general one-hop multi-terminal network with any number of transmitters and receivers.
%
The network is assumed to be memoryless and without feedback or causal cooperation among transmitters or receivers.
We consider a channel model in which messages can be encoded by multiple transmitters and decoded by multiple receivers.
This is a more general model than the channel in which each message is encoded at one transmitter and decoded at one receiver and it combines aspects of the BC,
the MAC and the IFC.
Additionally, in this general framework, splitting users into multiple virtual sub-users results in an enhanced channel which is still in the class of channels
under consideration.
%

\bigskip

More specifically, we consider a one-hop network in which $N_{\rm TX}$ transmitting nodes want to communicate with $N_{\rm RX}$ receiving nodes.
The encoding node $k \in [1\ldots N_{\rm TX}]$ has input $X_k$ to the channel while the decoding node $z \in [1 \ldots N_{\rm RX}]$
receives the channel output $Y_z$.
The channel is assumed to be memoryless with transition probability
\ea{
P_{\Yv|\Xv}=P_{Y_1  \ldots  Y_{N_{\rm RX}}|X_1  \ldots X_{N_{\rm TX}}}.
\label{eq:channel transition prob}
}
The subset of transmitting nodes $\iv$ is interested in reliably communicating the message
$W_{\iv \sgoes \jv}$ to the subset of receiving nodes $\jv$ over $N$ channel uses.
The message $W_{\iv \sgoes \jv}$, is uniformly distributed in the interval  $[1  \ldots 2^{N R_{\iv \sgoes \jv}}]$, where $N$ is the block-length and $R_{\iv \sgoes \jv}$ the message rate.
Each receiver $z \in \jv$ produces the estimate $\Wh_{\iv \sgoes \jv}^z$ of the transmitted message $W_{\iv \sgoes \jv}$.
The subset of transmitters $\iv$  and the subset of receivers $\jv$ are arbitrary but not empty.
The allocation of multiple messages $W_{\iv \sgoes \jv}$ between subsets of transmitters and subsets of receivers is defined by
\ea{
W_{\Vv}=\{ W_{\iv \sgoes \jv}, \ (\iv,\jv) \in \Vv\},
\label{eq:definition messages}
}
where $\Vv$
is any collection of arbitrary (non-empty) subsets of $[1\ldots N_{\rm TX}] \times [1\ldots N_{\rm RX}]$.
%

A rate vector $R_{\Vv}=\lcb R_{\iv \sgoes \jv},  \ (\iv,\jv) \in  \Vv \rcb$ is said to be achievable if, for all $(\iv,\jv) \in \Vv$,  there exists a sequence of encoding functions
\ea{
X_k^N = X_k^N
\lb \lcb
W_{\iv \sgoes \jv},  \ \forall \ (\iv, \jv) \in \Vv \ \ST \ k \in \iv \rcb \rb,
\label{eq:encoding function}
}
and a sequence of decoding  functions
\ea{
\widehat{W}_{\iv \sgoes \jv}^z =\widehat{W}_{\iv \sgoes \jv}^z \Big(Y_z^N\Big), \  \  \forall \ (\iv,\jv) \in \Vv \ \ST \ z  \in  \jv, \
\label{eq:decoding function}
}
such that
\ea{
\lim_{N \to \infty } \max_{z,\iv,\jv} \Pr\lsb  \widehat{W}_{\iv \sgoes \jv}^z \neq W_{\iv \sgoes \jv}  \rsb = 0.
}
The capacity region $\Cc(R_{\Vv})$ is the convex closure of the region of all achievable rates in the vector $R_{\Vv}$.

\begin{figure}
\centering
\includegraphics[width=.9 \textwidth ]{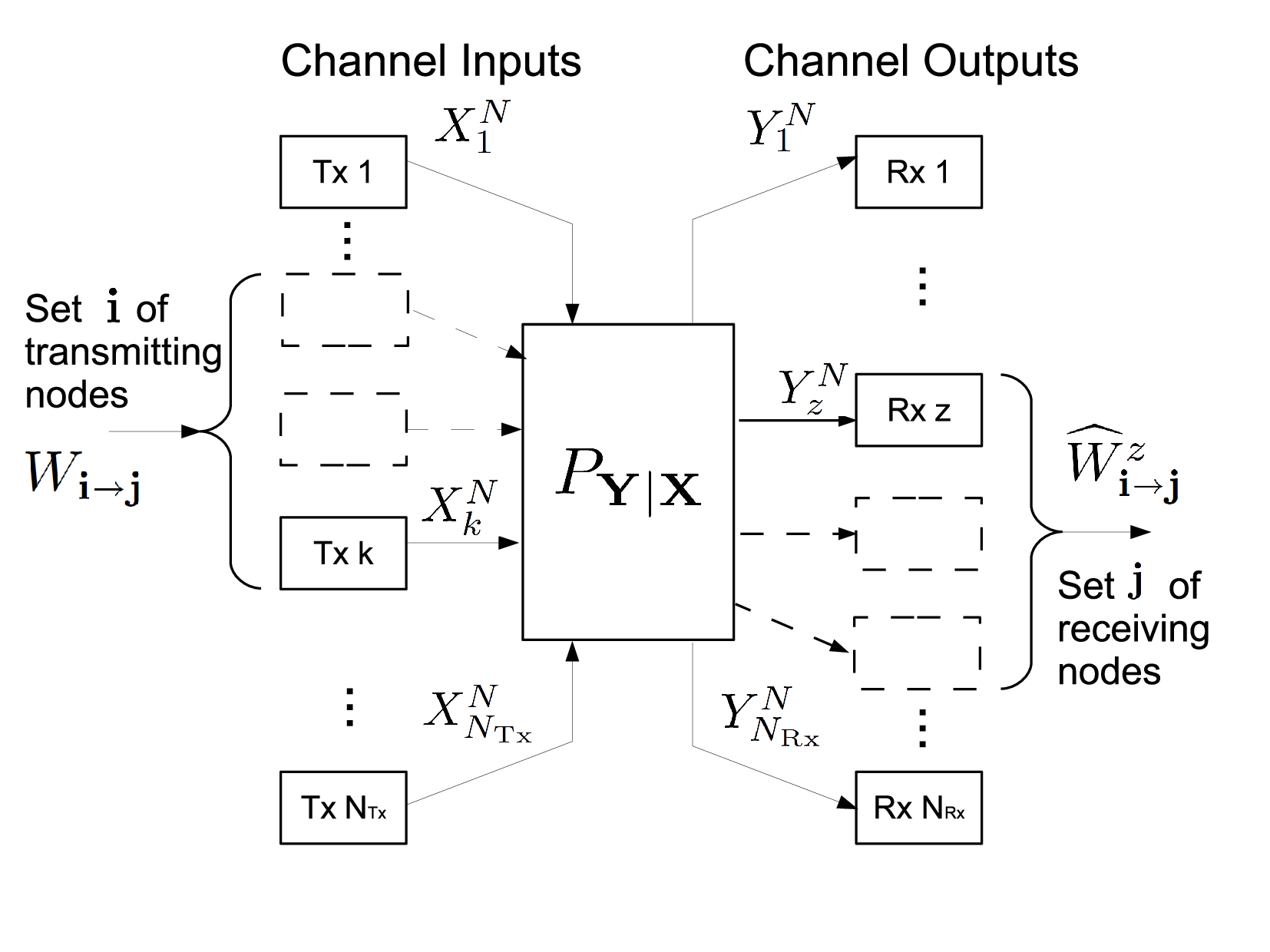}
\vspace{- 1 cm}
\caption{The general memoryless, one-hop multi-terminal network in Sec. \ref{sec:Network Model}.}
\label{fig:channelModel}
\vspace{-.5 cm}
\end{figure}

The channel model under consideration is depicted in Fig. \ref{fig:channelModel}: on the left side are the $N_{\rm TX}$
transmitting nodes while on the right are the $N_{\rm RX}$ receiving nodes.
A message $W_{\iv \sgoes \jv}$ is encoded by the set $\iv$ of transmitting nodes and decoded at the set $\jv$ of receiving nodes.
The channel input $X_k^N$ at each encoding node $k$ is obtained as a function of the messages available at this encoder according to \eqref{eq:encoding function}.
Receiver $z$ produces the estimate $W_{\iv \sgoes \jv}^z$  for all the messages $W_{\iv \sgoes \jv}$ such that $z \in \jv$ from the channel output $Y_z^N$ using the decoding function in \eqref{eq:decoding function}.
The channel under consideration is a variation of the network model in Cover and Thomas \cite[Ch. 15.10]{ThomasCoverBook}, but allows for messages to be allocated
to multiple users while not considering feedback and causal cooperation, that is, each node is either a transmitting nodes or a receiving nodes but not both.

\subsection{An example: the general two-user interference channel}
\label{sec:A brief example: the general interference channel}

\begin{figure}
\centering
\includegraphics[width=15 cm]{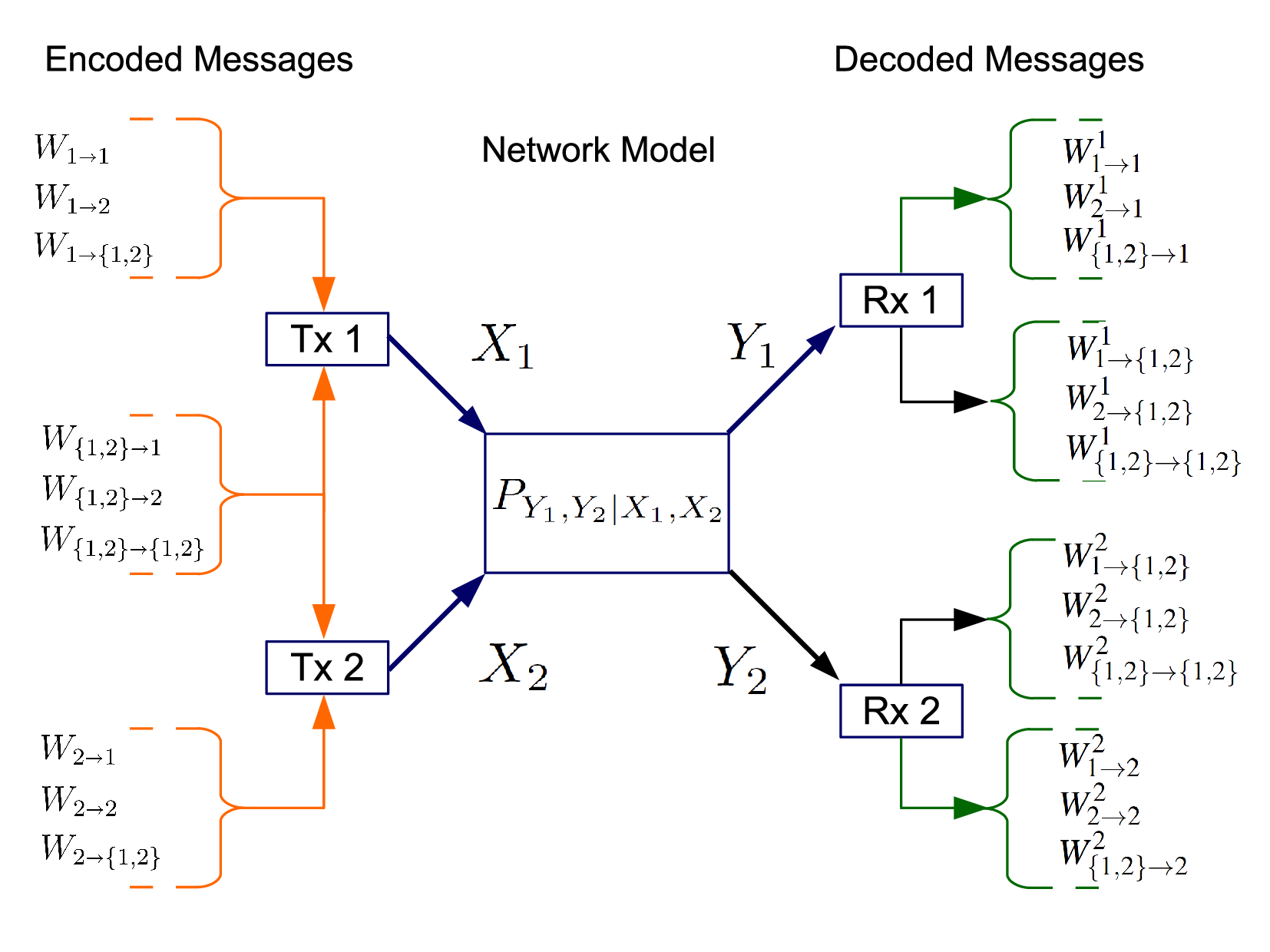}
\caption{The general IFC with the most general set of messages  to be exchanged amongst transmitters and receivers.}
\label{fig:generalIFC}
\end{figure}

To demonstrate the generality of our model, in this section we provide an example based on the general two-user IFC.
The channel model is depicted in Fig. \ref{fig:generalIFC}: two transmitter/receiver pairs ($N_{\rm TX}=N_{\rm RX}=2$) communicate through the memoryless channel $P_{Y_1,Y_2 | X_1, X_2}$.
The largest number of messages that can be sent over the channel is nine and is obtained by considering
all the possible ways in which a message can be encoded by a subset of transmitters and decoded by a subset of receivers.
Note that the IFC with all nine messages is not necessarily of interest to study in depth, and we will not consider it further in this paper;
the purpose of Fig. 4 is to illustrate the generality of our model to capture all possible message combinations that might be of interest in a given one-hop network.

 Tab. \ref{tab:example4nodes}: each column indicates the set of encoding nodes while each row a set of decoding nodes.
The messages $W_{1 \sgoes \jv}$ and $W_{2 \sgoes \jv}$  are the messages known only at transmitter~1 and~2, respectively, while message $W_{\{1,2\} \sgoes \jv}$ is a message known at both.
Similarly, $W_{\iv \sgoes 1}$ and $W_{\iv \sgoes 2}$  are the messages to be decoded only at receivers~1 and~2, while
the messages $W_{ \iv \sgoes \{1,2\} }$ are to be decoded at both.
\begin{table}
\centering
\caption{The messages for a general IFC.}
\label{tab:example4nodes}
\begin{tabular}{|l|lll|}
\hline
    & from Tx1 & from Tx2 & from Tx1 \& Tx2 \\
\hline
to Rx1        & $W_{1 \sgoes 1}$   & $W_{2 \sgoes  1 }$   & $W_{\{1,2\} \sgoes  1}$    \\
to Rx2        & $W_{1 \sgoes 2}$   & $W_{ 2 \sgoes  2 }$   & $W_{\{1,2\} \sgoes  2 }$    \\
to Rx1 \& Rx2  & $W_{ 1\sgoes  \{1,2\}}$ & $W_{2 \sgoes  \{1,2\}}$ & $W_{\{1,2\} \sgoes  \{1,2\}}$  \\
\hline
\end{tabular}
\end{table}
The general IFC encompasses a number of canonical channel models with and without common messages  as special cases including the BC, the MAC, the IFC and the CIFC  both with and without common messages.
Tab. \ref{tab:subcases4nodes} lists all special cases of the general two-users IFC that have been studied in the literature and the associated reference.
Note that in each such case a different proof was used to establish the achievability of the derived rate region.

\begin{table}
\centering
\caption {Specific subcases of the general interference channel}
\label{tab:subcases4nodes}
\begin{tabular}{|l|l|l|l|}
\hline
subcase & channel model & reference  \\
\hline
$\Cc\lb   R_{1 \sgoes 1}   \rb$   & point-to-point   & \cite{shannon48} \\
$\Cc\lb   R_{1 \sgoes 1},R_{2 \sgoes  1}    \rb$ & MAC &  \cite{ahlswede1971multi}  \\
$\Cc\lb   R_{1 \sgoes 1},R_{2 \sgoes  1},R_{ \{1,2\} \sgoes  1}     \rb$ & MAC with common message & \cite{cover1980multiple} \\
$\Cc\lb   R_{1 \sgoes 1},R_{1 \sgoes  2}     \rb$ & BC  & \cite{cover1972broadcast} \\
$\Cc\lb   R_{\{1,2\} \sgoes 1},R_{\{1,2\} \sgoes  2}     \rb$ & BC & \cite{cover1972broadcast}  \\
$\Cc\lb   R_{1 \sgoes 1},R_{1 \sgoes  2},R_{1 \sgoes  \{1,2\}}     \rb$ & BC with degraded message set  & \cite{kiirner1977general} \\
$\Cc\lb   R_{1 \sgoes 1},R_{2 \sgoes  2}     \rb$ & IFC & \cite{shannon1961two}\\
$\Cc\lb   R_{1 \sgoes 1},R_{2 \sgoes  2},R_{\{1,2\} \sgoes  \{1,2\}}     \rb$ & IFC with common information &  \cite{maric2005capacity}\\
$\Cc\lb   R_{1 \sgoes 1},R_{ \{1,2\} \sgoes  2}     \rb $ & CIFC & \cite{devroye_IEEE} \\
$\Cc\lb   R_{1 \sgoes 1},R_{\{1,2\} \sgoes  \{1,2\}}    \rb$ & CIFC with degraded message set & \cite{liang2009capacity} \\
$\Cc\lb   R_{1 \sgoes  \{1,2\}},R_{2 \sgoes  \{1,2\}}   \rb$ & compound MAC  & \cite{maric2005discrete} \\
$\Cc\lb   R_{1 \sgoes  \{1,2\}},R_{\{1,2\} \sgoes  \{1,2\}}   \rb$ & compound CIFC  & \cite{maric2005discrete} \\
\cline{1-3}
\end{tabular}
\end{table}


Some of the subcases of the general IFC have never been considered in the literature.
For instance the capacity $\Ccal(R_{1 \sgoes 1}, R_{2 \sgoes \{1,2\}})$ has never been investigated,
as well as $\Ccal(R_{1 \sgoes 1}, R_{2 \sgoes \{1,2\}},R_{2 \sgoes 2})$ and $\Ccal(R_{1 \sgoes 1}, R_{2 \sgoes \{1,2\}}, R_{2 \sgoes \{1,2\}})$  and
 many others models obtained by considering combinations of the messages in Tab. \ref{tab:example4nodes}.
%
%
Our approach allows  achievable rate regions for all subcases of the IFC, including those in Tab. \ref{tab:subcases4nodes}   and those not previously studied, to be derived in a unified manner.

%% file: UserVrtualization.tex
\section{User virtualization}
\label{sec:User Virtualization}
User virtualization consists of splitting users into multiple, virtual sub-users to produce an enhanced channel with a larger number of users.
This is obtained by splitting the message of each user into multiple sub-messages through rate-splitting, which guarantees that the rate of
the messages in the original channel is preserved in the enhanced model.
Moreover, since encoding capabilities  and decoding requirements in the original channel cannot be violated, a sub-message in the enhanced model can only be encoded by a smaller set of transmitters than the original message and decoded by a larger set of receivers.
%
\medskip

Having part of a message decoded at one or more receivers is a useful interference management strategy which arises naturally in many channel models.
In wireless systems, the transmissions of one user create interference at multiple receivers: by decoding part of the interfering signal, a receiver can cancel its effects on the intended signal.
The information decoded at multiple decoders is sometimes referred to as ``common information'', since it is shared by multiple receivers.
Restricting the set of nodes transmitting a message is another simple strategy to manage interference: when multiple encoders have knowledge of the same message, the node which creates the least amount of interference on the neighbouring users can be selected for transmission.
%

\medskip

After rate-splitting is applied, the sum of the rate of all the sub-messages must equal the rate of the original message: this guarantees that the same amount of information is being sent over the channel in the original and the enhanced model.
This requirement implies that the rate of each sub-message can be chosen in a number of ways, as long as the sum of their rates stays constant.
In other words, an achievable rate point in the original channel corresponds to a number of points in the enhanced model: we refer to this one-to-many mapping of the rate points as \emph{rate-sharing}, since the rate of one original user can be shared among all its virtual sub-users.

More specifically, user virtualization can be expressed through the \emph{user virtualization matrix} $\Gamma$, so that
\ea{
R_{\Vv^{O}} = \Gamma R_{\Vv},
\label{eq:projection}
}
where $\Vv^{\rm O}$ (\emph{O} for \emph{original}) is the original message allocation and $\Vv$ is the message allocation in the enhanced channel
\footnote{We use here the same notation $\Vv$ as in Sec. \ref{sec:Some graph-theoretic notions} since we later associate the message set $\Vv$ to a graph $\Gcal(\Vv,\Ev)$ in which nodes are codewords embedding the messages in the network. }.
Each term  in $\Gamma$
\ea{
\Gamma_{(\iv,\jv) \times (\lv,\mv) }, \quad \quad (\iv,\jv) \in \Vv^{\rm O}, \  (\lv,\mv) \in \Vv,
\label{eq:gamma definition}
}
indicates the portion of the message $W_{\iv \sgoes \jv}, \ (\iv,\jv) \in \Vv^{\rm O}$ in the original allocation that is embedded in the
message   $W_{\lv \sgoes \mv}, \ (\lv,\mv) \in \Vv$ in the enhanced channel.
%
Since encoding capabilities  and decoding requirements cannot be violated,
a message $W_{\iv \sgoes \jv}$ can be split into the messages $W_{\lv \sgoes \mv}$
only when  $\iv \supseteq \lv$ and $\jv \subseteq \mv$, that is, the new set of messages can only be encoded by a smaller set of transmitters or decoded by a larger set of receivers.
This implies
\ea{
\Gamma_{(\iv,\jv) \times (\lv,\mv)} \neq 0 \implies  \iv \supseteq \lv, \ \jv \subseteq \mv.
\label{eq:gamma zero elements}
}
Additionally, we have the constraint
\ea{
%
\sum_{(\lv,\mv)} \Gamma_{(\iv,\jv) \times (\lv,\mv)}=1,
\label{eq:sum one gamma elements}
}
since the rates of the original channels must be preserved.

Note that \eqref{eq:sum one gamma elements} implies that multiple (parts of) messages in $W_{\iv \sgoes \jv}, \ (\iv,\jv) \in \Vv^{\rm O}$ can be compounded
to form a single message $W_{\lv \sgoes \mv}, \ (\lv,\mv) \in \Vv$ in the enhanced channel.
This compounding of messages is rarely found in classical channel models, with the exception of the channel in \cite{rini2011capacityIFC-CR}.
In \cite{rini2011capacityIFC-CR}, an achievable rate region for the interference channel with a cognitive relay (IFC-CR) is derived: this channel is a variation of the classical IFC with an additional relay that has full, a priori knowledge of the messages of both users.
In this achievable strategy, the cognitive relay sends a common codeword which embeds part of the message of each user and which is decoded at both receivers.
This common codeword can be seen, in the formulation of \eqref{eq:projection}, as embedding two public sub-users in the IFC to one single common message transmitted by the cognitive relay.
%
%

\subsubsection{An example of  rate-splitting}
\label{sec:An example of  rate-splitting}
As an example of rates-splitting, consider the classical CIFC, equivalently indicated as $C \lb R_{1 \sgoes 1},  R_{ \{1,2\} \sgoes 2} \rb$: with rate-splitting we can  transform the problem of achieving the rate vector
\ea{
R_{\Vv^{\rm O}}=[\widehat{R}_{1 \sgoes 1} \ \Rh_{ \{1,2\} \sgoes 2}],
}
into the problem of achieving the rate vector
\ean{
R_{\Vv}=
[ R_{1 \sgoes  1} \ R_{1 \sgoes  2} \ R_{1 \sgoes  \{1,2\}} \  R_{2 \sgoes  2} \ R_{\{1,2\} \sgoes 2} \ R_{\{1,2\} \sgoes \{1,2\} }],
}
where the two vectors are related through the user virtualization matrix
\ea{
& \lsb
\p{
\Rh_{1 \sgoes  1}\\
\Rh_{\{1,2\} \sgoes  2}
}
\rsb
= \nonumber \\
& \quad \lsb
\p{
\Gamma_{1\sgoes 1 \times     1\sgoes 1} & \Gamma_{1\sgoes 1 \times 1\sgoes \{1,2\} } & 0 & 0 & 0 &  \\
0 & 0 &  \Gamma_{\{1,2\}\sgoes 2 \times  1\sgoes 2}&  \Gamma_{\{1,2\}\sgoes 2 \times \{1,2\} \sgoes  2} & \Gamma_{\{1,2\}\sgoes 2 \times  \{1,2\}\sgoes \{1,2\}}
%
}
\rsb
\cdot \nonumber \\
& \quad \quad \quad \quad \lsb \p{
R_{1 \sgoes  1} \
R_{1 \sgoes  \{1,2\}}  \
R_{1 \sgoes  2}  \
R_{\{1,2\} \sgoes 2}  \
R_{\{1,2\} \sgoes \{1,2\} }  \
}\rsb^{\rm T}.
\label{eq:rate-splitting CIFC}
}
Given the constraint in \eqref{eq:sum one gamma elements}, we have that all the non-zero elements of $\Gamma$ are equal to one and therefore \eqref{eq:projection}
reduces to
\ea{
\Rh_{1 \sgoes  1}       &= R_{1\sgoes 1} + R_{1\sgoes \{1,2\}} \nonumber \\
\Rh_{\{1,2\} \sgoes  2} &= R_{ 1\sgoes 2} +  R_{ \{1,2\} \sgoes  2} + R_{\{1,2\}\sgoes \{1,2\}}.
\label{eq:user virtualization CIFC}
}

\begin{figure}
\centering
\includegraphics[width=.8 \textwidth]{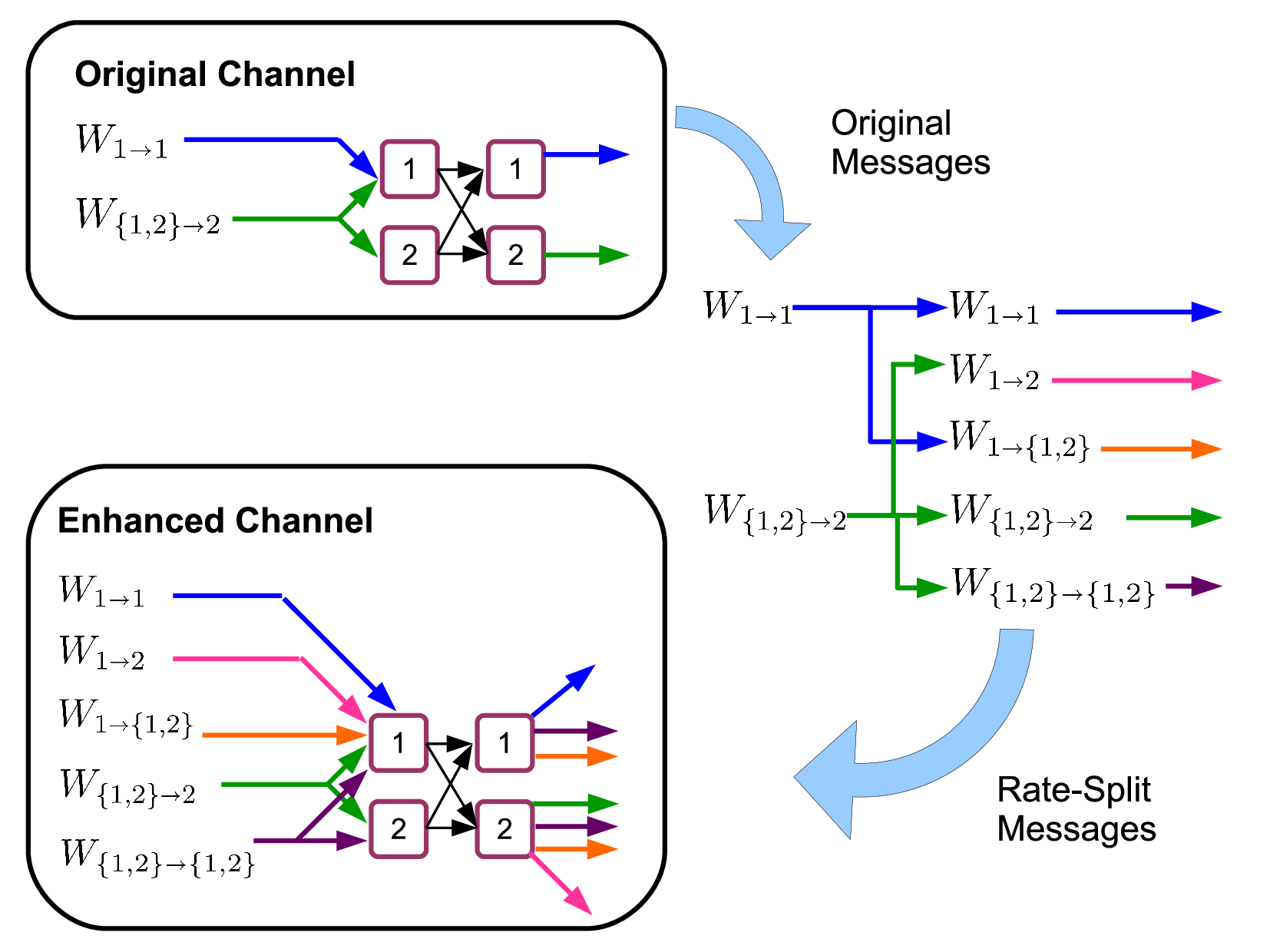}
\caption{A schematic representation of the rate-splitting example in Sec. \ref{sec:An example of  rate-splitting}.}
\label{fig:RateSplitFig}
\end{figure}

A graphical representation of this example is provided in Fig. \ref{fig:RateSplitFig}: on top of the figure is the original channel
$\Ccal(R_{1 \sgoes 1}, R_{\{1,2\} \sgoes 2})$ from the conceptual model in Fig. \ref{fig:channelModelIntro} while on the bottom is the channel after rate-splitting
\ean{
\Ccal(R_{1 \sgoes 1}, R_{1 \sgoes 2}, R_{1 \sgoes \{1,2\}} ,R_{\{1,2\} \sgoes 2},R_{\{1,2\} \sgoes 2}).
}
On the right of the figure is the mapping between the original channel and the rate-split channel.

\subsection{An example with rate-sharing}

The matrix $\Gamma$ effectively describes the mapping between the achievable points in $\Ccal(\Vv^{\rm O})$ and the achievable points in $\Ccal(\Vv)$.
%
%
%
When rate-sharing is not applied, there exists only one matrix $\Gamma$ which maps $\Vv^{\rm O}$ into $\Vv$ and this matrix
is binary, due to \eqref{eq:sum one gamma elements}.
When rate-sharing is applied, instead, $\Gamma$ is no longer unique and multiple matrices can be used to map $\Vv^{\rm O}$ into $\Vv$.
This implies that correspondence between the rate vectors $R_{\Vv^{O}}$ and the rate vectors  $R_{\Vv}$ in \eqref{eq:projection} is now a one-to-many correspondence in which the same vector $R_{\Vv^{O}}$ is obtained from multiple vectors $R_{\Vv}$ through different rate-splitting matrices.

%

Consider, for instance, the broadcast channel with a common message  \cite{liang2007rate} $\Ccal(R_{1 \sgoes 1}, R_{1 \sgoes 2}, R_{1 \sgoes \{1,2\}})$:
in this channel part of the private messages $W_{1\sgoes 1}$ and $W_{1 \sgoes 2}$ in the original channel
can be compounded with the common messages $W_{1 \sgoes \{1,2\}}$ in the enhanced channel.
More specifically, \eqref{eq:projection} takes the form
\ea{
\lsb \p{
\Rh_{1 \sgoes 1} \\
\Rh_{1 \sgoes 2} \\
\Rh_{1 \sgoes \{1,2\}}
}
\rsb
=
\lsb
\p{
1          & 0          & \Delta_1 \\
0          & 1         & \Delta_2 \\
0          &           0& 1- \Delta_1-\Delta_2
}
\rsb
\cdot
\lsb
\p{
R_{1 \sgoes 1} \\
R_{1 \sgoes 2} \\
R_{1 \sgoes \{1,2\}}
}
\rsb,
\label{eq:projection 2}
}
for any $\Delta_1,\Delta_2$ such that $\Delta_1+\Delta_2 \leq 1$.
Intuitively,  $\Delta_1$ is the part of $W_{1\sgoes 1}$ in the original channel embedded in $W_{1\sgoes \{1,2\}}$ in the enhanced channel and similarly for $\Delta_2$.
Given that $\Gamma$ is not unique, an achievable region $R_{\Vv}$  in the enhanced channel can be translated into the original problem
by considering the union over all the user virtualization matrices.
%

%% file: ChainGraphRepresentation.tex
\section{The chain graph representation of an achievable scheme}
\label{sec:The chain graph representation of an achievable region}

In this section we introduce a graph to represent a general transmission scheme involving superposition coding
and binning.
Given a channel model as described in Sec. \ref{sec:Network Model}, virtual sub-users can be created through the procedure in Sec. \ref{sec:User Virtualization}.
The resulting enhanced model has a larger number of users which is defined by the set $\Vv$ as in \eqref{eq:definition messages}.
For this enhanced model, we define a graphical representation of the coding operations by defining a graph in which each node is associated with a user in $\Vv$ in the enhanced channel.
More specifically, the set of nodes $\Vv$ in the graphical representation is the same set of users in the enhanced channel.
Two graphs are then defined over the set $\Vv$ which describes the coding operations.
A first graph, the  \emph{superposition coding graph}, $\Gcal(\Vv,\Sv)$, describes how superposition coding is applied to generate the codebook of each user.
Once the codebook to transmit each message has been generated, a second graph, the \emph{binning graph} $\Gcal(\Vv,\Bv)$, describes how binning is used to select the codewords in each codebook to encode a specific message set.
In superposition coding dependency among codewords is established by creating the codewords in the top codebook to be conditionally dependent on the bottom codebook.
In binning, dependency among codewords is established by looking for two codewords which belong to a jointly typical set, although generated conditionally independent.
For this reason, it is necessary first to create the codebook according to the superposition coding graph, then select conditionally typical codewords from the codebook according to the binning graph.

\subsection{Graph theory and chain graphs}
In the following we will assume some basic definitions and properties of graphs and graphical Markov
models.
For the reader's convenience, all such definitions and properties have been summarized in App. \ref{app:Graphical Markov Models} and App. \ref{app:Graphical Markov models 2} .
%

Graphical Markov models are used in the remainder of the paper to describe the distribution of the codewords in the codebook, we are particularly interested in those class of models in which the associated distribution factorizes in a convenient manner.
%
%
The UnDirected Graphs (UDGs) offer a factorization in terms of cliques, which are subsets of nodes in which every two nodes are connected by an edge.
For Directed Acyclic Graphs (DAGs), we have that the graph distribution $P$ factorizes in terms of parent nodes, i.e.
\ea{
P=\prod_{\al \in V} P_{\al | \pa(\al)},
\label{eq:factorization DAGs}
}
where $\pa(\al)$ indicates the parent nodes of $\al$.
%
The graph that we wish to construct contains both directed and undirected edges, which model joint binning, and is thus a Chain Graph (CG).
In general, a convenient and recursive factorization of $P$ for CGs is not available: the only case in which such
a simple factorization exists is when the chain graph is Markov-equivalent to a DAG.
%
%
DAGs also offer a convenient factorization of the marginal and conditional distribution for chosen subsets of nodes:
let $\Fv$ be a subset of $\Vv$ and $\Fvo$ be its complement in $\Vv$, i.e. $\Fvo=\Vv \setminus \Fv$.
If $\pa(\Fv) \subseteq \Fv$, we have that
\eas{
P( \{ \al \in \Fv \}) = \prod_{\al \in \Fv} P_{\al | \pa(\al)}
\label{eq:marginal DAG general} \\
P(  \{\be \in \Fvo \} | \{ \al \in \Fv \}) = \prod_{\be \in \Fvo} P_{\be | \pa(\be)}.
\label{eq:conditional DAG general}
}{\label{eq:convenient factorization}}
%
%
%

Note that \eqref{eq:marginal DAG general} and \eqref{eq:conditional DAG general} are particularly effective  ways of describing a marginal and a conditional
distribution  for a joint distribution $P$.
%
In particular, the entropy of the distributions in \eqref{eq:marginal DAG general}  can be written as
\ea{
H \lb P( \{ \al \in \Fv \}) \rb  = \sum_{\al \in \Fv} H( \al | \pa(\al)),
\label{eq:entropy marginal}
}
while for \eqref{eq:conditional DAG general} we have
\ea{
H \lb P\lb  \{\be \in \Fvo \} | \{ \al \in \Fv \}\rb \rb  = \sum_{\be \in \Fvo} H(\be | \pa(\be)).
\label{eq:entropy conditional}
}
For this reason, we refer to distributions with factorizations as given in \eqref{eq:convenient factorization}  as ``compact'', by which we mean that they offer a representation as a product of conditional distributions of single RVs and not as a marginalization of the joint distribution.
In the following derivation, the rate bounds will be expressed as the difference between entropy terms: here the factorizations in \eqref{eq:entropy marginal} and \eqref{eq:entropy conditional} will give rise to the usual mutual information expression for the rates bounds.

\subsection{Definition}

We refer to the set $\lsb \Vv,  \Sv, \Bv \rsb$ as the Chain Graph Representation of an Achievable Scheme (CGRAS).
This representation is useful in two ways: it formalizes graphically the coding constraints and it details precisely the codebook construction.
Superposition coding and joint binning can be applied only under certain conditions. For instance, when superimposing a codeword over another codeword,
 this top codeword must also be superimposed over the codewords over which the bottom codeword is superimposed.
This can be graphically expressed by requiring that parent nodes of the bottom codeword must also be parent nodes of the top codeword.

%
%
%
%

In addition to embedding the coding constraints, the CGRAS is also used to describe the codebook construction as well as encoding and decoding procedures.
By defining a GMM over the superposition coding and joint binning graph, we can define a distribution over these graphs in which a RV is associated to each user in the enhanced channel.
From this distribution, a codebook to embed a message can be obtained by randomly generating codewords with i.i.d. draws.
More specifically, the RV $U_{\iv \sgoes \jv}$ is associated with the user $(\iv,\jv) \in \Vv$ and is used to generate the codewords $U_{\iv \sgoes \jv}^N$ of length $N$ to embed the message $W_{\iv \sgoes \jv}$.

The superposition coding graph describes the conditional dependence among codewords, since the codebook embedding a given message is created conditionally dependent on the codebook of the parent nodes.
%
If binning is also applied, multiple codewords are created to transmit the same message: after the codebook has been generated using the superposition coding graph, the binning
 graph is used to select codewords for transmission according to a chosen conditional dependence among them.

In the unifying approach to the derivation of achievable rate regions we propose here, the CGRAS provides a simple structure which captures all the details of complex transmission strategies, including all possible combinations of superposition coding, binning, and coded time-sharing for both private and common information.

\subsubsection{Superposition coding graph}
\label{sec:Superposition coding graph}

In the superposition coding graph, $\Gcal(\Vv,\Sv)$, the nodes in $\Vv$ are associated with a message in the enhanced channel $\Ccal(R_{\Vv})$ and the edges $\Sv$ are the edges associated with superposition of the codewords embedding one message over the codeword embedding another.
Superposition coding can be thought of as stacking the codebook of one user over the codebook of another user.
For each base codeword, a new top codebook is created which is conditionally dependent on the given base codeword.
%
%
%
When a codeword from the bottom codeword is selected for transmission, the top codeword is selected from this conditionally dependent codebook.

At a receiver,  a top codeword cannot be correctly decoded unless the bottom codewords are also correctly decoded, since a different top codebook is associated to each bottom codeword.

In the superposition coding graph $\Gcal(\Vv,\Sv)$, the node $(\iv,\jv) \in \Vv$ is associated with the message $W_{\iv \sgoes \jv}$ embedded in the codeword $U_{\iv \sgoes \jv}^N$ obtained through i.i.d. draws from the RV $U_{\iv \sgoes \jv}$.
An edge $(\lv,\mv) \times (\iv,\jv) \in \Sv$ indicates that the codeword $U_{\iv\sgoes \jv}^N$ is superimposed over the codeword $U_{\lv \sgoes \mv}^N$.
The superposition of $U_{\iv \sgoes \jv}^N$ over $U_{\lv \sgoes \mv}^N$ is also indicated as $U_{\lv \sgoes \mv} \spc U_{\iv \sgoes \jv}$.

%
%
Superposition of two codewords can be performed  only under some restrictions, as we now define:
\begin{cond}{\bf Superposition Coding. \\}
\label{cond:Conditions for Superposition Coding}
The superposition  of  the codeword $U_{\iv \sgoes \jv}^N$ over another codeword  $U_{\lv \sgoes \mv}^N$ can be performed when the following two conditions hold:
\begin{itemize}
  \item $\lv \subseteq \iv$:
  that is, the bottom message is encoded by a larger set of encoders than the top message,
  \item $\mv \subseteq \jv$:
  that is, the bottom message is decoded by a larger set of  decoders than the top message.
\end{itemize}
Moreover, if $U_{\iv \sgoes \jv}^N$ is superimposed over $U_{\lv \sgoes \mv}^N$ and  $U_{\lv \sgoes \mv}^N$  over $U_{\vv \sgoes \tv}^N$,
then  $U_{\iv \sgoes \jv}^N$  is also superimposed over $U_{\vv \sgoes \tv}^N$.
%
\end{cond}

Given Condition \ref{cond:Conditions for Superposition Coding}, we conclude that
\ea{
U_{\lv \sgoes \mv} \spc U_{\iv \sgoes \jv} \implies \pa_{\Sv}(U_{\lv \sgoes \mv}) \subset \pa_{\Sv}(U_{\iv \sgoes \jv}).
}
Also, given Condition \ref{cond:Conditions for Superposition Coding}, $\Gcal(\Vv,\Sv)$ must be a    DAG: an undirected edge would occur only for two nodes for
which $\iv=\vv$ and $\jv=\tv$ which is not possible. Similarly, a cycle would occur only when there exists two messages encoded and decoded by the same set of transmitters and receivers.
%

The superposition coding graph and the conditions under which  superposition coding can be applied in Condition \ref{cond:Conditions for Superposition Coding}  are also illustrated in Fig. \ref{fig:schematic}.
The rounded squares in the figure represent the nodes $\Vv$ in the superposition coding graph while solid arrows represent the graph edges $\Sv$.
When a node is superimposed over another  node, it must also be superimposed over its parents in the superposition coding graph.

\subsubsection{Binning graph}
\label{sec:Binning graph}

The binning graph $\Gcal(\Vv,\Bv)$  describes how codewords are binned against each other after superposition coding has been considered in the codebook construction.
%
%
When the codebook is generated, codewords are created with conditionally dependent codebooks as prescribed by the superposition coding graph.
When binning is applied,  multiple codewords to transmit the same message are generated: one of these codewords is selected for transmission when it appears jointly typical with the chosen set of codewords.
%

%
%

An edge $(\iv,\jv) \times (\lv,\mv) \in \Bv$ indicates that the codeword $U_{\lv\sgoes \mv}^N$ is binned against the codeword $U_{\iv \sgoes \jv}^N$.
Binning of $U_{\lv \sgoes \mv}$ against  $U_{\iv \sgoes \jv}$ is also indicated as $U_{\iv \sgoes \jv} \bin U_{\lv \sgoes \mv}$
When $\iv=\lv$ two codewords can be binned against each other, as in Marton's region for the BC \cite{MartonBroadcastChannel}: we refer to this as \emph{joint binning}.
Joint binning of two codewords $U_{\lv \sgoes \mv}$ against  $U_{\iv \sgoes \jv}$ is indicated as $U_{\iv \sgoes \jv} \jbin U_{\lv \sgoes \mv}$.

As for superposition coding, binning can be applied only under some restrictions.
\begin{cond}{\bf Binning. \\}
\label{cond:Conditions for Binning}
Binning of the codeword $U_{\iv \sgoes \jv}^N$ against the codeword $U_{\lv \sgoes \mv}^N$ can be performed when the following condition holds:
\begin{itemize}
  \item $\iv \subseteq \lv$: that is, the set of encoders performing binning has knowledge of the interfering codeword.
\end{itemize}
%
Binning and superposition coding are mutually exclusive, that is two nodes can be adjacent either in $\Gcal(\Vv,\Sv)$ or $\Gcal(\Vv,\Bv)$ but not in both.
Lastly, binning does not form directed cycles, if a cycle exists it must be undirected.
%
\end{cond}

\begin{figure}
\centering
\includegraphics[width=.8 \textwidth]{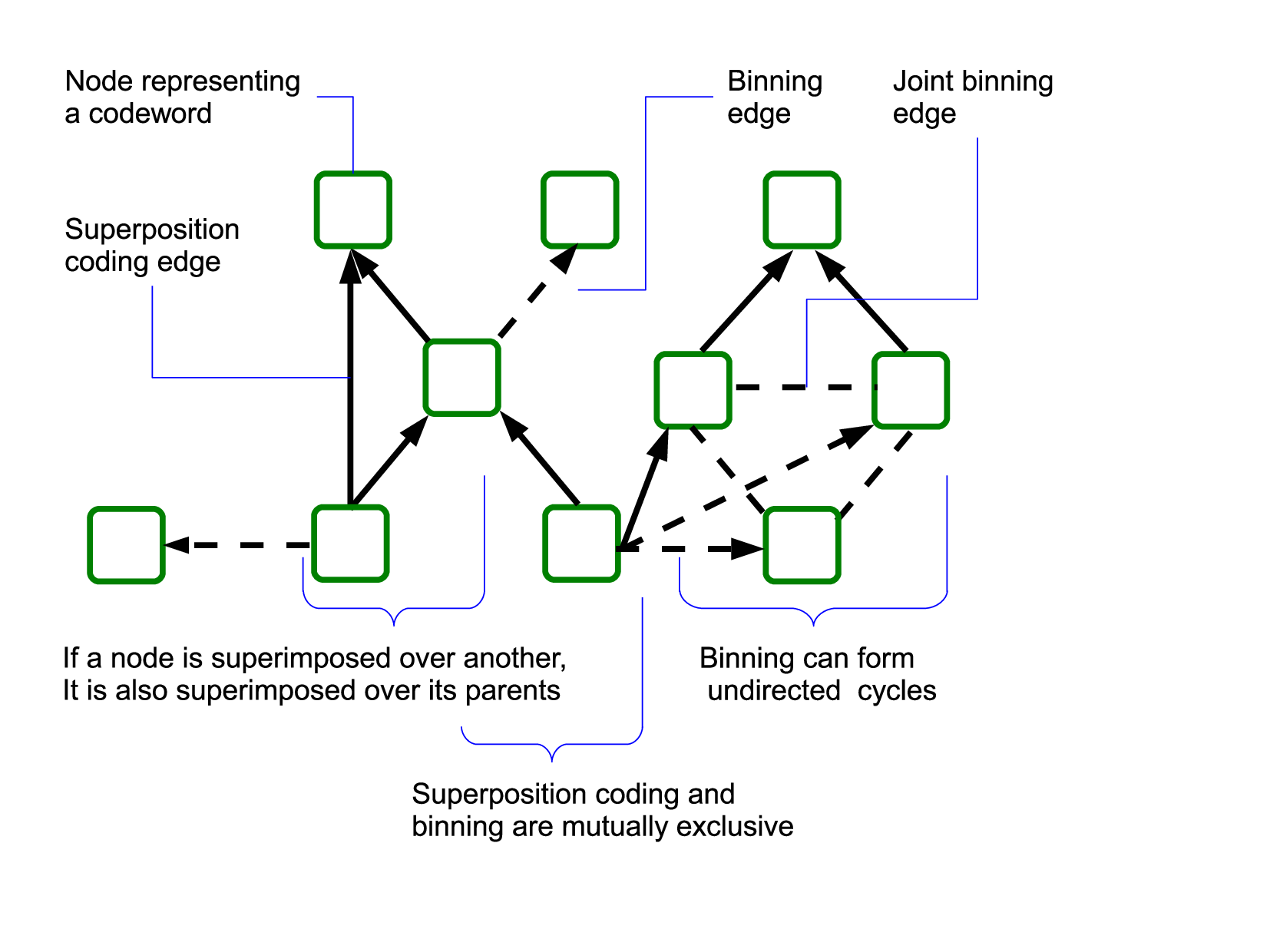}
\vspace{- 1 cm}
\caption{A schematic representation of the superposition coding graph $\Gcal(\Vv,\Sv)$ in Sec. \ref{sec:Superposition coding graph}
and the binning graph $\Gcal(\Vv,\Bv)$ in Sec. \ref{sec:Binning graph}.
The edges $\Sv$ are indicated with solid arrows while  the edges in $\Bv$ are indicated with dashed arrows and lines.
}
\label{fig:schematic}
\end{figure}

Given Condition \eqref{cond:Conditions for Binning}, it follows that $\Gcal(\Vv,\Bv)$ is a chain graph, since it has both directed and undirected edges and all cycles are undirected.
The binning graph and the conditions under which  binning can be applied in Condition \ref{cond:Conditions for Binning}  are also illustrated in Fig. \ref{fig:schematic}.
Rounded squares represent the nodes $\Vv$ in the binning graph while arrow and line edges represent binning edges.
Binning edges can be both directed and indirected; undirected binning edges can form cycle in $\Gcal(\Vv,\Bv)$, but no directed cycles can exist in $\Gcal(\Vv,\Bv)$. The graph $\Gcal(\Vv,\Bv)$ is a chain graph which does not possess directed cycles by definition since directed cycles cannot be associated with a well defined probability distributions.
Also, superposition coding edges and binning edges cannot connect two nodes, regardless of the direction of the edges.

\subsubsection{CGRAS}
\label{sec:CGRAS}

\bigskip
The CGRAS is then defined by the sets $\lsb \Vv, \Sv, \Bv\rsb $: since the superposition coding graph $\Gcal(\Vv,\Sv)$ and the binning code graph $\Gcal(\Vv,\Bv)$ are defined over the same set of nodes, the CGRAS can be represented through a graph with two types of edges as in Fig. \ref{fig:schematic}.
%
Each node of the graph is associated with a codeword encoding a specific message obtained after user virtualization.
Codewords can be superimposed and binned, respectively, only when Condition \ref{cond:Conditions for Superposition Coding}
and Condition \ref{cond:Conditions for Binning} are satisfied.
When a codeword is superimposed over another, this is indicated by a directed, solid arrow from the bottom to the top codeword.
Similarly, when a codeword is binned against another, this is indicated by a directed, dashed arrow from the first codeword to that it is binned against.
Joint binning is indicated with dashed lines in between nodes.

%% file: GMMassociated.tex
\section{GMMs associated with the CGRAS}
\label{eq:GMM associated with the CGRAS}

The CGRAS in Sec. \ref{sec:The chain graph representation of an achievable region} compactly describes how codewords are coded and graphically expresses the condition under which superposition coding and joint binning are feasible.
These two coding operations have been presented so far from a high level perspective as an in-depth description of the encoding and decoding procedures will follow in Sec. \ref{sec:Codebook construction, encoding and decoding operations}.
%
%
In both superposition coding and joint binning, codewords are generated through i.i.d. draws from some prescribed distribution.
A codeword is then selected at the encoder depending on what message is being transmitted.
When combining these two coding strategies, the distribution according to which codeword is generated and selected for transmission can be difficult to describe.
For this reason, we show in this section how GMMs can be associated to the CGRAS in Sec. \ref{sec:The chain graph representation of an achievable region} to describe the distribution of the codewords.
%
%
%
%
%
%

Both superposition coding and joint binning are used to introduce conditional dependence among codewords.
In superposition coding,  a different top codebook is created for each bottom codeword and this top codebook is generated conditionally dependent on the bottom codeword.
In binning, on the other hand, multiple codewords are generated to encode the same message and one of these codewords is selected when it is conditionally dependent on the given realization of the interfering codeword.

Given these two mechanisms to impose conditional dependence among codewords, a transmission strategy involving these two techniques is obtained in two steps.
First, the overall codebook is generated by applying superposition coding and then it is distributed to all nodes in the network.
Successively,  when a message is selected for transmission,  binning determines which codewords are selected to embed this message.

For the first phase, the distribution of the codewords in the codebook can be described using a GMM associated with the superposition coding graph, the
\emph{codebook GMM}.
For the second phase, the distribution of the codewords after both superposition coding and joint binning are applied is associated with the \emph{encoding GMM}.
Accordingly, we refer to the distribution associated to the first GMM as the \emph{codebook distribution} and to the distribution associated with the second GMM as the \emph{encoding distribution}.
 %
%

\subsection{Codebook GMM}
\label{sec:Codebook GMM}

The codebook GMM describes the distribution from which codewords are obtained through i.i.d. draws.
The conditional dependence among codewords in the codebook is determined only by superposition coding, for this reason only the graph $\Gcal(\Vv,\Sv)$ is necessary when defining the codebook GMM.

A GMM over the graph $\Gcal(\Vv,\Sv)$ is readily obtained: the graph $\Gcal(\Vv,\Sv)$ is a DAG and this class of graphs satisfies the global Markov property in Def. \ref{def:global markov property}.
Additionally, this GMM possesses a convenient factorization of the associated distribution as in \eqref{eq:factorization DAGs}.
%
For this reason, the \emph{codebook distribution} factorizes as
\ea{
P^{\rm codebook}= \prod_{ ( \iv,\jv) \in \Vv} P_{U_{\iv \sgoes \jv} | \pa_{\Sv}(U_{\iv \sgoes \jv})},
\label{eq:codebook distribution}
}
%
where $\pa_{\Sv}$ indicates the parents of the node $(\iv,\jv)$ in the superposition coding graph.
This GMM is used to generate the codebook associated with a CGRAS as detailed in the next section.

\subsection{Encoding GMM}

%
When a message is selected for transmission, the associated codeword is distributed according to the codebook distribution:
binning can be used to impose additional dependency among codewords which is not originally present in the codebook.
This is done by creating multiple codewords to transmit the same message and selecting one such codeword so as to appear conditionally dependent on other codewords selected for transmission.
For this reason, after binning is applied, the codewords selected for transmission have a joint distribution which is more general than the codebook distribution, that is, it includes more conditional dependencies among codewords.
This distribution, which we refer to as the encoding distribution, can be described by a GMM constructed over the graph  $\Gcal(\Vv,\Sv \cup \Bv)$.
While the superposition coding graph $\Gcal(\Vv,\Sv)$ can be used to construct a GMM with a convenient factorization of the associated distribution, the same does not hold for the graph $\Gcal(\Vv,\Sv \cup \Bv)$.

The graph $\Gcal(\Vv,\Sv \cup \Bv)$ is a CG and thus this is a well-defined GMM.
On the other hand, the CG contains both directed and undirected cycles and a factorization in the form of \eqref{eq:convenient factorization} is not possible in such a complex graph.
Being able to express the distribution of the codewords after encoding as in \eqref{eq:convenient factorization} is particularly important since this will, in turn, provide a simple representation of the CGRAS achievable rate region.
For this reason we now introduce some further restriction on the binning steps so that the GMM constructed over the graph $\Gcal(\Vv,\Sv \cup \Bv)$ can be made Markov equivalent with a DAG.

\begin{assumption}
{\bf Transitive Binning Restriction (TB-restriction)\\}
\label{ass:Binning is transitive}
The following holds
\ea{
 U_{\vv \sgoes \tv} \bin U_{\lv \sgoes \mv} , \ U_{\lv \sgoes \mv} \bin U_{\iv \sgoes \jv}   \Rightarrow U_{\vv \sgoes \tv} \bin U_{\iv \sgoes \jv}.
\label{eq:binnig transitive}
}
\end{assumption}

\begin{assumption}
{\bf Connected Subset Joint Binning Restriction (CSJB-restriction) \\}
\label{ass:Joint binning forms cliques}
Nodes in the binning graph that are connected by an undirected edge form fully connected sets, that is
\ea{
U_{\iv \sgoes \jv} \jbin U_{\iv \sgoes \mv} , \ U_{\iv \sgoes \jv} \jbin U_{\iv \sgoes \tv}   \Rightarrow U_{\iv \sgoes \mv} \jbin U_{\iv \sgoes \tv}.
\label{eq:assumption joint binning}
}
Moreover, jointly binned codewords have the same parent nodes in $\Gcal(\Vv, \Sv \cup \Bv)$:
\ea{
U_{\iv \sgoes \jv} \jbin U_{\iv \sgoes \tv} \implies \pa_{\Sv \cup \Bv}(U_{\iv \sgoes \jv})=\pa_{\Sv \cup \Bv}(U_{\iv \sgoes \tv}).
\label{eq:joint binning parents}
}
\end{assumption}

Using the  TB-restriction and the CSJB-restriction, we are now able to obtain a Markov equivalent DAG from the encoding CG.

\begin{thm}
\label{thm: DAG equivalence of graph representation}
If Assumption \ref{ass:Binning is transitive} and Assumption \ref{ass:Joint binning forms cliques} hold for the CGRAS $\lsb \Vv, \Sv, \Bv \rsb $, the GMM $\Gcal(\Vv, \Sv \cup \Bv)$ is Markov-equivalent to any DAG obtained from a non-cyclic orientation of the binning edges, either directed or undirected, which are not connected to the source node in $\Gcal(\Vv,\Bv)$, as defined in Sec. \ref{sec:Some graph-theoretic notions}.
\end{thm}

\begin{IEEEproof}
The assumptions of theorem not only assure the existence of a Markov-equivalent DAG, but also that an equivalent DAG can be obtained with a different orientation of the binning edges.
In particular, the direction of the jointly binned edges can be chosen at will, provided that it does not result in a cycle.
For the directed edges, a change of direction is possible only when both nodes are binned against another node.
The complete proof is presented in Appendix \ref{app: DAG equivalence of graph representation}.
\end{IEEEproof}

\begin{figure}
\centering
\includegraphics[width=.8 \textwidth]{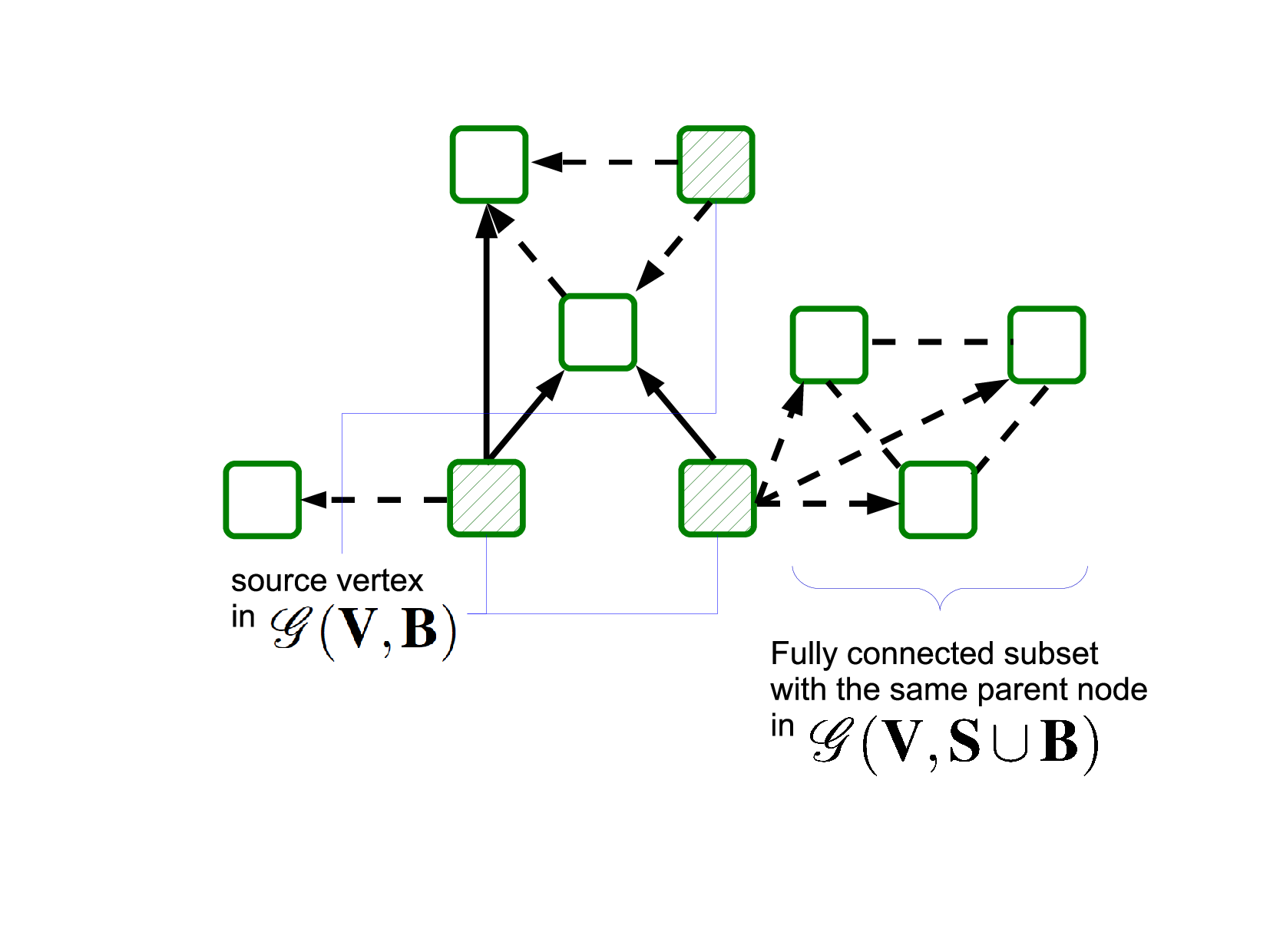}
\vspace{- 2 cm}
\caption{A schematic representation of the assumptions in Th. \ref{thm: DAG equivalence of graph representation}, the TB-restriction and the CSJB-restriction.}
\label{fig:graph-equivalentDAG}
\end{figure}

\begin{figure}
\centering
\includegraphics[width=.8 \textwidth]{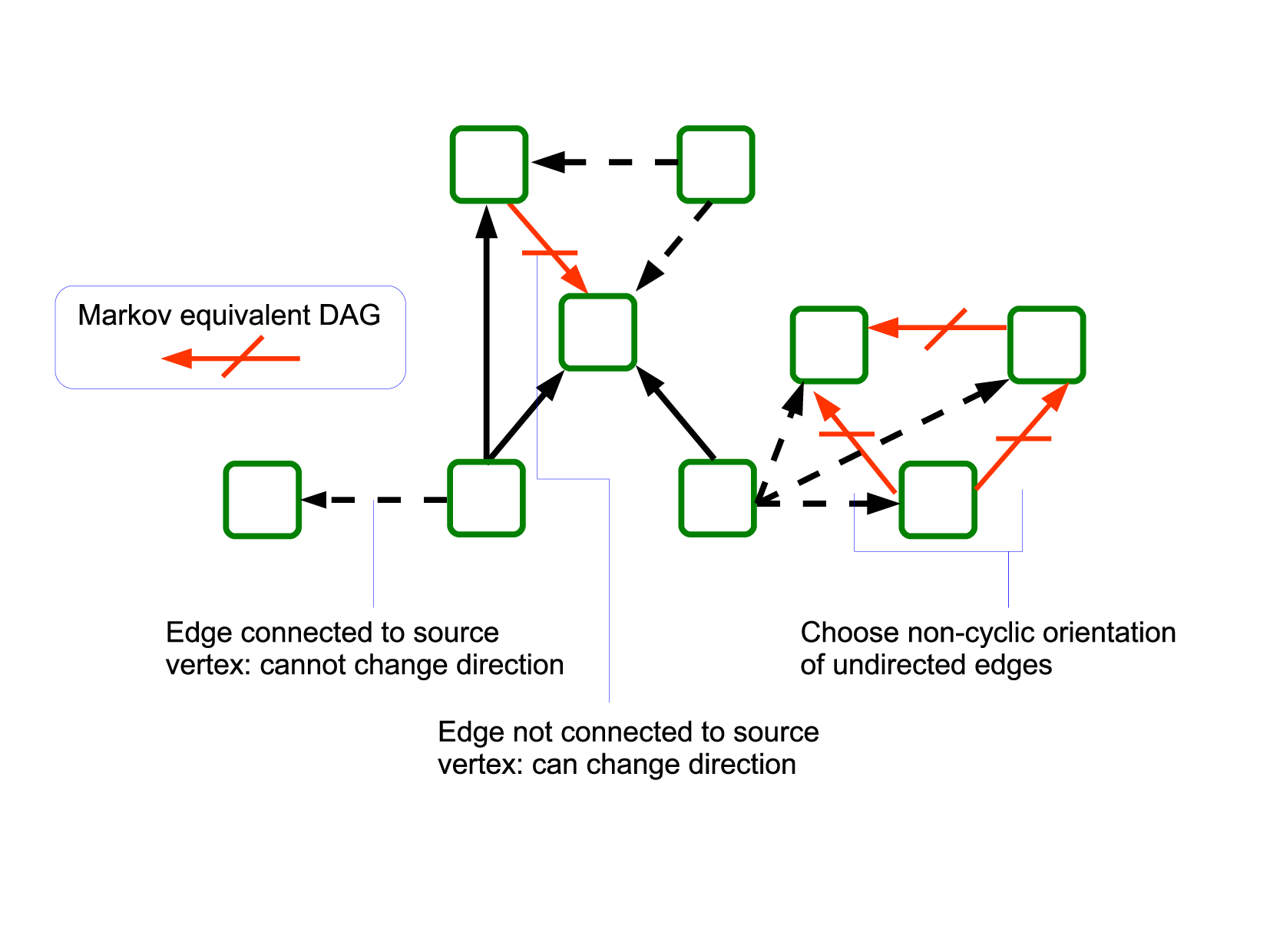}
\vspace{- 1.5 cm}
\caption{A schematic representation of the conclusions in  Th. \ref{thm: DAG equivalence of graph representation}. }
\label{fig:graph-equivalentDAGAFTER}
\end{figure}

A pictorial illustration of Th. \ref{thm: DAG equivalence of graph representation} is provided in Fig. \ref{fig:graph-equivalentDAG} and Fig. \ref{fig:graph-equivalentDAGAFTER}.

Fig. \ref{fig:graph-equivalentDAG} shows a valid CGRAS in which the source nodes in $\Gcal(\Vv,\Bv)$ are indicated as hatched rounded boxes.
In the figure, a connected subset of nodes with the same parent nodes in $\Gcal(\Vv,\Sv \cup \Bv)$ is also indicated: for this set of nodes a convenient factorization is not available since we cannot use the form \eqref{eq:convenient factorization} to describe their distribution.

By applying the result in Th. \ref{thm: DAG equivalence of graph representation} we conclude that the graph in Fig. \ref{fig:graph-equivalentDAG}  is Markov equivalent to the graph in Fig. \ref{fig:graph-equivalentDAGAFTER}, that is, the two graphs express the same factorization of the associated joint distribution.
In Fig. \ref{fig:graph-equivalentDAGAFTER} the orientation of the undirected edge in Fig. \ref{fig:graph-equivalentDAG} has been chosen in a way that does not introduce cycles: the additional edges are indicated in orange and with a slanted mark.
In the CGRAS of Fig. \ref{fig:graph-equivalentDAGAFTER}  we can express the factorization of the distribution of these nodes as in \eqref{eq:convenient factorization}.
Additionally, the orientation of a binning node has also been changed: this node is also indicated in orange and with a slanted mark.
This edge is not connected to a source node and its orientation can be changed without altering the factorization of the joint distribution of the associated GMM.
%

%

Given any graph $\Gcal(\Vv,\Sv \cup \Bv)$, Assumption \ref{ass:Joint binning forms cliques} (CSJB-restriction) can be satisfied by adding additional binning steps.
For this reason this assumption does not restrict the generality of our result, since binning steps can only enlarge the achievable rate region.
%
On the other hand, given any graph $\Gcal(\Vv,\Sv \cup \Bv)$, Assumption \ref{ass:Binning is transitive} (the TB-restriction) can be made to hold only by substituting some superposition coding edges with binning edges.
This implies a certain loss of generality in our approach but this assumption is necessary to  obtain a convenient factorization of the mutual information expression and hence to compactly express the achievable rate region.

Using Th. \ref{thm: DAG equivalence of graph representation} we can now define the (not-necessarily unique) Markov-equivalent DAG $\Gcal(\Vv,\Sv \cup \Bvt)$ which is obtained through a non-cyclic orientation of the binning edges in $\Gcal(\Vv,\Sv \cup \Bvt)$ that are not connected to sink nodes.
%
%
%
Through  this graph, we can now write the distribution associated with the encoding GMM, the \emph{encoding distribution}, as
\ea{
P^{\rm encoding} = \prod_{ ( \iv,\jv) \in \Vv} P_{U_{\iv \sgoes \jv} | \pa_{\Sv \cup \Bvt}(U_{\iv \sgoes \jv})}.
\label{eq:encoding distribution}
}
%

The assumptions required by Th. \ref{thm: DAG equivalence of graph representation} are quite specific, but theorem establishes a fairly large class of Markov-equivalent DAGs to the graph $\Gcal(\Vv,\Sv \cup \Bv)$.
Although looser conditions can be considered to obtain a Markov-equivalent DAG, the stronger assumptions in Th. \ref{thm: DAG equivalence of graph representation} are instrumental in the following when deriving the achievable rate region associated with the CGRAS.

This theorem  makes it possible to obtain a number of different Markov equivalent DAGs which can be used to bound the probability of different error events.
Through this ease of analysis,  it is possible to obtain a compact expression of the achievable region.

%

Note that the codebook and  encoding distributions in \eqref{eq:codebook distribution} and \eqref{eq:encoding distribution}, respectively,
have an identical factorization among the RVs except for the RVs connected by a binning edge.
In other words, $P^{\rm encoding}$ is a more general distribution than $P^{\rm codebook}$: RVs which are conditionally independent in $P^{\rm codebook}$
are conditionally dependent in $P^{\rm encoding}$.

%% file: CodebookConstruction.tex
\section{Codebook construction, encoding and decoding operations}
\label{sec:Codebook construction, encoding and decoding operations}

The CGRAS, as defined in Sec. \ref{sec:The chain graph representation of an achievable region}, describes a series of coding operations through the graphs
$\Gcal(\Vv, \Sv)$ and $\Gcal(\Vv, \Bv)$ which indicate superposition coding and joint binning, respectively.
Superposition coding introduces conditional dependence among codewords as described by the codebook GMM constructed over the graph $\Gcal(\Vv,\Sv)$.
Binning is applied after superposition coding and it further introduces conditional dependence across codewords: the graph $\Gcal(\Vv,\Sv\cup \Bv)$ can be used to describe the conditional dependency across codewords after binning is applied.
%

%
In this section, we combine the description of the coding operation in Sec. \ref{sec:The chain graph representation of an achievable region}
and the distribution of the codewords in Sec. \ref{eq:GMM associated with the CGRAS}  to obtain a general transmission strategy.

In particular, we specify:
\begin{itemize}
  \item {\bf codebook selection through coded time-sharing:\\}
  %
%
%
  %
  Time-sharing utilizes a transmission strategy for a portion of the time and another transmission strategy for the remainder of the time.
  This strategy can be generalized and improved upon by selecting among multiple transmission strategies according to a random sequence which is made available at all the nodes.
  This strategy is referred to as coded time-sharing and it attains the convex closure of the union of the rate regions corresponding to each transmission strategy.

  \item {\bf codebook generation through superposition coding:\\}
  Superposition coding entails stacking the codebook of one user over the codebook of another.
  %
  This can be obtained in a sequential manner by generating the codeword of the bottom user first and subsequently generating the codeword of the top users.
  In the resulting codebook, codewords are conditionally dependent according to the codebook distribution in \eqref{eq:codebook distribution}.

  \item {\bf encoding of the messages through binning:\\}
  Once a set of messages has been chosen for transmission, binning is used to determine the set of codewords to embed each message.
  This is done by selecting the set of codewords which is in the typical set of the encoding distribution in \eqref{eq:encoding distribution} although generated according to the codebook distribution in \eqref{eq:codebook distribution}.

  \item {\bf input generation:\\}
after binning has been applied, the channel input at each encoder is obtained through a deterministic function of the codewords known at this encoder.

  \item {\bf decoding of the messages at the receivers using typicality:\\}
  Each receiver decodes the subset of the transmitted codewords which are destined for it.
  Codewords are determined through a typicality decoder, that is, by identifying a set of codewords in the codebook that look jointly typical with the given channel output.

\end{itemize}

%
%
%
%
%
%
In this section, we will also better motivate Condition  \ref{cond:Conditions for Superposition Coding} and Condition \ref{cond:Conditions for Binning} %
which were introduced above as conditions on the set of edges in the CGRAS and not motivated from the point of view of the coding scheme itself.

\subsection{Codebook selection through coded time-sharing}
\label{sec:Codebook selection through coded time-sharing}
%
%
%
In coded time-sharing all the codewords in the codebook are generated conditionally dependent on an i.i.d. sequence $Q^N$ with distribution $P_Q$.
%
Before transmission begins, a random realization $q^N$ is produced and distributed at all nodes to select a transmission codebook.
Coded time-sharing outperforms both time and frequency division multiplexing (TDM/FDM respectively) and is used to convexify the achievable region of a transmission strategy.
%
%
\subsection{Codebook generation through superposition coding}
\label{sec:Codebook generation through superposition coding}
In this phase, the transmission codewords are created by stacking the codebooks of users in the enhanced channel one over the other according to the superposition coding steps in the CGRAS.
%
%
%
For any codebook distribution that factorizes as in  \eqref{eq:codebook distribution}, the codebook GMM $\Gcal(\Vv,\Sv)$ in Sec. \ref{sec:Codebook GMM} can be used to obtain a transmission codebook by recursively applying the following procedure:

\begin{itemize}
\item
  Consider the node $U_{\iv \sgoes \jv}$ in $\Gcal(\Vv,\Sv)$, let $\pa_S$ indicate the parents of $U_{\iv  \sgoes \jv}$ in the graph $\Gcal(\Vv,\Sv)$ and assume that it has no parent nodes  or that the codebook of all the parent nodes has been generated and indexed by
  $l_{\lv \sgoes \mv}$,
  i.e.
  \ea{
  U_{\lv \sgoes \mv}^N(l_{\lv \sgoes \mv}),  \  \forall \ U_{\lv \sgoes \mv}  \in \pa_{\Sv}(U_{\iv \sgoes \jv}) ,
  }
  %
  %
  then, for each possible set of base codewords
\ea{
\lcb  U_{\lv \sgoes \mv}^N(l_{\lv \sgoes \mv}), \ U_{\lv \sgoes \mv}  \in \pa_{\Sv}(U_{\iv \sgoes \jv}) \} \rcb,
\label{eq:all base codewords}
}
  repeat the following:
\begin{enumerate}
\item
generate $2^{N L_{\iv \sgoes \jv}}$ codewords, for
\eas{
L_{\iv \sgoes \jv} & = R_{\iv \sgoes \jv}+R'_{\iv \sgoes \jv}  \\
R'_{\iv \sgoes \jv} &  \lcb  \p{
 \geq 0 & \exists \ (\vv,\tv) \ U_{\vv \sgoes \tv} \bin U_{\iv \sgoes \jv}  \\
= 0                       & \rm{otherwise},
}\rnone
}
with  i.i.d. symbols drawn from the distribution $P_{U_{\iv \sgoes \jv} | \pa_{\Sv}(U_{\iv \sgoes \jv}), Q}$ conditioned on the set of base codewords in \eqref{eq:all base codewords} and the coded time-sharing sequence.
\smallskip
In the following we refer to $R_{\iv \sgoes \jv}$ as the \emph{message rate} while we refer to $R'_{\iv \sgoes \jv}$ as the \emph{binning rate}.
\item
    If $R'_{\iv \sgoes \jv} \neq 0$,  place each codeword $U_{\iv \sgoes \jv}^N$ in $2^{N R_{\iv \sgoes \jv} }$ bins of size $2^{N R'_{\iv \sgoes \jv}}$ indexed by $b_{\iv \sgoes \jv} \in [1...2^{N R'_{\iv \sgoes \jv} }]$.

    If $R'_{\iv \sgoes \jv} = 0$, simply set $b_{\iv \sgoes \jv}=1$.
\item

    Index each codebook of size $2^{N L_{\iv \sgoes \jv}}$  using the set
    $\{ l_{\lv \sgoes \mv}, \ \forall \ (\lv,\mv) \ST  \ \in \pa_{\Sv}(U_{\iv \sgoes \jv})\}$
    so that
    \ea{
& U_{\iv \sgoes \jv}^N( l_{\iv \sgoes \jv})  =
\label{eq:codeword indexing} \\
& \quad U_{\iv \sgoes \jv}^N \lb w_{\iv \sgoes \jv} , b_{\iv \sgoes \jv}  , \{ l_{\lv \sgoes \mv}, U_{\lv \sgoes \mv}  \in \pa_{\Sv}(U_{\iv \sgoes \jv}) \}   \rb.
\nonumber
}
The index $w_{\iv \sgoes \jv}$ is referred to as the \emph{message index} while the index $b_{\iv \sgoes \jv}$ is referred to as the \emph{binning index}.
%
The message index $w_{\iv \sgoes \jv}$ selects the bin while the binning index $b_{\iv \sgoes \jv}$ selects a codeword inside each bin.

\end{enumerate}
\end{itemize}

\begin{figure}
\centering
\includegraphics[width=.78 \textwidth]{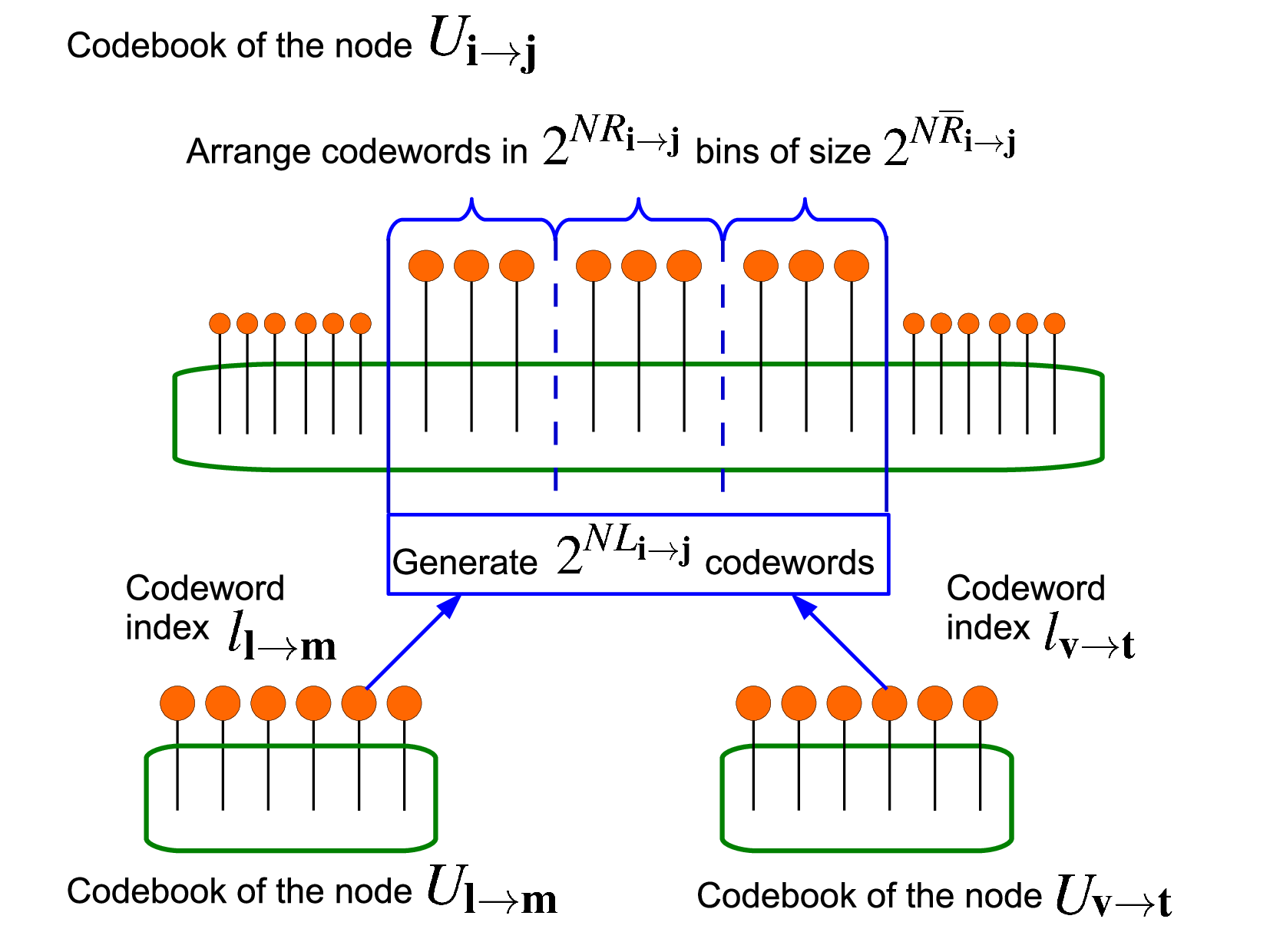}
\vspace{-.5 cm}
\caption{A graphical representation of the codebook generation in Sec. \ref{sec:Codebook generation through superposition coding}. }
\label{fig:codebook_generation}
\vspace{-.5 cm}
\end{figure}

A graphical representation of the codebook generation is provided in Fig. \ref{fig:codebook_generation}: the nodes $U_{\lv \sgoes \mv}$ and $U_{\vv \sgoes \tv}$ are parents of the node $U_{\iv \sgoes \jv}$. For each codeword $U_{\lv \sgoes \mv}^N(l_{\lv \sgoes \mv})$ and $U_{\vv \sgoes \tv}^N(l_{\vv \sgoes \tv})$ a new set of codewords for $U_{\iv \sgoes \jv}$ is generated.
More specifically, $2^{N L_{\iv\sgoes \jv}}$ codewords are generated conditionally dependent on the selected parent codewords and placed in $2^{N R_{\iv \sgoes \jv}}$ bins of size $2^{N R'_{\iv \sgoes \jv}}$.
Codewords in the same bin are used to encode the same message: a specific codeword in the bin is selected in the next step.

\subsection{Encoding of the messages through binning}
\label{sec:Encoding procedure}

In the previous step, multiple codewords are generated to encode the same message at the nodes involved in binning.
One of these codewords is chosen so that, overall,  the transmitted codewords appear to be conditionally dependent when actually generated conditionally independent.
The codewords are generated according to the codebook distribution and the codewords are selected so as to look as if generated according to the encoding distribution.
The codebook distribution is described by a GMM over $\Gcal(\Vv,\Sv)$  while the encoding distribution is described by the more general GMM $\Gcal(\Vv,\Sv \cup \Bv)$.

More precisely, the encoding procedure is as follows: given a set of messages to be transmitted $w_{\Vv}=\{ w_{\iv \sgoes \jv}, \ (\iv,\jv) \in \Vv\}$,  the message index at all the nodes is set to the corresponding transmitted message.
The binning indices in
\ea{
\Vv^{\Bv} = \{ (\iv,\jv),  \ \exists \ (\lv,\mv), \ U_{\lv \sgoes \mv} \bin  U_{\iv \sgoes \jv} \}
\label{eq:definition VB}
}
are jointly chosen so that the selected codewords appear to have been generated with i.i.d. symbols drawn from the encoding distribution \eqref{eq:encoding distribution} despite being generated according to the codebook distribution in \eqref{eq:codebook distribution}.
If such an index does not exist, encoding fails.

\medskip

\begin{figure}
\centering
\includegraphics[width=.78 \textwidth]{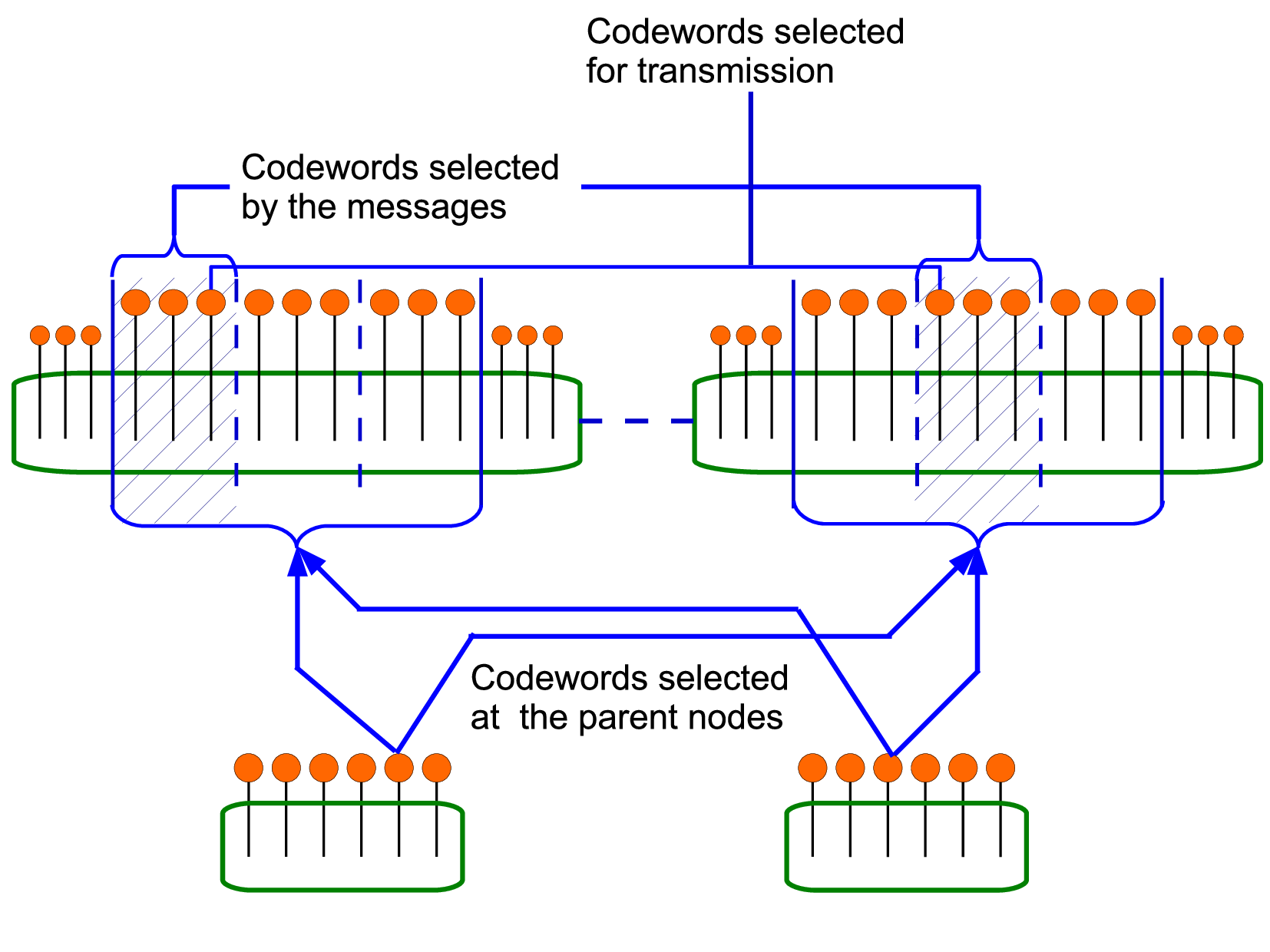}
\caption{A graphical representation of encoding of the messages through binning in Sec. \ref{sec:Encoding procedure}. }
\label{fig:encoding}
\vspace{-.5 cm}
\end{figure}

A graphical representation of the encoding procedure is provided in Fig. \ref{fig:encoding}: the codeword chosen by the parent nodes selects a set of codewords in the codebook of the jointly binned nodes. In these sets, the transmitted message selects the bin $w_{\iv \sgoes \jv}$. Among the codewords inside the bin, a codeword is selected that looks as if generated according to the encoding distribution despite being generated according to the codebook distribution.

\medskip

\subsection{Input generation}
%


The $k^{th}$ encoder produces the channel input $X_k^N$ as a deterministic function of its codebook(s) and the time sharing sequence, i.e.
\ea{
X_k^N= X_k^N \lb \lcb U_{\iv \sgoes \jv}^N , \ \forall \ (\iv,\jv) \ \ST \ k \in \iv \rcb, Q^N \rb.
\label{eq:deterministic encoding functions}
}

Restricting the class of encoding functions to deterministic functions instead of random functions can be done
 without loss of generality \cite{willems1985discrete}.

\subsection{Decoding of the messages at the receivers using typicality}
\label{sec:Decoding procedure}

Receiver $z$  is required to decode the transmitted messages $W_{\iv \sgoes \jv}$  for
\ea{
\Vv^z & = \{ (\iv,\jv) \in \Vv, \ \jv \in z \},
\label{eq:definition Vz}
}
and it does so by employing a typicality decoder which determines the set of indices
\ea{
\{ \wh_{\iv \sgoes \jv},\bh_{\iv \sgoes \jv}, \ (\iv,\jv) \in \Vv^z \},
}
such that
\ea{
\lcb Y_z^N,  \lcb U_{\iv \sgoes \jv}^N(\lh_{\iv \sgoes \jv}), \ (\iv,\jv) \in \Vv^z \rcb \rcb  \nonumber \\
\in \Tcal_{\ep}^n \lb  P_{Y_z, \rm encoding} \rb,
}
where
 \ea{
& P_{Y_z, \rm encoding,Q} = P_{Y_z|\{ U_{\iv \sgoes \jv},  \ (\iv,\jv) \in \Vv^z \}} \nonumber \\
& \quad \quad \cdot P_Q  \prod_{(\iv, \jv)  \in  \Vv^z } P_{U_{\iv \sgoes \jv} | \pa_{\Evt^z}(U_{\iv \sgoes \jv}),Q},
\label{eq:decoding step ditribution}
}
for
\ea{
\Evt^z=\lb \Sv \cup \Bvt \rb \cap (\Vv^z \times \Vv^z),
\label{eq:Evt z}
}
and for each known coded time-sharing sequence $q^N$ and for
$\Bvt$ obtained through Th. \ref{thm: DAG equivalence of graph representation}.

If no such set of indices can be found, an error is declared at receiver $z$.


\medskip

Each receiver only decodes a portion of the CGRAS:  more precisely, receiver $z$ decodes the codewords $U_{\iv \sgoes \jv}$ for which $z \in \jv$.
Accordingly, the nodes of the graph decoded at $z$ are the nodes in the set $\Vv^z$ and the Markov-equivalent DAG to this portion of the graph is $\Gcal(\Vv^z, \Evt^z)$ as defined in \eqref{eq:Evt z}.
%
%

A transmission error occurs if any receiver decodes any index  incorrectly, either message index or binning index.

\begin{remark}{\bf Condition \ref{cond:Conditions for Superposition Coding}.\\}
The codebook  construction Sec. \ref{sec:Codebook generation through superposition coding} motivates the condition on the superposition coding edges in Condition \ref{cond:Conditions for Superposition Coding}.
Consider the case in which $U_{\lv \sgoes \mv} \spc U_{\iv \sgoes \jv}$ and $U_{\vv \sgoes \tv} \spc U_{\lv \sgoes \mv}$.
The total number of codewords  $U_{\vv \sgoes \tv}^N$ is $2^{N R_{\vv \sgoes \tv}}$: since a codebook of size $2^{N R_{\lv \sgoes \mv}}$ is generated for each codeword $U_{\vv \sgoes \tv}^N$, the overall number of codewords $U_{\lv \sgoes \mv}^N$ is $2^{N (R_{\lv \sgoes \mv}+R_{\vv \sgoes \tv})}$.
Since a codebook for $U_{\iv \sgoes \jv}^N$ is generated for each codeword $U_{\lv \sgoes \mv}^N$ as in \eqref{eq:all base codewords}, the overall  number of codewords for $U_{\iv \sgoes \jv}^N$ is $2^{N (R_{\iv \sgoes \jv} + R_{\lv \sgoes \mv}+R_{\vv \sgoes \tv})}$ and each base codeword $U_{\lv \sgoes \mv}^N$
also uniquely identifies a base codeword $U_{\vv \sgoes \tv}^N$.
This situation implies that $U_{\vv \sgoes \tv} \spc U_{\iv \sgoes \jv}$, since a top codeword is generated for each set of bottom codewords.
%
\end{remark}

\begin{remark}{\bf Condition \ref{cond:Conditions for Binning}.\\}
In Sec. \ref{sec:Encoding procedure} we have seen that a binning index is chosen so that a codeword appears to be conditionally dependent on a set of random variables despite being generated independently from these.
If two codewords have already been superimposed, then they are already conditionally dependent and binning cannot meaningfully be applied in this scenario.
For this reason binning and superposition coding cannot both be applied between two codewords.
%
%
%
%
%
%
%
Similarly, a directed cycle in the binning graph does not correspond to a well-defined operation, since the choice of  binning index cannot depend on itself.
On the other hand, undirected cycles are feasible, since this indicates that codewords are jointly chosen so that they appear jointly typical according to some joint distribution.
\end{remark}

%

%% file: AchievableRateRegion.tex
\section{The achievable rate region of a CGRAS}
\label{sec:The Achievable Rate Region of the CGRAS}

In this section we derive the achievable rate region associated with the achievable scheme in Sec. \ref{sec:The chain graph representation of an achievable region}.
As the achievable scheme is compactly described through the CGRAS, so the achievable region is also using the CGRAS.
In particular, we associate encoding and decoding error events to this graphical structure using the encoding and the codebook GMM and derive the conditions under which the probability of these events vanishes when the block-length goes to infinity.

We present this main result in three steps, considering three classes of CGRAS, with an increasing level of generality:
\begin{itemize}
  \item we first consider the CGRAS with only superposition coding, then
  \item the case with superposition coding and binning
  \item finally, the most general case with superposition coding, binning and joint binning.
\end{itemize}

%

We begin by considering the case with only superposition coding to illustrate the error analysis associated with decoding error, which is the only type of error in this case.
The case with binning and superposition coding is used to explain the encoding error analysis in this case, since now both encoding and decoding errors are  possible.
In the most general case  we focus on the effects of joint binning on the encoding and decoding error probability.

\bigskip

The achievable scheme in Sec. \ref{sec:Codebook construction, encoding and decoding operations} produces an error in two situations:
\begin{itemize}
  \item {\bf Encoding errors:\\}
  A set of encoders cannot successfully determine a set of binning indices that satisfy the desired conditional typicality conditions
  \item {\bf Decoding error:\\}
  One of the receivers cannot determine a typical set of codewords or the selected codewords are different from the transmitted ones.
\end{itemize}

An encoding error occurs only under binning, when the number of codewords that encode the same message is too small and thus a codeword that satisfied the required typicality condition cannot be found.
For this reason, the probability of an encoding error can be set to zero by taking the binning rates $R'_{\iv \sgoes \jv}$ to be sufficiently large.
Consequently, the achievable region is then expressed as lower bounds on $R_{\iv \sgoes \jv}'$.
The difficulty in finding a codeword that satisfies the desired conditional typicality condition also depends on how similar the encoding and codebook distributions are.
The codewords are generated according to the codebook distribution in \eqref{eq:codebook distribution} but binning chooses a set of codewords which belongs to the typical set of the encoding distribution in \eqref{eq:encoding distribution}.
%
%

A decoding error occurs when one of the receivers decodes an incorrect codeword, which can happen under superposition coding alone or superposition coding and joint binning.
In particular, a decoding error happens when the overall codebook contains too many codewords and the typicality decoder cannot correctly identify the transmitted codeword.
The probability of decoding errors can therefore  be set to zero by lower bounding the message rates plus the binning rates $L_{\iv \sgoes \jv}=R_{\iv \sgoes \jv}+ R'_{\iv \sgoes \jv}$.
%
%
Since, in superposition coding, top codewords are created conditionally dependent on the bottom codewords, a codeword cannot be correctly decoded unless all
the codewords beneath it have also been correctly decoded as well.
%
For this reason, an incorrectly decoded codeword is still conditionally dependent on the correctly decoded parent codewords.
%
%
The same does not occur with binning: when a codeword which is binned against a correctly decoded codeword is incorrectly decoded, it is conditionally independent on the correctly decoded codeword.
This provides a ``decoding boost'' in binning with respect to superposition coding, since error events can be more easily recognized at the typicality decoder.

\subsection{Achievable region of a CGRAS with superposition coding only}
\label{sec:Superposition Coding}

We begin by considering the CGRAS involving only superposition coding.
In this case the achievable rate region is expressed as a series of upper bounds on the message rates
under which correct decoding occurs with high probability.
%
%
%
Each bound relates to the probability that a set of codewords is incorrectly decoded at receiver $z$ and the probability of this event is bounded using the
packing lemma \cite[Sec. 3.2]{el2011network}.

\begin{thm}{\bf Achievable region with superposition coding\\}
\label{th:Decoding errors only supeperposition}%
Consider any CGRAS employing only superposition coding
 and let the region $\Rcal$ be defined as
\ea{
&\sum_{(\iv,\jv) \in \Fvo^z} R_{\iv \sgoes \jv}\leq I( Y_z;  U_{\iv \sgoes \jv}, \ (\iv,\jv) \in \Fvo^z  |  U_{\iv \sgoes \jv}, \ (\iv,\jv) \in \Fv^z  , Q),
\label{eq:Decoding errors only supeperposition rate bound}
}
for every $z$ and every set $\Fv^z \subseteq \Vv^z$,  $\Fvo^z=\Vv \setminus \Fv^z $ and such that
\ea{
\pa_{\Sv}( \Fv^z) \subseteq \Fv^z \ {\rm or} \ \pa_{\Sv}( \Fv^z)= \emptyset.
\label{eq:condition decoding errors superposition}
}
%
%
%
Then, for any distribution of the terms $\{U_{\iv \sgoes \jv}\}$ that factors as in \eqref{eq:codebook distribution},
the rate region $\Rcal$ is achievable.
\end{thm}

\begin{IEEEproof}
The complete proof is provided in  Appendix \ref{app:Decoding errors only supeperposition}.
Each bound in \eqref{eq:Decoding errors only supeperposition rate bound}
relates to the probability that the codewords in $\Fv$ are correctly decoded while the ones in $\Fvo$ are incorrectly decoded.
Since the codebooks are stacked one over the other through superposition coding, a top codeword can be correctly decoded only when all the codewords
beneath are also correctly decoded.
This condition is expressed by \eqref{eq:condition decoding errors superposition}.
%
%
A decoding error occurs only if the rate of the incorrectly decoded codewords is high enough for the typicality decoder to find a set of incorrect codewords that appears jointly typical with the channel output.
The probability of this event relates to the mutual information term in the RHS of \eqref{eq:Decoding errors only supeperposition rate bound}
through the packing lemma \cite[Sec. 3.2]{el2011network}.
\end{IEEEproof}

\begin{figure}
\centering
\includegraphics[width=.78 \textwidth]{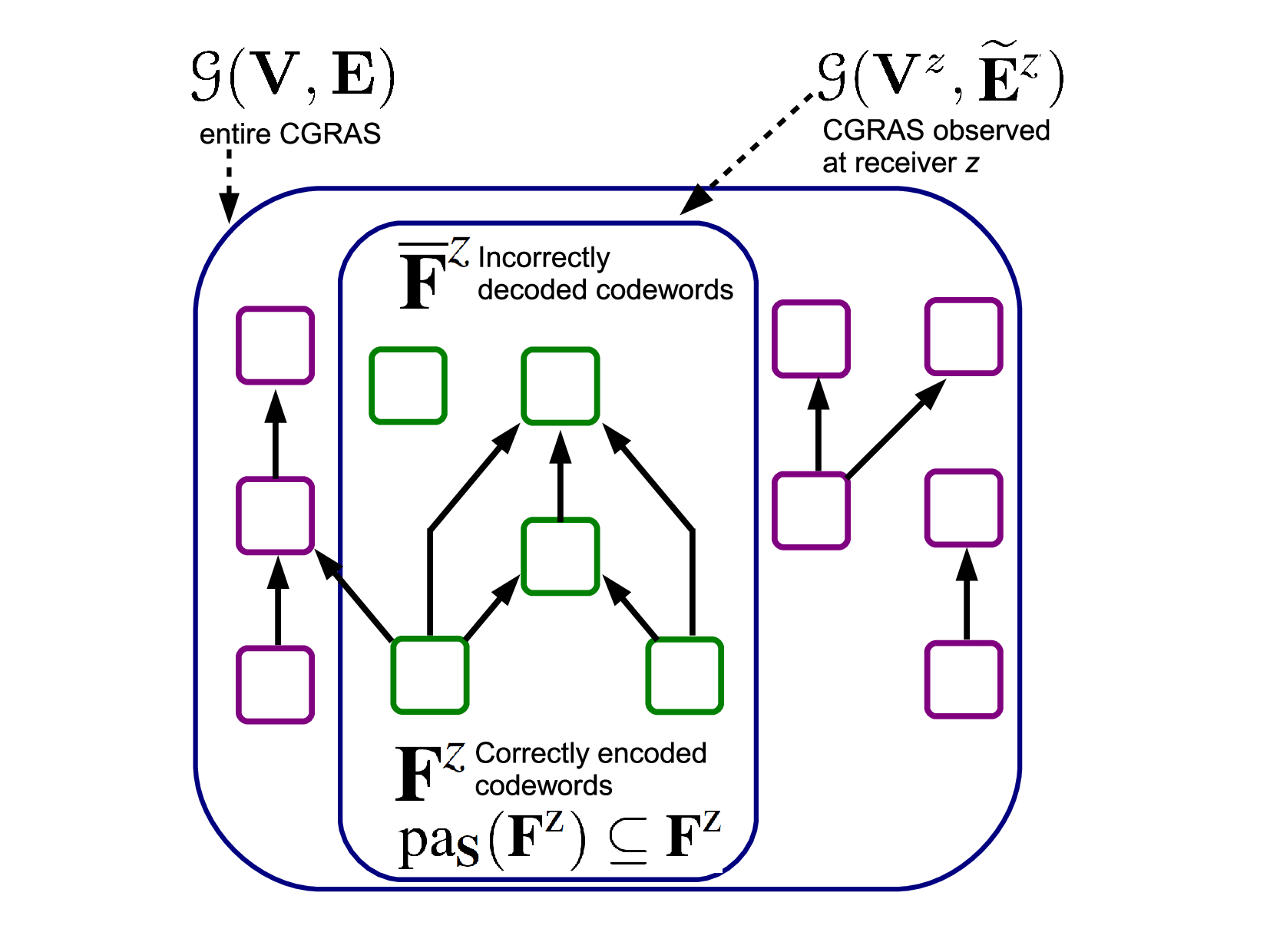}
\caption{A graphical representation of Th. \ref{th:Decoding errors only supeperposition}. }
\label{fig:superpositionOnly}
\end{figure}

A graphical representation of  Th. \ref{th:Decoding errors only supeperposition} is provided in Fig. \ref{fig:superpositionOnly}:
the channel output $Y_z^N$ is used at receiver $z$ to decode the set of codeword in $\Vv^z$ in \eqref{eq:definition Vz}, which is the portion of the CGRAS decoded at receiver $z$.
A lower bound on the message rates can be obtained by considering all the possible error patterns at all the decoders and bounding the probability of each
event using the packing lemma.
Since, in superposition coding, a top codeword is generated conditionally dependent on the bottom codeword, a codeword can be correctly decoded only when all the parent codewords have also been correctly decoded.
For this reason, each rate bound is obtained by considering a set $\Fv^z$ of correctly decoded codewords such that all the parent nodes of elements in $\Fv^z$ in the superposition coding graph are in $\Fv^z$ as well.
%
%

\subsubsection{An example with superposition coding}
\label{sec:An Example with superposition coding}

Let's return now to the example in Sec. \ref{sec:An example of  rate-splitting} of rate-splitting for the classical CIFC and construct an achievable region involving only superposition coding.
Consider the enhanced channel obtained with the rate-splitting matrix $\Gamma$ in \eqref{eq:rate-splitting CIFC}.
%
\begin{figure}
\centering
\includegraphics[width=.78 \textwidth]{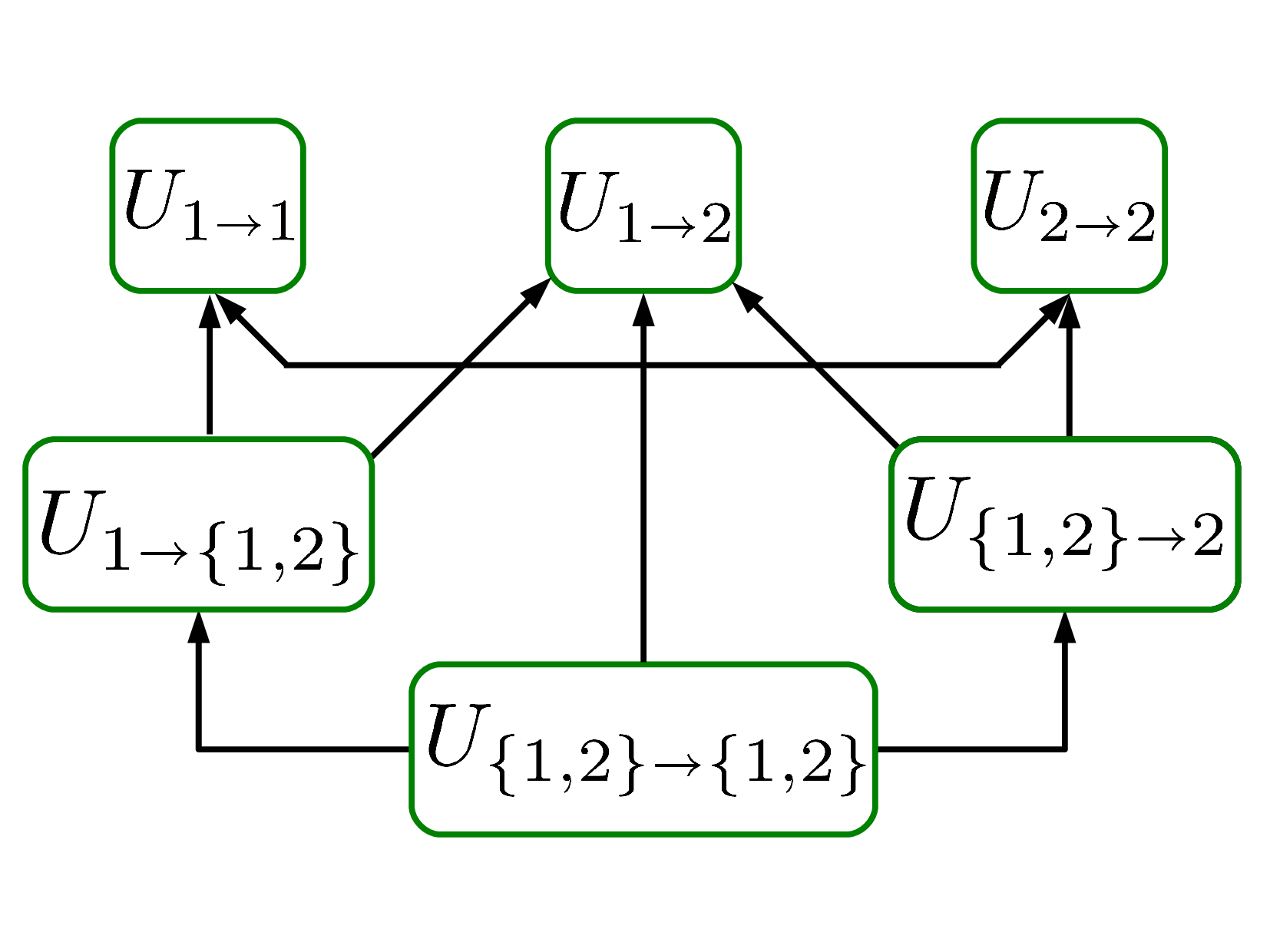}
\vspace{-.8 cm}
\caption{The chain graph for the achievable scheme in Sec. \ref{sec:An Example with superposition coding}.}
\label{fig:superpositionOnlyExample}
\vspace{-.5 cm}
\end{figure}
For this graph the codebook distribution is
\eas{
P^{\rm codebook}&= P_{U_{ \{1,2\} \sgoes \{1,2\} },Q} \\
& \quad  P_{U_{ 1 \sgoes \{1,2\} } | U_{ \{1,2\} \sgoes \{1,2\}},Q} \\
& \quad  P_{U_{ \{1,2\} \sgoes 2} | U_{ \{1,2\} \sgoes \{1,2\}},Q}  \\
& \quad  P_{ U_{1 \sgoes 1} | U_{ 1 \sgoes \{1,2\} } ,  U_{ \{1,2\} \sgoes \{1,2\}},Q} \\
& \quad  P_{ U_{2 \sgoes 2} | U_{ \{1,2\} \sgoes  2 } ,  U_{ \{1,2\} \sgoes \{1,2\}},Q}  \\
& \quad  P_{ U_{1 \sgoes 2} | U_{ 1 \sgoes   \{1,2\}  } , U_{ \{1,2\} \sgoes  2 } ,  U_{ \{1,2\} \sgoes \{1,2\}},Q}.
\label{eq:factorization codeword superposition only}
}
Using the result in Th. \ref{th:Decoding errors only supeperposition}, we can easily obtain the achievable region by deriving  the sets of $\Fv^z$ such that
\eqref{eq:condition decoding errors superposition} holds.
The graphs observed at each decoder are
\eas{
\Vv^1 &= \lcb \{1,2\} \sgoes \{1,2\} , \ 1 \sgoes \{1,2\} , \ 1 \sgoes 1 \rcb  \\
\Vv^2 &= \lcb \{1,2\} \sgoes \{1,2\} , \ \{1,2\} \sgoes 2 , \ 1 \sgoes 2, 2\sgoes 2 \rcb.
}
We list all possible subsets of $\Fv^1$ in Table \ref{tab:superpositionOnly D1}: each row in the table corresponds to a possible set $\Fv$.
In each row, a $1$ indicates that $(\iv,\jv)\in \Fv$ while a $0$ indicates that $(\iv,\jv) \not\in \Fv^1$.
The last column indicates whether the set $\Fv^1$ satisfies \eqref{eq:condition decoding errors superposition} or not.
Since it decodes the three messages, there are $8$ possible subsets of $\Vv^z$.
\begin{table}
\centering
\caption{Decoding error events table for decoder~1 in for the graph in Fig. \ref{fig:superpositionOnly}.}
\label{tab:superpositionOnly D1}
\begin{tabular}{|l|l|l|l|}
\hline
 $U_{ \{1,2\}  \goes \{1,2\}}$ & $U_{1  \goes \{1,2\}}$ & $U_{1  \goes 1}$ & valid $\Fv$ ?\\
 \hline
 0 & 0 & 0 & $\times$ \\
\hline
 0 & 0 & 1 & $\times$ \\
\hline
 0 & 1 & 0 & $\times$ \\
\hline
 0 & 1 & 1 & $\times$ \\
\hline
 1 & 0 & 0 & $\checkmark$ \\
\hline
 1 & 0 & 1 & $\times$ \\
\hline
 1 & 1 & 0 & $\checkmark$ \\
\hline
 1 & 1 & 1 & $\checkmark$ \\
\hline
\end{tabular}
\end{table}
For decoder~2, instead, we only list the valid $\Fv^2$ in Table \ref{tab:superpositionOnly D2}.
\begin{table}
\centering
\caption{Decoding error events table for decoder~2 in for the graph in Fig. \ref{fig:superpositionOnly}.}
\label{tab:superpositionOnly D2}
\begin{tabular}{|l|l|l|l|l|}
\hline
$U_{ \{1,2\}  \goes \{1,2\}}$ & $U_{1  \goes \{1,2\}}$ &  $U_{\{1,2\}   \goes  2}$ & $U_{1  \goes 2}$ & $U_{2  \goes 2}$\\
\hline
1 & 0 & 0 & 0 & 0 \\
\hline
 1 & 1 & 0 & 0 & 0 \\
\hline
1 & 0 & 1 & 0 & 0 \\
\hline
1 & 1 & 1 & 0 & 0 \\
\hline
1 & 1 & 1 & 1 & 0 \\
\hline
1 & 1 & 1 & 0 & 1 \\
\hline
1 & 1 & 1 & 1 & 1 \\
\hline
\end{tabular}
\end{table}
The achievable rate region is obtained using \eqref{eq:Decoding errors only supeperposition rate bound}  for each $\Fv^1$ and $\Fv^2$.

\subsection{Superposition coding and binning}
\label{sec:Superposition Coding and One-way Binning}

We now consider the case of a CGRAS employing both superposition coding and binning but not joint binning.
In binning,  multiple codewords are generated to encode the same message: one of these codewords is selected for transmission when it appears jointly typical with the codewords against which it is binned.

For a scheme that includes both superposition coding and binning, two types of error can occur: encoding errors and decoding errors.
An encoding error is committed when a set of transmitters cannot determine a set of codewords which looks conditionally typical according to the encoding distribution.
%
%
Since the amount of extra codewords is controlled by the binning rates $R'_{\iv \sgoes \jv}$, the probability of encoding error can be made arbitrarily small
by imposing a lower bound on the binning rates.

As for the scheme in Sec. \ref{sec:Superposition Coding}, the decoding error events are made small by reducing the overall number of codewords
in the codebook.
This translates into a lower bound on the messages and binning rates, that is a lower bound on the rates $L_{\iv \sgoes \jv}$.

\begin{thm}{\bf Achievable region with superposition coding and binning \\}
\label{th:Decoding errors only supeperposition and one-way binning}
Consider any CGRAS employing only superposition coding, binning but not joint binning and for which Assumption \ref{ass:Binning is transitive}, the TB-restriction, holds.
Let moreover the region $\Rcal'$ be defined as
\eas{
\sum_{(\iv,\jv) \in \Fv^{\Bv}}  R'_{\iv \sgoes \jv} & \geq  \sum_{(\iv,\jv) \in \Vv^{\Bv}} I(U_{\iv \sgoes \jv} ; \pa_{\Bv} \{ U_{\iv \sgoes \jv}\} | \pa_{\Sv} \{ U_{\iv \sgoes \jv}\} , Q)
\label{eq:binnig rates one way binning 1}\\
& \quad \quad -\sum_{(\iv,\jv) \in \Fvo^{\Bv}}  I( U_{\iv \sgoes \jv} ; A_{\iv \sgoes \jv}(\Fv^{\Bv})| \pa_{\Sv} (U_{\iv \sgoes \jv}),Q),
\label{eq:binnig rates one way binning 3}
}{\label{eq:binnig rates one way binning}}
with
\ea{
A_{\iv \sgoes \jv}(\Fv^{\Bv}) =  \pa_{\Bv}(U_{\iv \sgoes \jv}) \cup \ch_{\Bv\cap (\Fvo^{\Bv} \times \Fv^{\Bv})}(U_{\iv \sgoes \jv}),
\label{eq:Aij definition binnig}
}
for all the subsets $\Fv^{\Bv} \subseteq \Vv^{\Bv}$  and $\Fvo^{\Bv}=\Vv^{\Bv} \setminus \Fv^{\Bv}$  such that
\ea{
\pa_{\Sv}(\Fv^{\Bv}) \cap \Vv^{\Bv} \subseteq \Fv^{\Bv} \  {\rm or} \ \pa_{\Sv}(\Fv^{\Bv}) = \emptyset.
\label{eq:condition enocding errors one way binning}
}
Moreover let the region $\Rcal$ be defined as
\eas{
& \sum_{(\iv,\jv) \in \Fvo^z} L_{\iv \sgoes \jv}\leq  \nonumber \\
&  \quad I( Y_z;  U_{\iv \sgoes \jv}, \ (\iv,\jv) \in \Fvo^z | U_{\iv \sgoes \jv}, \ (\iv,\jv) \in \Fv^z , Q)  \\
& \quad \quad + \sum_{(\iv,\jv) \in \Fv^z\cap \Fv^{\Bv} } I( U_{\iv \sgoes \jv} ; A_{\iv \sgoes \jv}(\Fv^z\cap \Fv^{\Bv}) |  \pa_{\Sv}(U_{\iv \sgoes \jv}),Q ),
\label{eq:binning decoding boost one way}
}{\label{eq:one way binning decoding}}
for every $z$ and every set $\Fv^z \subseteq \Vv^z$ such that
\ea{
\pa_{\Sv}(\Fv^z) \subseteq \Fv^z.
\label{eq:condition decoding errors supeperposition and one-way binning}
}
%
%
Then, for any distribution of the terms $\{U_{\iv \sgoes \jv}\}$ that factors as in \eqref{eq:encoding distribution},
the rate region $\Rcal'\cap \Rcal$ is achievable.
\end{thm}

\begin{IEEEproof}
The complete proof is provided in  Appendix \ref{app:Decoding errors only supeperposition and one-way binning}.
The bounds in \eqref{eq:binnig rates one way binning} guarantee that the encoding error probability is vanishing while the bounds in
\eqref{eq:one way binning decoding} guarantee that the decoding error probability goes to zero as the block-length goes to infinity.

The codewords which possess a binning index are the codewords in $\Vv^{\Bv}$ as defined in \eqref{eq:definition VB}: these nodes are the non-source nodes in the binning graph.
Each term in \eqref{eq:binnig rates one way binning}  corresponds to the event that the binning indices in $\Fv^{\Bv}$
have been correctly determined while the binning indices in $\Fvo^{\Bv}$ are not.
A node can be correctly encoded only when its parents in the superposition coding graph have also been correctly encoded:
this condition is expressed by \eqref{eq:condition enocding errors one way binning}.
The probability of these error events is evaluated by constructing a Markov equivalent DAG GMM to the DAG GMM $\Gcal(\Vv,\Sv \cup \Bv)$ in which
the conditional distribution of the incorrectly encoded codewords factors as in \eqref{eq:conditional DAG general}.
This is the case when the parents of the nodes in $\Fv^{\Bv}$ are in $\Fv^{\Bv}$  as well, which is the case in the Markov equivalent DAG we have chosen.
%

As for Th. \ref{th:Decoding errors only supeperposition}, the possible decoding errors are those in which the parents in the superposition coding graph
of the correctly decoded nodes are also correctly decoded.
%
%
%
%
%
For this reason,  decoding error events are analyzed in a similar manner as in Th. \ref{th:Decoding errors only supeperposition} with the exception of
the additional term \eqref{eq:binning decoding boost one way}.
This term corresponds to a ``decoding boost'' provided by  binning.
In superposition coding the joint distribution between a codeword and its parents is the same whether the top codeword is correctly or incorrectly decoded.
%
%
This does not occur in binning.
To incorrectly decode a binned codeword, there must exist another codeword which looks as if generated conditionally
dependent with the correctly decoded parent codewords.
Unfortunately  this boost comes at the cost of having to decode both binning and message indices, which reduces the attainable message rate.
%
\end{IEEEproof}

The novel ingredient in Th. \ref{th:Decoding errors only supeperposition and one-way binning} with respect to Th. \ref{th:Decoding errors only supeperposition} is the encoding error analysis which results in the lower bound on the binning rates in \eqref{eq:binnig rates one way binning}.
The term in the RHS of \eqref{eq:binnig rates one way binning 1} relates to the overall difference between the encoding distribution and the codebook distribution.
The term in \eqref{eq:one way binning decoding} intuitively represents the distance between encoding and codebook distribution
for the incorrectly encoded codewords given the correctly encoded ones.
%
%
%
This conditional distribution cannot be easily evaluated in general: it can be compactly expressed only when $\pa_{\Bv}(\Fv^{\Bv}) \cap \Vv^{\Bv} \in \Fv^{\Bv}$, in which case
\eqref{eq:conditional DAG general} applies.
For this reason, Th. \ref{thm: DAG equivalence of graph representation} is used to produce an equivalent DAG for which this condition holds:
the existence of such a DAG for each $\Fv^{\Bv}$ can be guaranteed  only when Assumption \ref{ass:Binning is transitive}, the TB-restriction, holds.

\begin{figure}
\centering
\includegraphics[width=.78 \textwidth]{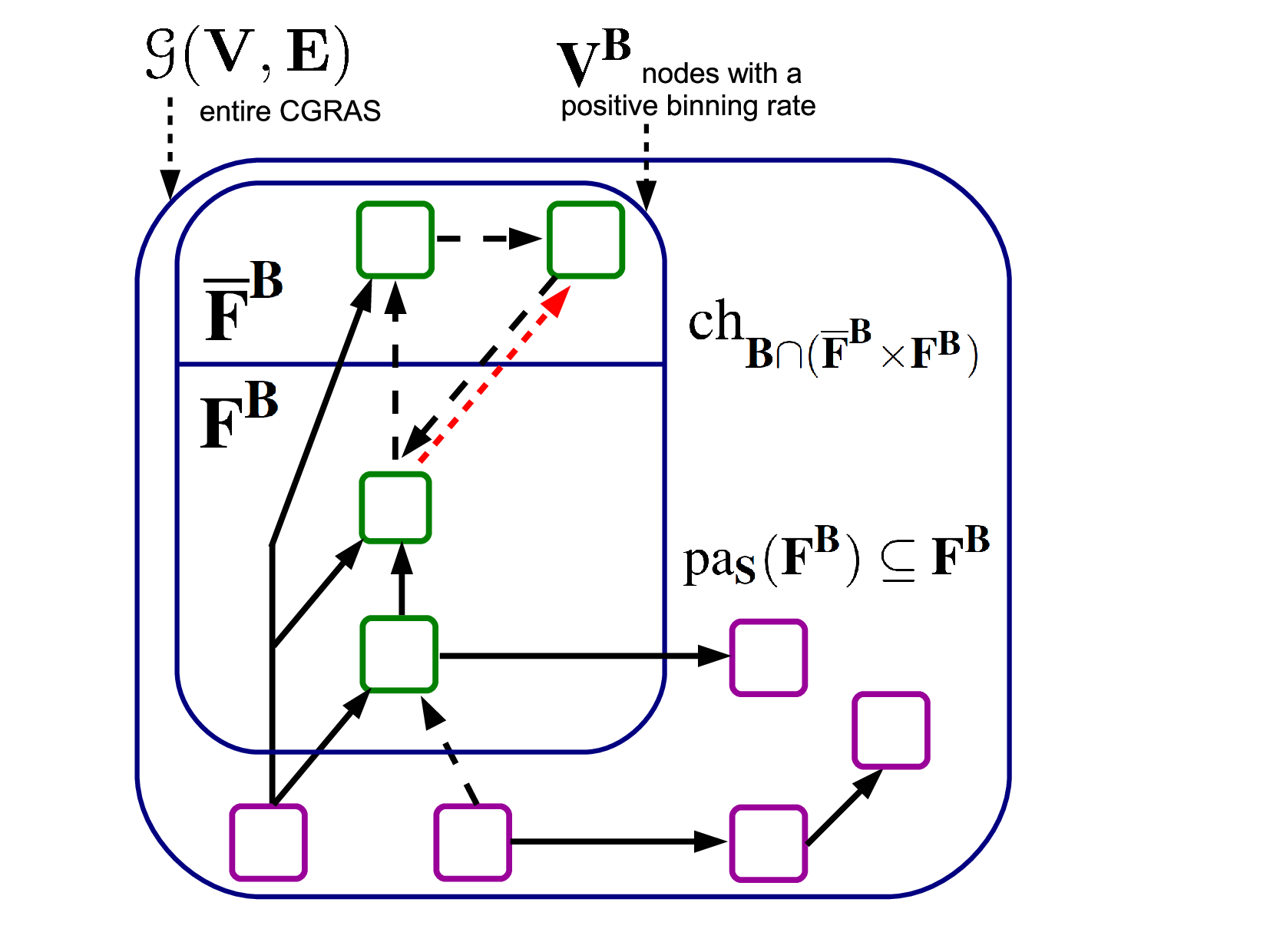}
\caption{A graphical representation of the binning rate bounds in Th. \ref{th:Decoding errors only supeperposition and one-way binning}. }
\label{fig:binningRep}
\end{figure}

A graphical representation of the construction of the Markov equivalent GMM to the encoding GMM is provided in  Fig. \ref{fig:binningRep}.
The set of valid $\Fv^{\Bv}$ are the ones for which parents of correctly encoded codewords are also correctly encoded.
The Markov equivalent DAG GMM  is constructed by reversing the direction of the edges from $\Fvo^{\Bv}$ to $\Fv^{\Bv}$: this set of edges is identified by the term $A_{\iv \sgoes \jv}$ in \eqref{eq:Aij definition binnig}.
This produces an equivalent DAG only under Assumption \ref{ass:Binning is transitive} (the TB-restriction).

%
%

\subsubsection{An example with superposition coding and binning}
\label{sec:An Example with superposition coding and binning}
\begin{figure}
\centering
\includegraphics[width=.78 \textwidth]{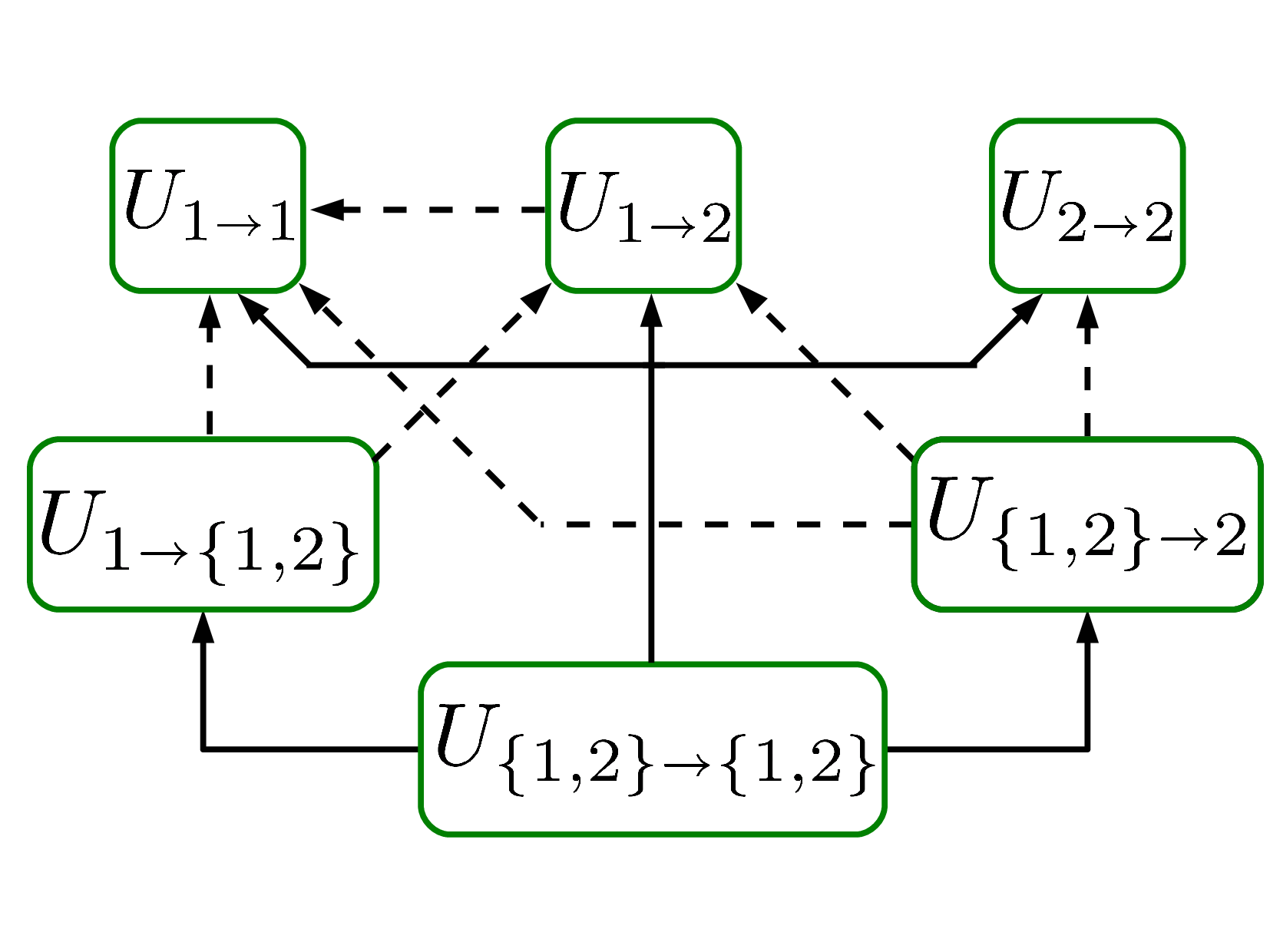}
\vspace{- 1 cm}
\caption{The chain graph for the achievable scheme in Sec. \ref{sec:An Example with superposition coding and binning}. }
\label{fig:superpositionOneWayBinning}
\vspace{-.5 cm}
\end{figure}

We next expand on the example already considered in Sec. \ref{sec:An Example with superposition coding} to include binning.
We again consider the user virtualization matrix in \eqref{eq:user virtualization CIFC} for the channel model in Sec. \ref{sec:An example of  rate-splitting}.
For this enhanced model, we consider the transmission strategy in the CGRAS of Fig. \ref{fig:superpositionOneWayBinning}.
The CGRAS in Fig. \ref{fig:superpositionOneWayBinning} involves less superposition coding steps than the graph in Fig. \ref{fig:superpositionOnly} in order to allow for more binning steps for us to analyze.
The codebook distribution (plus time-sharing) factorizes as
\ea{
& P^{\rm codebook}
\label{eq:codebook dist sup and binn}\\
&= P_{U_{ \{1,2\} \sgoes \{1,2\} },Q} \nonumber \\
& \quad  P_{U_{ 1 \sgoes \{1,2\} } | U_{ \{1,2\} \sgoes \{1,2\}},Q} \nonumber \\
& \quad  P_{U_{ \{1,2\} \sgoes 2} | U_{ \{1,2\} \sgoes \{1,2\}},Q}  \nonumber \\
& \quad  P_{ U_{1 \sgoes 1} | U_{ \{1,2\} \sgoes \{1,2\}},Q} \nonumber \\
& \quad  P_{ U_{2 \sgoes 2} | U_{ \{1,2\} \sgoes \{1,2\}},Q}  \nonumber \\
& \quad  P_{ U_{1 \sgoes 2} | U_{ \{1,2\} \sgoes \{1,2\}},Q}, \nonumber
}
while the encoding distribution (with coded time-sharing) factorizes as
\ea{
&P^{\rm encoding}
\label{eq: encoding dist sup and binn}\\
&= P_{U_{ \{1,2\} \sgoes \{1,2\} },Q} \nonumber \\
& \quad  P_{U_{ 1 \sgoes \{1,2\} } | U_{ \{1,2\} \sgoes \{1,2\}},Q} \nonumber \\
& \quad  P_{U_{ \{1,2\} \sgoes 2} | U_{ \{1,2\} \sgoes \{1,2\}},Q}  \nonumber \\
& \quad  P_{ U_{1 \sgoes 1} | U_{ 1 \sgoes \{1,2\} } ,  U_{ \{1,2\} \sgoes 2} , U_{ \{1,2\} \sgoes \{1,2\}},Q} \nonumber \\
& \quad  P_{ U_{2 \sgoes 2} | U_{ \{1,2\} \sgoes  2 } ,  U_{ \{1,2\} \sgoes \{1,2\}},Q}  \nonumber \\
& \quad  P_{ U_{1 \sgoes 2} | U_{ 1 \sgoes   \{1,2\}  } , U_{ \{1,2\} \sgoes  2 } ,  U_{ \{1,2\} \sgoes \{1,2\}},Q}. \nonumber
}
To obtain the achievable rate region of the scheme in Fig. \ref{fig:superpositionOneWayBinning} we need to identify all the possible sets $\Fv^{\Bv}$, $\Fv^1$ and $\Fv^2$.
Given one such set, we have to consider a distinct Markov equivalent GMM to the encoding distribution in which the edges cross from $\Fv$ to $\Fvo$.
For brevity, we consider here only some sets $\Fv^{\Bv}$ and evaluate \eqref{eq:binnig rates one way binning} in these examples.
%

We begin by evaluating the term \eqref{eq:binnig rates one way binning 1} which quantifies the overall distance between encoding and decoding distributions.
The nodes $U_{\{1,2\}\sgoes \{1,2\}}, U_{\{1,2\}\sgoes \{1,2\}}$ and $U_{\{1,2\}\sgoes \{1,2\}}$  are not involved in binning, so they don't appear in $\Vv^{\Bv}$.
\eas{
& \sum_{(\iv,\jv) \in \Vv^{\Bv}} I(U_{\iv \sgoes \jv} ; \pa_{\Bv} \{ U_{\iv \sgoes \jv}\} | \pa_{\Sv} \{ U_{\iv \sgoes \jv}\} , Q)\\
& = I(U_{1 \sgoes 1}; U_{1 \sgoes \{1,2\}}, Q)  \\
&  \quad +I(U_{1 \sgoes 1}; U_{1 \sgoes \{1,2\}},U_{ \{1,2\} \sgoes 2} | U_{\{1,2\}\sgoes \{1,2\}},Q)  \\
&  \quad +I(U_{2 \sgoes 2}; U_{\{1,2\} \sgoes 2} | U_{\{1,2\}\sgoes \{1,2\}},Q) \\
&  \quad +I(U_{1 \sgoes 2}; U_{\{1,2\}\sgoes 2}, U_{\{1,2\} \sgoes 2} | U_{\{1,2\}\sgoes \{1,2\}},Q).
}{\label{eq:distance enrcoding codebook dist sup and binning}}
Next, we evaluate terms in \eqref{eq:binnig rates one way binning 3} for three possible $\Fvo^{\Bv}$
\eas{
\Fvo^{\Bv}& =\{ 1\sgoes 1, 1 \sgoes 2 \},
\label{eq:example set sup and binn 1}\\
\Fvo^{\Bv}& =\{ 1\sgoes 1 \}
\label{eq:example set sup and binn 2}\\
\Fvo^{\Bv}& =\{ 1\sgoes 2 \}
\label{eq:example set sup and binn 3},
}{\label{eq:example set sup and binn}}
For the case $\Fvo^{\Bv}=\{ 1\sgoes 1, 1 \sgoes 2 \}$ the term in \eqref{eq:binnig rates one way binning 3} is obtained as
\ea{
\eqref{eq:binnig rates one way binning 3}
&=-I(U_{1 \sgoes 1}; U_{1 \sgoes 2} ,U_{ \{12\} \sgoes 2}, U_{1 \sgoes \{1,2\}} | U_{ \{12\} \sgoes \{1,2\}},Q) \nonumber \\
& \quad -I(U_{1 \sgoes 2}; U_{ \{12\} \sgoes 2}, U_{1 \sgoes \{1,2\}} | U_{ \{12\} \sgoes \{1,2\}},Q).
}
Since there are no edges crossing from $\Fvo^{\Bv}$ into $\Fv^{\Bv}$, $A_{\iv \sgoes \jv}(\Fv^{\Bv}) =  \pa_{\Bv}(U_{\iv \sgoes \jv})$.

\smallskip

For the case $\Fv^{\rm \Bv}=\{ 1\sgoes 1, 1 \sgoes 2 \}$: we have
\eas{
\eqref{eq:binnig rates one way binning 3}
&=-I(U_{1 \sgoes 1}; U_{1 \sgoes 2} ,U_{ \{12\} \sgoes 2}, U_{1 \sgoes \{1,2\}} | U_{ \{12\} \sgoes \{1,2\}},Q)  \\
&\quad -I (U_{1 \sgoes 2}; U_{ \{12\} \sgoes 2}, U_{1 \sgoes \{1,2\}} | U_{ \{12\} \sgoes \{1,2\}},Q).
\label{eq: F 1sgoes1 and 1sgoes2}
}
Again, since there are no edges crossing from $\Fvo^{\Bv}$ into $\Fv^{\Bv}$, $A_{\iv \sgoes \jv}(\Fv^{\Bv}) =  \pa_{\Bv}(U_{\iv \sgoes \jv})$.

\smallskip

Finally, for the case $\Fv^{\Bv}=\{1 \sgoes 2 \}$ we have
\ea{
\eqref{eq:binnig rates one way binning 3} & = -I( U_{1 \sgoes 2}; U_{1\sgoes 1}, U_{ 1 \sgoes \{1,2\}}, U_{ \{1,2\} \sgoes 2}  | U_{ \{1,2\} \sgoes \{1,2\}},Q),
}
since, in this case, we take the Markov equivalent GMM in which the direction of the edge $U_{1 \sgoes 2} \bin U_{1\sgoes 1}$ is switched to
 $U_{1 \sgoes 1} \bin U_{1\sgoes 2}$, so that all the edges cross from $\Fvo^{\Bv}$ to $\Fv^{\Bv}$.


\subsection{Superposition coding, binning and joint binning}
\label{sec:Superposition Coding and Joint Binning}

We  now consider the most general CGRAS which encompasses superposition coding, binning and joint binning
%
%
In Sec. \ref{sec:Superposition Coding}, the CGRAS with only superposition coding is studied: in this case, the achievable rate region is obtained from the
decoding error analysis and depends solely on the codebook GMM.
%
In Sec. \ref{sec:Superposition Coding and One-way Binning} we consider the CGRAS with both superposition coding and binning.
For this scenario, it is necessary to analyze both encoding and decoding error events and the achievable rate region
depends on the properties of both the codebook and the encoding GMMs.
The achievable rate region for this class of CGRASs is obtained by considering a series of Markov equivalent GMMs to the encoding GMM which are both DAGs.
Assumption \ref{ass:Binning is transitive}, the TB-restriction, is necessary to guarantee that an equivalent DAG exists, which makes it possible to compactly express the achievable rate region.

For the general case we consider next, the encoding GMM is no longer a DAG, as in the previous two sections, but rather a CG.
For this more general class of GMMs,  both Assumption \ref{ass:Binning is transitive}  and Assumption \ref{ass:Joint binning forms cliques}, the TB-restriction and the CSJB-restriction respectively, are necessary
to construct a series of Markov equivalent DAGs to the encoding GMM and obtain an expression of the rate bounds of the achievable region as in Th. \ref{th:Decoding errors only supeperposition and one-way binning}.

\begin{thm}{\bf Achievable region with superposition coding, binning and joint binning \\}
\label{th:Decoding errors only supeperposition and joint binning}
Consider any CGRAS for which Assumption \ref{ass:Binning is transitive}  and Assumption \ref{ass:Joint binning forms cliques}, the TB-restriction and the CSJB-restriction respectively, holds and let
the region $\Rcal'$ be defined as
\eas{
& \sum_{(\iv,\jv) \in \Fv^{\Bv}}  R'_{\iv \sgoes \jv}  \geq  \sum_{(\iv,\jv) \in \Vv^{\Bv}} I(U_{\iv \sgoes \jv} ; \pa_{\Bvt} \{ U_{\iv \sgoes \jv}\} | \pa_{\Sv} \{ U_{\iv \sgoes \jv}\} ,Q)
\label{eq:binnig rates joint term 1}  \\
& - \sum_{(\iv,\jv) \in \Fv^{\Bv}} I(U_{\iv \sgoes \jv} ; A_{\iv \sgoes \jv}(\Fv^{\Bv}) | \pa_{\Sv} (U_{\iv \sgoes \jv}) , Q),
\label{eq:binnig rates joint term 2}
}{\label{eq:binnig rates joint}}
with
\ea{
A_{\iv \sgoes \jv}(\Fv^{\Bv}) =\pa_{\Bvt} (U_{\iv \sgoes \jv}) \cup \ad_{\Bv \cap (\Fvo^{\Bv} \times \Fv^{\Bv})} (U_{\iv \sgoes \jv}),
\label{eq:A ij}
}
for some non-cyclic orientation $\Bvt$ of the edges in $\Bv$ in $\Gcal(\Vv,\Sv \cup \Bv)$
and for any $\Fv^{\Bv} \subseteq \Vv^{\Bv}$ such that
\ea{
\pa_{\Sv}(\Fv^{\Bv}) \cap  \Fv^{\Bv} \subseteq \Fv^{\Bv}.
\label{eq:condition enocding errors one way binning}
}
Also, let the region $\Rcal$ be defined as
\ea{
& \sum_{(\iv,\jv) \in \Fvo^z} L_{\iv \sgoes \jv}\leq   \nonumber \\
& \quad I( Y_z; \{ U_{\iv \sgoes \jv}, \ (\iv,\jv) \in \Fvo^z \} | \{ U_{\iv \sgoes \jv}, \ (\iv,\jv) \in \Fv^z \} , Q) \nonumber \\
& \quad \quad + \sum_{(\iv,\jv) \in \Fv^z\cap \Fv^{\Bv} } I( U_{\iv \sgoes \jv} ;  A_{\iv \sgoes \jv}(\Fv^z\cap \Fv^{\Bv}) |  \pa_{\Sv}(U_{\iv \sgoes \jv}), Q ),
\label{eq:message rates}
}
for all $z$ and all the sets $\Fv^z \subseteq \Vv^z$ such that
\ea{
\pa(\Fv^z) \subseteq \Fv^z.
}
Then, for any distribution of the terms $\{U_{\iv \sgoes \jv}\}$ that factors as in \eqref{eq:encoding distribution}, the rate region $\Rcal'\cap \Rcal$ is achievable.
\end{thm}
\begin{IEEEproof}
The proof, provided in Appendix \ref{app:Decoding errors only supeperposition and joint binning}, is similar to the proof of Th. \ref{th:Decoding errors only supeperposition and one-way binning}, although a more detailed analysis of the encoding and decoding error event is necessary.
In joint binning, the choice of one binning index depends on the choice of another binning index.
%
%
In particular, the region $\Rcal'$ in \eqref{eq:binnig rates joint} is similar to the region $\Rcal'$ in \eqref{eq:binnig rates one way binning}
in that, in both regions,  the set $\Fv^{\Bv}$ intuitively relates to the set of correctly encoded codewords while $\Fvo^{\Bv}$ relates to the set of incorrectly encoded codewords.
The difference in the two regions \eqref{eq:binnig rates joint} and \eqref{eq:binnig rates one way binning} is how the Markov equivalent DAG is constructed from the encoding CG GMM.
In \eqref{eq:binnig rates joint}, a Markov equivalent DAG is constructing from the encoding CG GMM  for each given set $\Fv^{\Bv}$.
This DAG is constructed so that the direction of the undirected edges in $\Gcal(\Vv,\Sv \cup \Bv)$ is chosen to cross from the set $\Fv^{\Bv}$ to the set $\Fvo^{\Bv}$.
%
%
%
The region in  \eqref{eq:message rates} is obtained from the decoding error analysis and uses a similar construction of the Markov equivalent DAG
 as in \eqref{eq:binnig rates one way binning}  but restricted to $\Vv^z$, the subset of graph observed at receiver $z$.
\end{IEEEproof}

\begin{figure}
\centering
\includegraphics[width=.78 \textwidth]{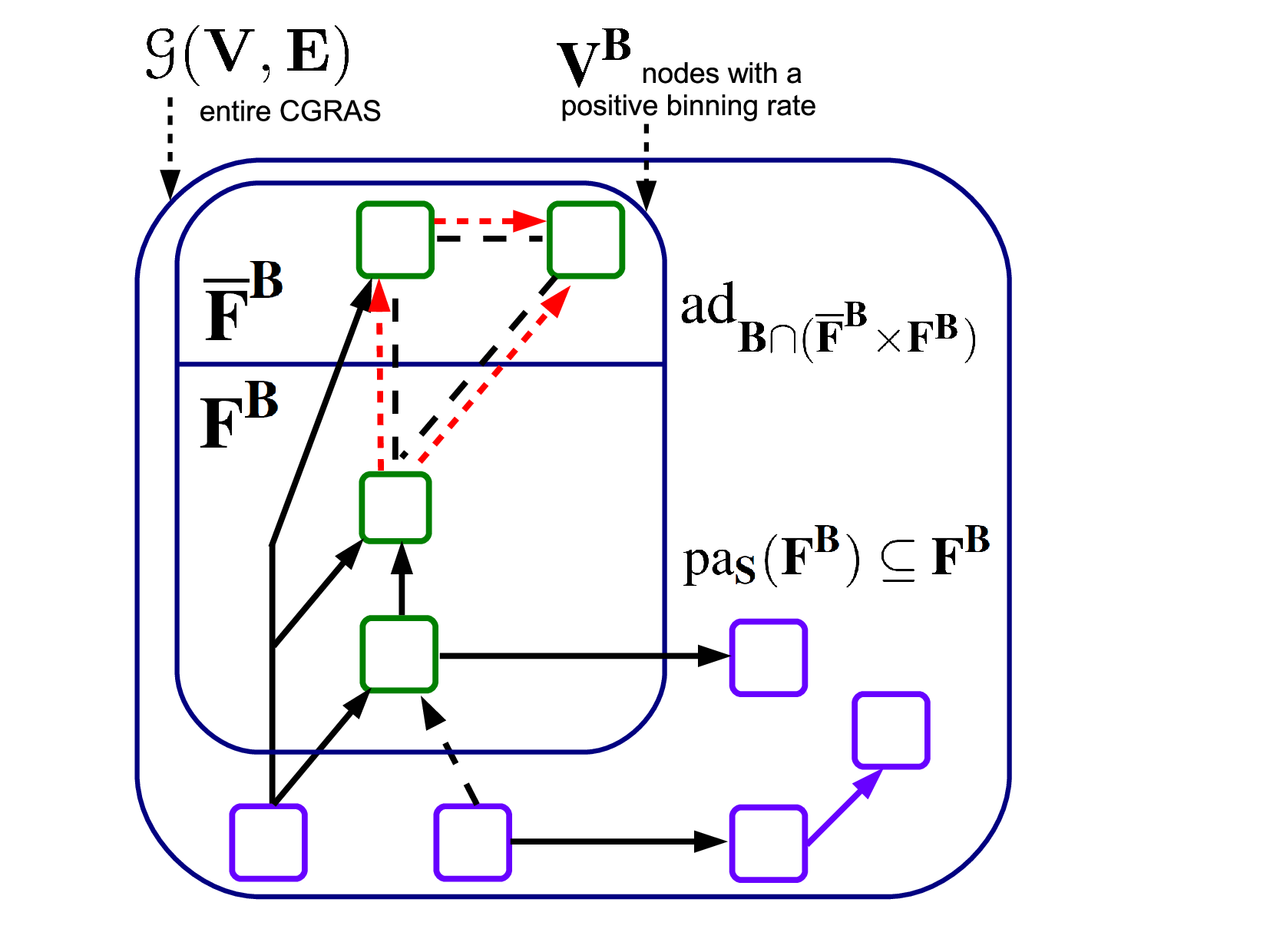}
\caption{A graphical representation of the binning rate bounds with joint binning in Th. \ref{th:Decoding errors only supeperposition and joint binning}. }
\label{fig:JointBinningRep}
\end{figure}

A graph representation of Th.  \ref{th:Decoding errors only supeperposition and joint binning} is provided in  Fig. \ref{fig:JointBinningRep}:
as in the proof of Th. \ref{th:Decoding errors only supeperposition and one-way binning}, the term in the RHS of \eqref{eq:binnig rates joint term 1} relates to the overall distance between encoding and codebook distribution.
The term in \eqref{eq:A ij} is also similar to the term in \eqref{eq:Aij definition binnig} and is obtained by choosing the Markov-equivalent DAG in which the direction of the undirected edges is chosen so as to cross from $\Fvo$ to $\Fv$.
%
%

A graphical representation of the construction of the Markov equivalent GMM to the encoding GMM is provided in Fig. \ref{fig:JointBinningRep}.
The set of valid $\Fv^{\Bv}$ are the ones for which the parents of the correctly encoded codewords is also correctly encoded.
The Markov equivalent DAG GMM  is constructed by choosing an acyclic orientation of the binning edges, directed or undirected, such that edges cross from $\Fvo^{\Bv}$ to $\Fv^{\Bv}$: this set of edges is identified by the term $A_{\iv \sgoes \jv}$ in \eqref{eq:A ij}.
This produces an equivalent DAG only under Assumption \ref{ass:Binning is transitive} and Assumption \ref{ass:Joint binning forms cliques}, the TB-restriction and the CSJB-restriction respectively.

%
%
%

\subsubsection{An example with superposition coding, binning and joint binning}
\label{sec:An Example with superposition coding and joint binning}

\begin{figure}
\centering
\includegraphics[width=.78 \textwidth]{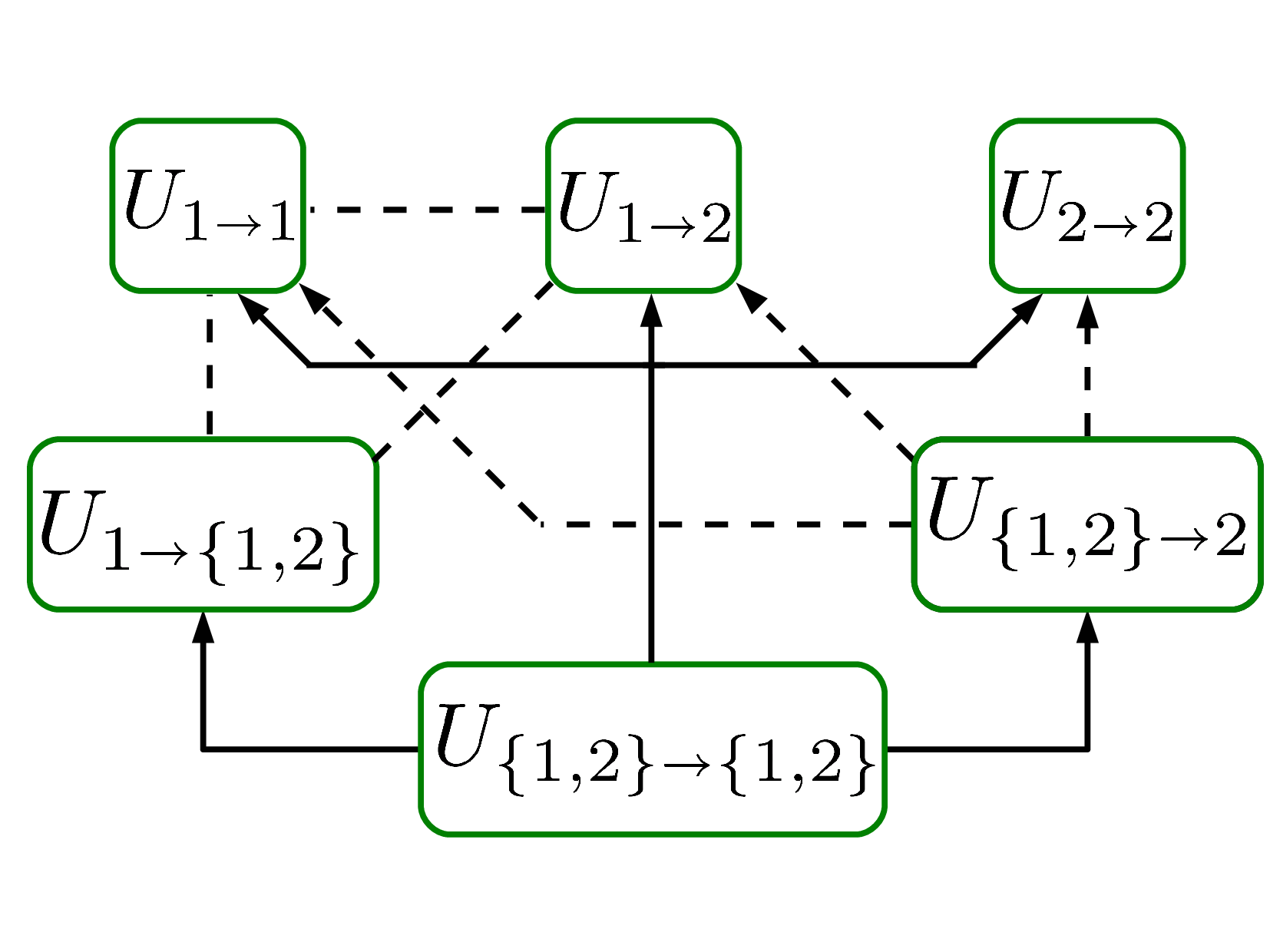}
\vspace{-.5 cm}
\caption{The chain graph for the achievable scheme in Sec. \ref{sec:An Example with superposition coding and joint binning}. }
\label{fig:superpositionJointBinning}
\vspace{-.5 cm}
\end{figure}

We next expand on the example  in Sec. \ref{sec:An Example with superposition coding and binning} to include joint binning.
Consider the CGRAS in Fig. \ref{fig:superpositionJointBinning}: with respect to the CGRAS in Fig. \ref{fig:superpositionOneWayBinning}, this CGRAS has been enhanced with further binning edges.
In this graph there is a set of jointly binned nodes which forms a complete subset, that is: $U_{1 \sgoes 1} \jbin U_{1 \sgoes \{1,2\}}$, $U_{1 \sgoes 1} \jbin U_{1 \sgoes 2}$ and $U_{1 \sgoes 2} \jbin U_{1 \sgoes \{1,2\}}$.
%
%
%
%
The encoding and decoding distributions  of the scheme in Fig. \ref{fig:superpositionJointBinning} are the same as in Fig. \ref{fig:superpositionOneWayBinning},
in  \eqref{eq:codebook dist sup and binn} and \eqref{eq: encoding dist sup and binn} respectively.
This is because the DAG in Fig. \ref{fig:superpositionOneWayBinning} is a non-cyclic orientation of the chain graph in Fig. \ref{fig:superpositionJointBinning}.
Given this consideration, we also have that the term in the RHS of \eqref{eq:binnig rates joint term 1} is equal to the term in
\eqref{eq:distance enrcoding codebook dist sup and binning}.
We return now to the subsets in \eqref{eq:binnig rates joint term 2} for the scheme with joint binning in Fig. \ref{fig:superpositionJointBinning}.
For the subset in \eqref{eq:example set sup and binn 1}, $\Fvo=\{1\sgoes 2, 1 \sgoes 1\}$.
We choose the orientation of the jointly binned edges as
\eas{
U_{1 \sgoes \{1,2\}} & \bin U_{1 \sgoes 2} \\
U_{1 \sgoes \{1,2\}} & \bin U_{1 \sgoes 1} \\
      U_{1 \sgoes 1} & \bin U_{1 \sgoes 2},
}
and this term can therefore be evaluated as
\ean{
\eqref{eq:binnig rates joint term 2} & = - I(U_{1 \sgoes 1}; U_{1 \sgoes \{1,2\}}, U_{\{1,2\} \sgoes 2}| U_{ \{1,2\} \sgoes \{1,2\}},Q) \\
& \quad - I(U_{1 \sgoes 2}; U_{1 \sgoes 1}, U_{1 \sgoes \{1,2\}}, U_{\{1,2\} \sgoes 2} |  U_{ \{1,2\} \sgoes \{1,2\}},Q).
}

For the term $\Fvo=\{ 1 \sgoes 1\}$ we choose the orientation of the edges as in Fig. \ref{fig:superpositionOneWayBinning}:
\eas{
U_{1 \sgoes \{1,2\}} & \bin U_{1 \sgoes 2} \\
U_{1 \sgoes \{1,2\}} & \bin U_{1 \sgoes 1} \\
      U_{1 \sgoes 1} & \bin U_{1 \sgoes 2}.
}
For this reason the term in \eqref{eq:binnig rates joint term 2} is equal to the term in \eqref{eq: F 1sgoes1 and 1sgoes2}.

For the last term, $\Fvo=\{1 \sgoes 2\}$,  we choose the following orientation of the edges:
\eas{
U_{1 \sgoes \{1,2\}} & \bin U_{1 \sgoes 2} \\
U_{1 \sgoes \{1,2\}} & \bin U_{1 \sgoes 1} \\
      U_{1 \sgoes 1} & \bin U_{1 \sgoes 2},
}
and the term in \eqref{eq:binnig rates joint term 2}  is equal to
\ea{
\eqref{eq:binnig rates joint term 2}=-I( U_{1 \sgoes 2}; U_{1\sgoes 1}, U_{ 1 \sgoes \{1,2\}}, U_{ \{1,2\} \sgoes 2}  | U_{ \{1,2\} \sgoes \{1,2\}},Q).
}

The derivation of the remaining terms and, therefore, of the overall achievable region follows similarly from the derivation of the terms above.
\subsection{Non-unique decoding}
\label{sec:Non-unique decoding}

In \cite{chong2008han} the concept of non-unique decoding is introduced:  in the Han and Kobayashi region for the IFC \cite{Han_Kobayashi81} each receiver decodes the common message from the other, interfering user.
In the original analysis of \cite{Han_Kobayashi81}, the rate of the common messages was bounded so that the correct decoding of the common parts was possible at both decoders.
In fact, the correct decoding of these common messages  is not necessary at the non-intended receivers and this error event can be disregarded.
This, in turn, implies that some of the bounds in the achievable rate region can be dropped.

This approach can be applied to the general framework we consider as follows.

\begin{thm}{ \bf Non-unique decoding\\}
\label{th:Non-unique decoding}
Consider a user virtualization matrix $\Gamma$ and a  CGRAS $\Gcal(\Vv, \Ev)$.
Let moreover $W_{\Vv^{\rm O}}$ be the message allocation in the original channel model and $W_{\Vv}$ the message allocation in the enhanced model.
The achievable regions in Th. \ref{th:Decoding errors only supeperposition},  Th. \ref{th:Decoding errors only supeperposition and one-way binning} and Th.\ref{th:Decoding errors only supeperposition and joint binning} can be enlarged by considering only those subsets $\Fv^z$ such that
\ea{
\exists \ (\lv,\mv) \in \Fvo^z \ \ST \ \Gamma_{ (\iv,\jv) \times (\lv,\mv)} \neq 0  \ {\rm for \ some} \ \jv,  z \in \jv , \ (\iv,\jv) \in \Vv^{\rm O},
\label{eq:non unique decoding condition}
}
%
for the region $\Rcal$ in  \eqref{eq:condition decoding errors superposition}, \eqref{eq:one way binning decoding} and \eqref{eq:message rates}.
\end{thm}

\begin{IEEEproof}
The set $\Fv^z$ is used in deriving the decoding error analysis in the achievability theorems for all theorems above.
Each decoder $z$ is only interested in the codewords that carry part of the messages $W_{\iv \sgoes \jv}, \ (\iv,\jv) \in \Vv^{\rm O}, \ z \in \jv$.
When a message $W_{\lv \sgoes \mv}, \ (\lv,\mv) \in \Vv, \ z \in \mv$ is obtained through user virtualization and it does not carry part of the message $W_{\iv \sgoes \jv}$, its incorrect decoding has no influence on the error performance of the decoder.
%
%
For this reason, when $\Gamma_{ (\iv,\jv) \times (\lv,\mv)}=0$ for all the $\jv$ such that $z \in \jv$, this decoding error event can be neglected.
\end{IEEEproof}

Th. \ref{th:Non-unique decoding} states  that a decoding error in the enhanced channel can be disregarded when all the incorrectly decoded messages at receiver $z$ do not carry information regarding the messages decoded at this receiver.
This occurs when the set of incorrectly decoded messages $\Fvo^z$ contains random variables which do not embed part of the messages to be decoded at $z$ in the original problem $\Wv^{\rm O}$.
As a consequence, the corresponding upper bound on message and binning rate can be disregarded.

Non unique decoding is not likely to enlarge the achievable region when both the codeword $U_{\iv \sgoes \jv}$  and the message $U_{\iv \sgoes \jv\setminus z}$ for all the
$z$ in \eqref{eq:non unique decoding condition}.
In this case, the rate of the message $U_{\iv \sgoes \jv}$ decoded at receiver $z$ can be shifted to the codeword $U_{\iv \sgoes \jv\setminus z}$  so that the corresponding rate bound is no longer active.
This essentially corresponds to the approach in \cite{bidokhti2012non} where an alternative interpretation and proof techniques for joint decoding are proposed.
In \cite{bidokhti2012non}, the problem of non-unique decoding is translated to the problem of unique decoding in two achievable schemes: in one the codeword is decoded at a given decoder while, in the second, it is treated as noise.
Nonetheless, the achievable rate expression in Th. \ref{th:Non-unique decoding} remains very useful as it makes it possible to reduce the number of rate bounds before considering the union over the possible rate splitting matrices.

%
%
%
%

%% file: ConcludingRemark.tex
\section{New Achievable Regions based on the Proposed Framework}
\label{sec:ConcludingRemark}

This section provides examples of using the framework of
 Sec. \ref{sec:The Achievable Rate Region of the CGRAS} to improve upon or derive new achievable rate regions for  two channel models:
 the interference channel with a common message and the relay assisted down-link cellular network.
The aim of the first example is to illustrate the role of user virtualization and rate-sharing while the aim of the second example is to delineate the numerical derivation of achievable rate regions for a large network.
Our aim in this section is to highlight some relevant aspects of our result, rather than a systematic study of these models.
For this reason we shall present them from a high level perspective, while details are deferred to our publications \cite{rini2013interference} and \cite{rini2013rate} respectively.

\subsection{The interference channel with a common messages}

\begin{figure}
\centering
\includegraphics[width=.75 \textwidth ]{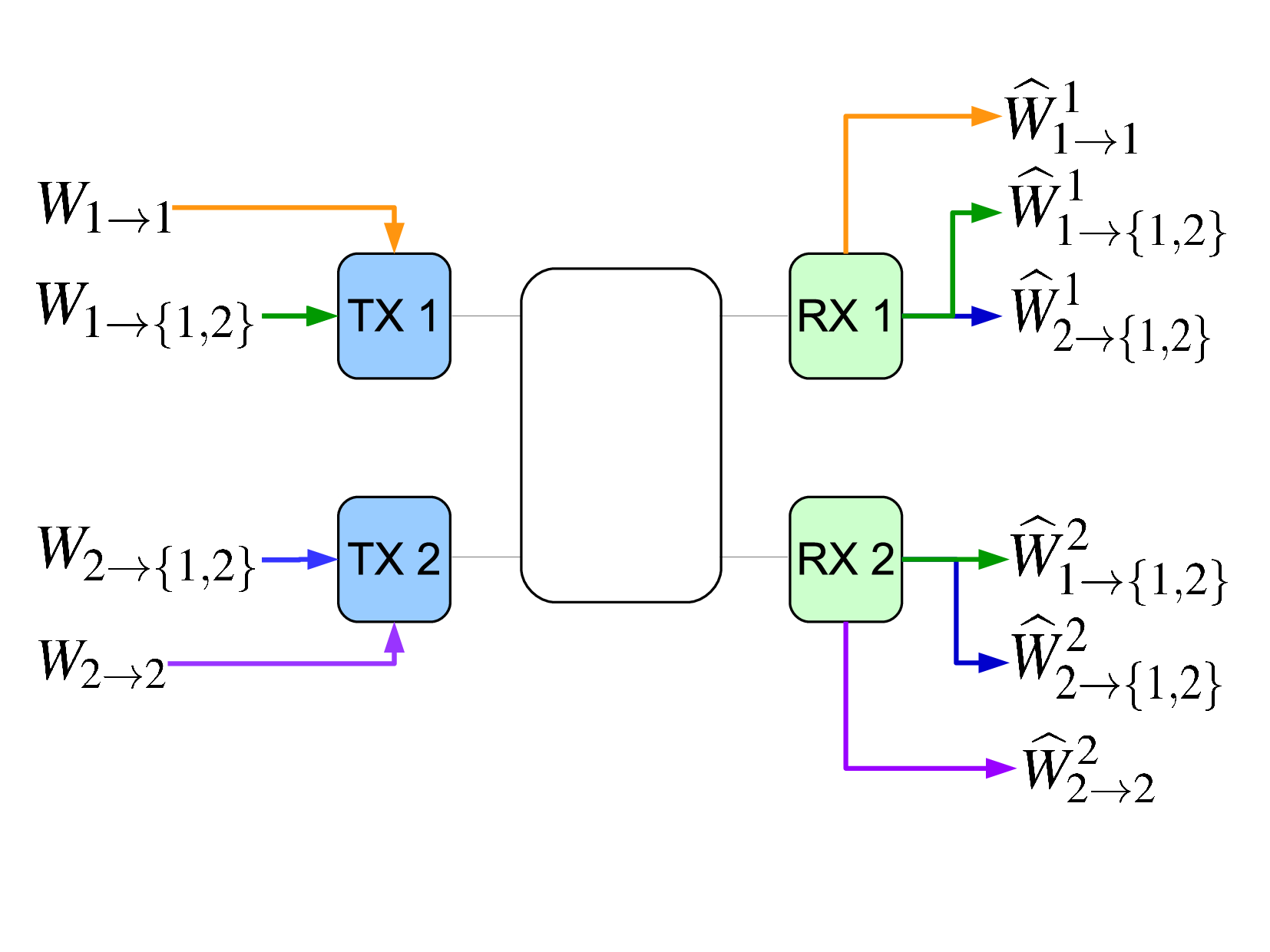}
\vspace{- 1 cm}
\caption{The Interference Channel with Common Messages (IFC-CMs).}
\label{fig:IFC-CMs}
\vspace{-.5 cm}
\end{figure}

Consider the classical IFC $\Ccal(W_{1 \sgoes 1},W_{2 \sgoes 2})$: when deriving an achievable region for this model user virtualization produces
the enhanced model $\Ccal(W_{1 \sgoes 1},W_{2 \sgoes 2},W_{1 \sgoes \{1,2\}},W_{2 \sgoes \{1,2\}})$, also depicted in Fig. \ref{fig:IFC-CMs}.
In the model of Fig. \ref{fig:IFC-CMs},  each transmitter also possesses a common message to be decoded at both receiver, for this reason we term it IFC with Common Messages (IFC-CMs).
%
%
%
The mapping between original and enhanced model is
\ea{
\lsb\p{
R'_{1 \sgoes 1 } \\
R'_{2 \sgoes 2 }
}\rsb
=
\lsb \p{
1  & 1  & 0 & 0 \\
0  & 0  & 1 & 1
}\rsb
\cdot
\lsb \p{
R_{1 \sgoes 1 } \\
R_{1 \sgoes \{1,2\}} \\
R_{2 \sgoes 2 }  \\
R_{2 \sgoes \{1,2\}}
}\rsb.
\label{eq:mapping IFC}
}
The mapping in \eqref{eq:mapping IFC} is rather straightforward: the original private rates are the sum of the common and private rates is the enhanced model.
This is  a one-to-one mapping, unlike the mapping in \eqref{eq:projection 2}, and one does not need to consider the union over all possible rate-sharing matrices.
Given the region involving superposition coding from \cite{Han_Kobayashi81}, \cite{ChongMotaniSemiDet} showed that the Fourier Motzkin Elimination (FME) can be used to obtain
a corresponding region for the IFC.
Unfortunately the expression that one obtains with the FME  contains some redundant bounds and in \cite{ChongMotaniSemiDet} a proof is developed to discard such redundant bounds.
Unfortunately this proof is tailored to this model and is not clear how one would identify redundant bounds for other models.

We provide an alternative and simpler proof to that of \cite{Han_Kobayashi81} by noticing that there exists a one-to-many mapping for the channel
 $\Ccal(W_{1 \sgoes 1},W_{2 \sgoes 2},W_{1 \sgoes \{1,2\}},W_{2 \sgoes \{1,2\}})$ onto itself that can be used to enhance an achievable region for this model.
 This mapping is
\ea{
\lsb \p{
R_{1 \sgoes 1} \\
R_{1 \sgoes \{1,2\}}\\
R_{2 \sgoes 2}\\
 R_{2 \sgoes \{1,2\}}
}\rsb
=
\lsb \p{
1 & \Delta _1    & 0 & 0 \\
0 & 1 -\Delta _1 & 0 & 0 \\
0 & 0 & 1 & \Delta _2    \\
0 & 0 & 0 & 1-\Delta _2    \\
} \rsb
\cdot
\lsb \p{
R_{1 \sgoes 1} \\
R_{1 \sgoes \{1,2\}}\\
R_{2 \sgoes 2}\\
 R_{2 \sgoes \{1,2\}}
}\rsb.
\label{eq:mapping IFC-CMs}
}
for some $0 \leq \Delta_i \leq 1$.
As for the mapping in \eqref{eq:mapping IFC}, the mapping in \eqref{eq:mapping IFC-CMs} is also rather intuitive: given any achievable rate vector for the IFC-CMs,
it is always possible to let the common codeword carry part of the private message.
This self-mapping  is unlikely to provide an improvement of the Han and Kobayashi region for the IFC after considering the union over
all the possible distributions.
On the other hand, performing the Fourier-Motzkin elimination of the variable $\Delta_1$ and $\Delta_2$ from the Han and Kobayashi region
can result in a larger achievable region  for a fixed input distribution.
This is particularly useful in proving capacity since the union over distributions is often not explicitly evaluated,
but rather inner and outer bounds are compared for fixed distributions.
Indeed in \cite{rini2013interference} we show that, with this approach, one can obtain the simplified region of \cite{ChongMotaniSemiDet} \
without making use of the tailored approach of  \cite{ChongMotaniSemiDet}.
This proof simply requires one to consider all possible rate-sharing strategies of a given model, so it can be easily extended to other channels.
%

%
%
%
%

\subsection{The relay assisted down-link cellular network}

\begin{figure}
\centering
\includegraphics[width=.75 \textwidth ]{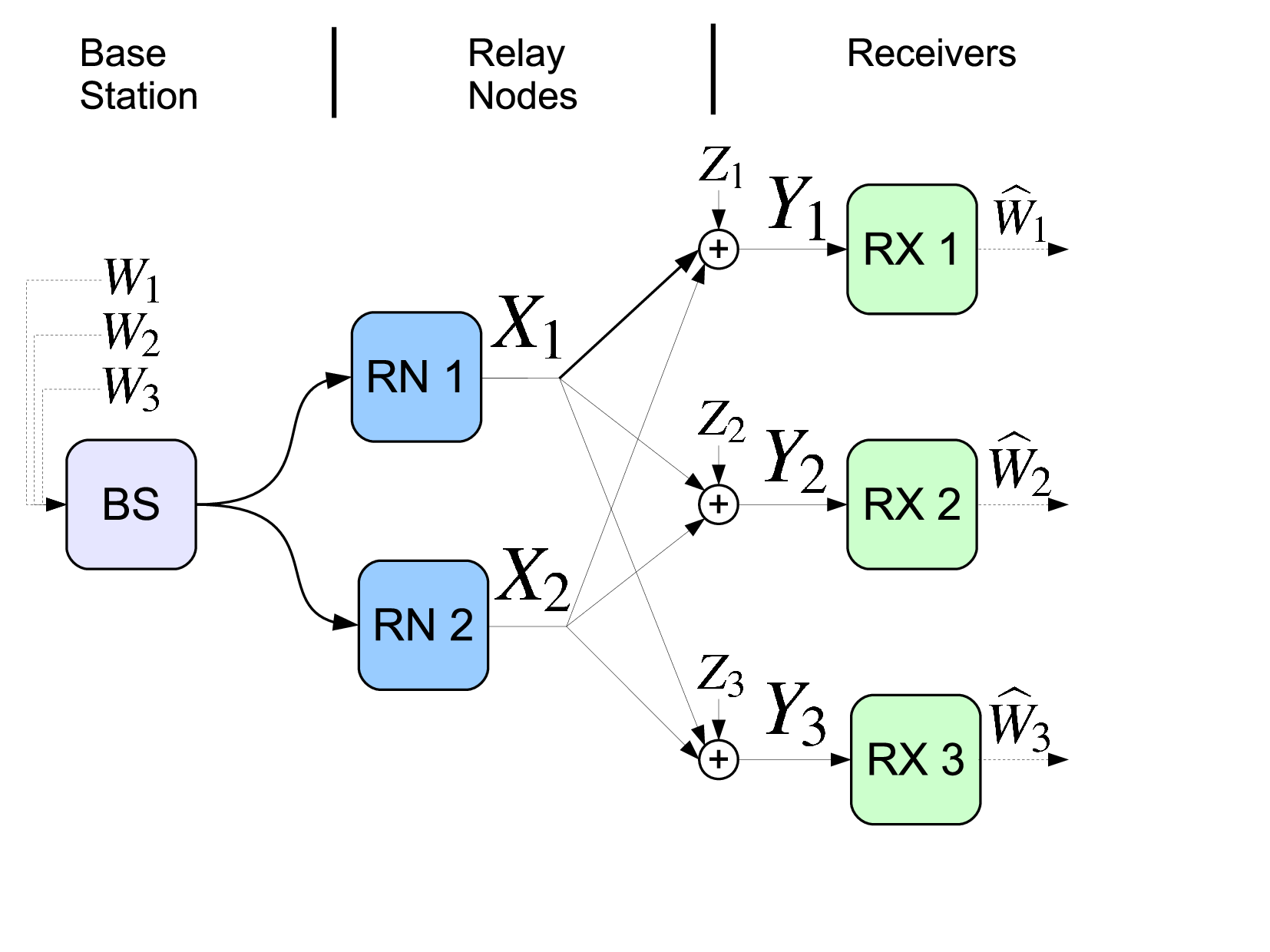}
\vspace{- 1 cm}
\caption{The relay assisted down-link cellular system with 2 relays and 3 receivers.}
\label{fig:3channel}
\vspace{-.5 cm}
\end{figure}

As an example of the numerical derivation of attainable rate regions, consider the relay assisted down-link cellular system with 2 relays and 3 receivers in Fig. \ref{fig:3channel}.
This is the model which we study in \cite{rini2013rate} and well illustrates how numerical methods can be applied to the proposed approach to investigate the capacity of large networks.
%

%
%

In the model of Fig. \ref{fig:3channel}, a base station wishes to communicate a message to each of three receivers via two relays.
The base station has a dedicated  channel toward each relay node while the transmissions between relay nodes and receivers are
cross- interfering and affected by additive Gaussian noise.
%
Given the values of the channel gains
and the power constraint at both the base station and the relays,
we would like to choose the message allocation  at the relays  and their transmission strategy which yield the largest throughput.
Although extremely useful from a practical perspective, it is not clear how this problem can be tackled in an effective manner.
Each possible message allocations corresponds to a different set of possible transmission strategies between relays and receivers.
There are in total $2^3=8$ possible message allocations and different transmission strategies can be developed in each scenario.
%

%
Unfortunately there are only a small number of transmission strategies at the relay that be conveniently expressed or evaluated numerically.
For instance,  we can consider the case in which a message is allocated only to one relay node and each receiver treats the interference as noise.
This strategy minimizes the rate of communication between base station and relay nodes but does not allow for any interference management.
The achievable region of this strategy is described by three bounds, each of the type $1/2 \log(1+SINR_i)$ for $i \in \{1,2,3\}$ and one needs to compare ${3 \choose 2} =3$
 possible message allocations at the relays.
This strategy is conceptually simple and can be easily extended to a channel with  any number of relays, $n_{\rm RN}$, and receivers, $n_{\rm  RN}$:
the number of bounds defining the achievable region is equal to $n_{\rm  RX}$, and the number of possible message allocations  is  equal to $n_{\rm  RX} \choose n_{\rm  RN}$.
%

Another strategy that can be considered is the virtual MIMO BC approach: the base station distributes all the messages to all the relay nodes and all
the receivers decode all the messages.
In this case the achievable region has the form of $R_{\sum} \leq \min_{i} I(Y;X)$ where $R_{\sum}$ is the sum rate and $i$ identifies each receiver.
As for the previous strategy, this strategy  can also be easily generalized to the case of any number of relays and receivers and evaluated numerically.

%
%
We now see the value of the result in Sec. \ref{sec:The Achievable Rate Region of the CGRAS}:  it provides a way to numerically derive rate expressions for any message
allocation and search for the best throughput in a large class of possible coding schemes.
%
%
Explicitly deriving these attainable regions for each possible message allocation would require a considerable effort and would not be feasible for a larger number of nodes in the network.
%
In \cite{rini2013rate}  we show that the total number of schemes involving superposition coding for the model in Fig. \ref{fig:3channel} is on the order of a few hundreds
 and that significant rate gains can be obtained over naive strategies of treating the interference as noise and virtual MIMO BC.
%
Although computationally intensive, this approach results in the best transmission strategy from an achievable-rate perspective for any channel condition.

Note, additionally, that it is possible to consider any function of the rates as optimization objectives, since the results in Th. \ref{sec:The Achievable Rate Region of the CGRAS}
provide the complete achievable region.
For example, in \cite{kurniawan2012energy,rini2014energy} we apply the same approach as in \cite{rini2013rate} to optimize over the energy efficiency instead of maximum throughput.

%% file: Conclusion.tex
\section{Conclusions and future work}
\label{sec:Conclusion}

This paper presents a unified approach  to the derivation of achievable rate regions based on random coding and valid for a wide class of single-hop wireless networks.
Given any single-hop, memoryless channel with any number of users and common information, a general transmission scheme is derived in two steps.
First, each user is divided into a set of virtual sub-users: user virtualization increases the number of users in the network and offers more coding opportunities.
In this phase, rate-splitting is employed to map the messages in the original channel to the virtual sub-messages in the enhanced model.
Successively, an achievable scheme for this enhanced channel is considered which contains any combination of coded time-sharing, superposition coding, and binning.
A graphical model is used to represent this general strategy in which nodes represent codewords and edges represent coding operations.
Although conceptually simple, this graph can be used to precisely describe the codebook construction as well as the encoding and decoding operations for any scheme.
Additionally, a graphical Markov model is constructed on this graph to describe the distribution of the codewords and the effect of superposition coding and binning on the conditional dependence across codewords.
By also linking this graphical Markov model to the encoding and decoding error probability, it is possible to derive the achievable rate region of this general scheme.

\bigskip

A subject of ongoing research is whether there exists a combination of encoding strategies that yields the largest achievable rate region among all  possible transmission strategies within the proposed framework.
It is commonly believed that superposition coding produces a larger achievable rate region as compared to conditionally independent codewords.
Joint binning is also thought to outperform different combinations of (one-way) binning.
In both cases, it is not easy to show that this is indeed the case, since one needs to consider different distributions of the codewords which are used to express the achievable rate region.
%
We believe that the general formulation proposed here provides a powerful tool to resolve these conjectures.

A further extension of this work would consider multi-hop networks and the generalization of transmission techniques such as Markov encoding, amplify-and-forward , decode-and-forward and  compress-and-forward to this general model.
These techniques were originally developed for the relay channel and have successively been extended to the multiple relay channel and some simple relay networks, such as the unicast relay network.
We believe that it would be possible to extend these strategies to even more complex scenarios through a systematic approach to the error analysis as the one employed in this paper.
Such an extension will make it possible to investigate the optimal interference management strategies in multi-hop transmissions, which has yet to be adequately addressed.
%

\section*{Acknowledgements}
The authors would like to thank the Associate Editor and the reviewers for their detailed comments and suggestions on the manuscript, which served to greatly improve its presentation and clarity. They would also like to thank Professor Emre Teletar for his suggestions to include non-unique decoding in our analysis.

%% file: GraphicalMarkovModels.tex
\section{A brief introduction to graph theory }
\label{app:Graphical Markov Models}

As described in the main body of the paper, we use a graph representation to describe coding schemes for the general network model
in Sec. \ref{sec:Network Model} after user virtualization.
%
In particular, codewords are associated with graph nodes while coding operations are associate with graph edges.
One type of edge describes superposition coding while another type describes binning.
This graph is used to detail how the codebook encoding of a given message is generated as a function of the codebooks of the other messages.

We successively obtain the achievable rate region of each such scheme by linking the graph representing the transmission strategy to the encoding and decoding
error probabilities. This is done by defining a GMM over the graph representing the transmission scheme, that is, by associating a joint
distribution to the graph by letting nodes represent Random Variables (RVs) while edges represent conditional dependence.
This section presents the graph-theoretic notions and the GMMs that will be used for this encoding representation.

\subsubsection{Some graph-theoretic notions}
\label{sec:Some graph-theoretic notions}

A \emph{graph} $\Gcal(\Vv,\Ev)$ is defined by a finite set of \emph{nodes} $\Vv$ and a set of \emph{edges} $\Ev \subseteq \Vv \times \Vv$
i.e. a set of ordered pairs of distinct nodes.
An edge $(\al,\be) \in \Ev$  whose opposite $(\be,\al) \in \Ev$  is called an \emph{undirected edge},
whereas an edge $(\al,\be) \in \Ev$  whose opposite $(\be,\al) \not\in \Ev$  is a \emph{directed edge}.
Two nodes $\al$ and $\be$ are \emph{adjacent} in $\Gcal$ if they are connected by an edge, that is
\ea{
\ad(\al) = \lcb \be \in \Vv | (\al,\be) \in \Ev, \ {\rm or} \ (\be,\al) \in \Ev \rcb.
}
An undirected graph is \emph{complete} if all pairs of nodes are joined by an edge. A subset $\Av \subset \Vv$ is complete in $\Gcal$ if it induces a complete
subgraph.
A \emph{source node} is a node that has no incoming edges while a \emph{sink node} is a node with no outgoing edges.

If $\Av \subseteq \Vv$ is a subset of nodes, it induces a \emph{subgraph} $\Gcal_{\Av}=(\Av, \Ev_{\Av})$, where $\Ev_{\Av}=\Ev \cap (\Av \times \Av)$.
The \emph{parents} of a node $\al \in \Vv$ in $\Av$ are those nodes linked to $\al$  by a directed edge in $\Ev_{\Av}$, i.e.
\ea{
\pa_{\Ev_\Av}(\be)= \lcb \al \in \Av  | \ (\al,\be) \in \Ev_{\Av}, \ (\be,\al) \not\in \Ev_{\Av} \rcb.
\nonumber
}

\begin{figure}
\centering
\includegraphics[width=15 cm]{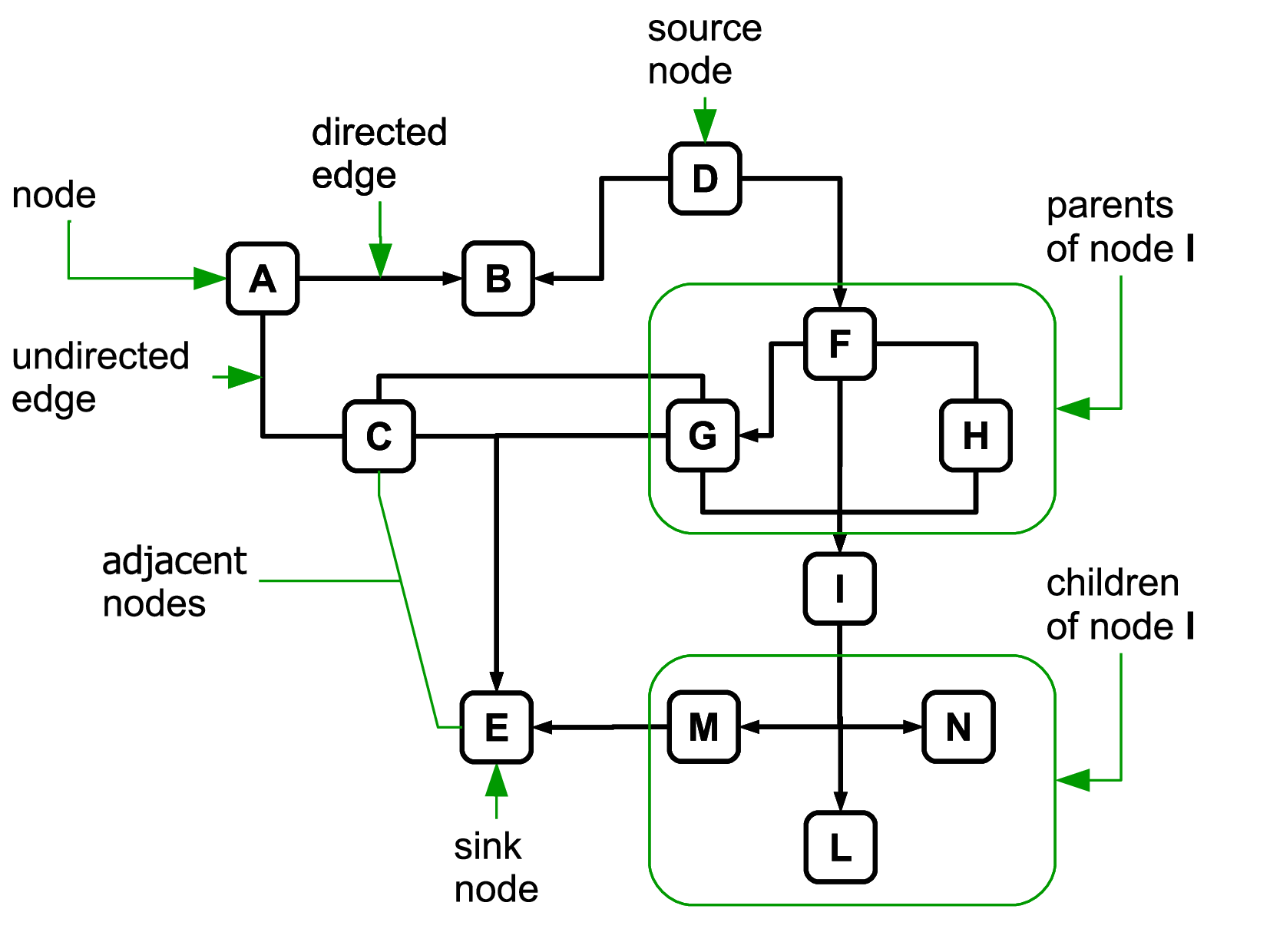}
\vspace{-.5 cm}
\caption{A graphical representation of the graph theoretic notions in Sec.\ref{sec:Some graph-theoretic notions}.}
\label{fig:Markov1}
\end{figure}

Similarly, the  \emph{children} of a node $\al \in \Vv$ in $\Av$ are those nodes that can be reached from $\al$  by a directed edge in $\Ev_{\Av}$, i.e.
\ea{
\ch_{\Ev_\Av}(\al)= \lcb \be \in \Av  | \ (\al,\be) \in \Ev_{\Av}, \ (\be,\al) \not\in \Ev_{\Av} \rcb.
\nonumber
}
These definitions, as illustrated in Fig. \ref{fig:Markov1}, readily extend to subsets of nodes.

\begin{figure}
\centering
\includegraphics[width=15 cm]{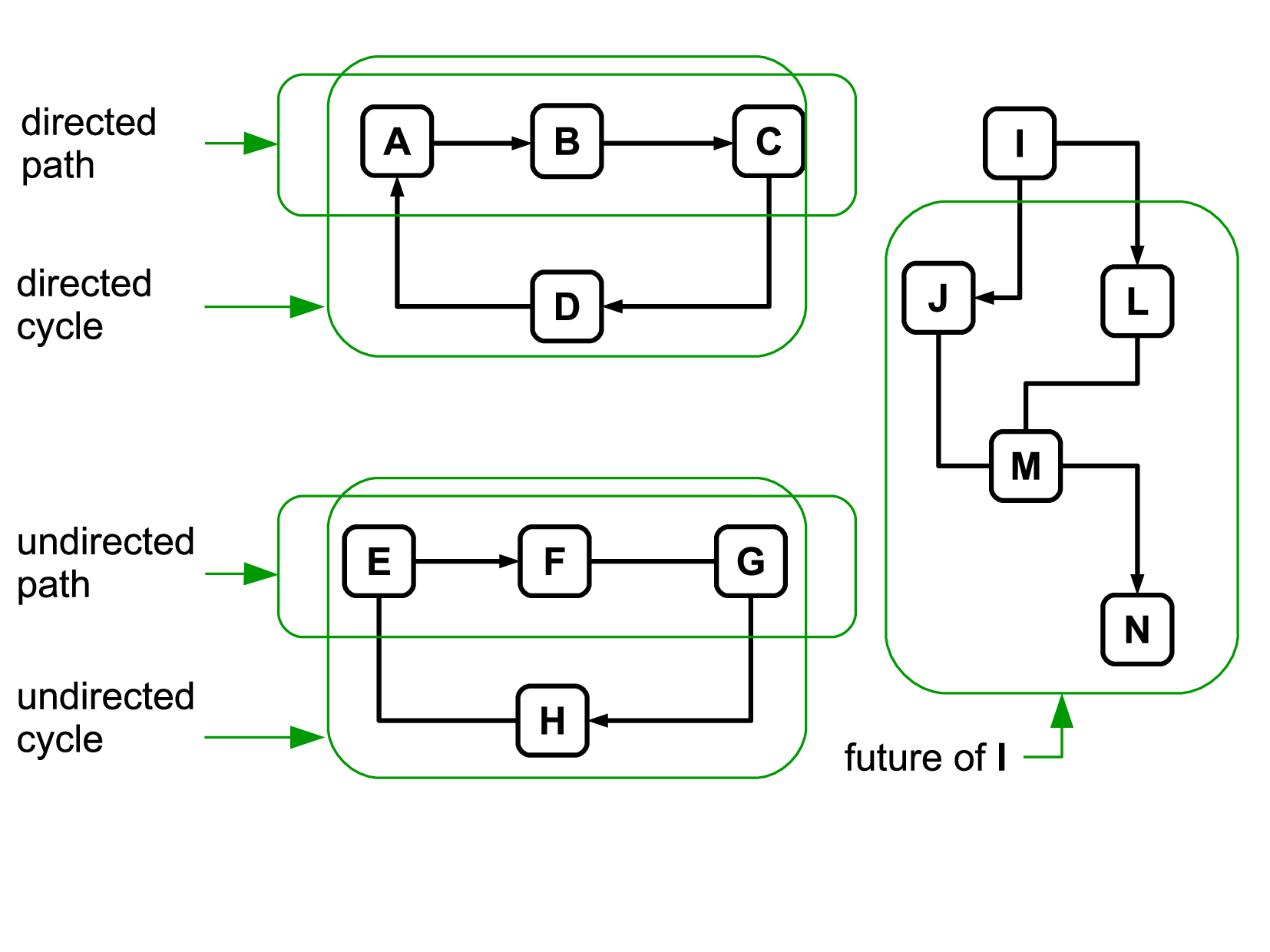}
\vspace{-2 cm}
\caption{A graphical representation of the graph theoretic notions in Sec.\ref{sec:Some graph-theoretic notions}.}
\label{fig:Markov2}
\end{figure}

A path $\pi$ of length $n$ from $\al_0$ to $\al_n$ is a sequence $\pi = \{ \al_0, \al_1...\al_n \} \subseteq \Vv$ of distinct nodes such that
$(\al_{n-1}, \al_n) \in \Ev$  for all $i=1...n$.
If the edge $(\al_{n-1}, \al_n)$ is directed for at least one of the nodes $i$, we call the path \emph{directed}.
If none of the edges are directed, the path is called \emph{undirected}.
A cycle is a path in which $\al_0=\al_n$.
We define the \emph{future} of a node $\al$ in $\Gcal$, denoted by $\phi(\al)$, as the set of nodes that can be reached by $\al$ through a directed path.
These path definitions are illustrated in Fig. \ref{fig:Markov2}.
\smallskip

Graphs are generally classified into three categories:

\begin{itemize}
\item
{\bf UDG:}
if all the edges are undirected, the graph is said to be an \emph{UnDirected Graph}.

\item
{\bf DAG:}
if all the edges are directed and the graph contains no cycles, the graph is said to be a \emph{Directed Acyclic Graph}.

\item
{\bf CG:}
if edges are both directed and undirected and the graph does not contain directed cycles, the graph is called a \emph{Chain Graph}.
\end{itemize}


\section{A brief introduction to graphical Markov models}
\label{app:Graphical Markov models 2}
%
In  GMMs, graphs are used to represent the factorization of a multivariate distribution.
GMMs were introduced by Perl in 1988 \cite{pearl1988probabilistic} and have enjoyed a surge of interest in the last two decades.
Although conceptually simple, GMMs can be used to represent a highly varied and complex system of multivariate dependencies by means of the
global structure of the graph, thereby obtaining efficiency in modeling, inference, and probabilistic calculations
 \cite{lauritzen1996graphical,whittaker1990graphical,dawid1999probabilistic}.
%
%
%
A GMM is constructed using the graph  $\Gcal(\Vv, \Ev)$  and associating the set of nodes in the graph, $\Vv$, to a set of Random Variables (RVs)
and the set of edges $\Ev$ to the conditional dependencies among these RVs.

In GMMs, dependencies between Random Variables (RVs) are represented through a graph: each node is associated with a random variable and an edge between two RVs indicates
conditional dependency.
For simple scenarios, this implies that the distribution of the random variable (RV) associated with a node depends only on the distribution of the RVs associated with neighboring nodes and it is conditionally independent from the RVs associated with the remaining nodes in the graph.

In order to define a joint distribution for the RVs associated with a general graph, a more rigorous formulation is necessary.
In particular, it is necessary to avoid recursive dependencies among variables that can be established through global properties of the graph, such as cycles.
That is, a random variable cannot be conditionally dependent on itself, since this does not correspond to a meaningful conditional distribution.

\begin{definition}{\bf Global G-Markov Property: \\}
\label{def:global markov property}
Let the notation $A \perp B  \ | C \ \  [P]$ indicate that $A$  is conditionally independent of $B$ given $C$ under the distribution $P$.
Also, consider a graph $\Gcal(\Vv, \Ev)$ and a probability measure $P$ on $\Vv$ obtained as the product probability measure
\ea{
\Ucal= \bigtimes_{\al \in \Vv} \Ucal_{\al}.
\label{eq:product space probability}
}
The distribution $P$ is said to be \emph{global G-Markovian} \cite{hammersley1971markov} if
\ea{
\al \perp \be \ | \lb \Vv \backslash \phi(\al) \rb \backslash \{ \al , \be \} \ \  [P],
\label{eq:conditionally independent set global markov}
}
for $\al$ and $\be$ not adjacent, $\be \not \in \phi(\al)$, and if, given four disjoint subsets $A,B,C$ and $D$, the following holds:
\ea{
& A \perp B  \ | C \cup D \ \ [P]  \ {\rm and} \ A \perp C \ | B \cup D \ \  [P]
\nonumber \\
&
\quad \quad  \implies A \perp B \cup C \ | D \ \ [P ].
\label{eq:condition CI5}
}
\end{definition}
Definition \ref{def:global markov property} can be interpreted as follows: the distribution $P$ is global Markov if two nodes $\al$ and $\be$ that are not adjacent and such that $\be$ is not in the future of $\al$ are conditionally independent given all the nodes in $\Vv$ minus the future of $\al$ plus  $\al$ and $\be$ themselves
\footnote{
The formulation in Def. \ref{def:global markov property} of the global Markov property is not the most general,  but we refer to this definition for simplicity.
A more detailed discussion on Markov properties for graphs is provided in \cite[Sec. 3]{andersson1997markov}.
}
.
%
%
%

The global Markov property is necessary to ensure that the distribution $P$ over the graph $\Gcal(\Vv,\Ev)$ is well-defined, that is, $P$ can be factorized into a product of conditional distributions, one for each variable.
This definition is also crucial to establish a notion of equivalence among graphs.
In general, graphs with a different sets of edges may describe the same factorization of the joint distribution, in the same way that a joint distribution
can be described as the product of different conditional distributions (e.g. $P_X P_{Y|X} P_{Z|X}= P_{Z} P_{Z|X} P_{Y|X}$).
For this reason, it is not trivial to determine which graphs describe the same dependency structure.
%
On the other hand, being able to change the representation of a joint distribution from one graph to another can be very useful when evaluating functionals of this distribution.

\begin{definition}{\bf Markov Equivalence, \cite[Th. 3.1]{andersson1997markov}: \\ }
\label{def:Markov Equivalence}
Two chain graphs $\Gcal_1(\Vv, \Ev_1)$ and  $\Gcal_2(\Vv, \Ev_2)$ are called \emph{Markov-equivalent} if,
 for every product space $\Xcal$ indexed by $\Vv$, the classes of probability measures that are global G-Markovian on $\Gcal_1$ and $\Gcal_2$ are equivalent.
\end{definition}

\noindent
In \cite{andersson1997markov} a rigorous theory to establish the equivalence between GMMs is developed.

%% file: Appendix.tex
\subsection{Proof of Theorem \ref{thm: DAG equivalence of graph representation}}
\label{app: DAG equivalence of graph representation}

The proof is structured as follow: first we will establish that under Assumption \ref{ass:Joint binning forms cliques}, the CSJB-restriction, there exist a DAG which is Markov-equivalent to the chain graph describing the CGRAS. Successively we argue that any orientation of the edges which produces a DAG corresponds to an equivalent DAG.
After this first part of the proof, we show that under Assumption \ref{ass:Binning is transitive}, the TB-restriction,  any orientation of the binning edges in the Markov-equivalent DAG produces yet another Markov-equivalent DAG.
This properties of the CGRAS will be extremely convenient when analyzing the probability of an error in the encoding and decoding operations as described in Sec. \ref{sec:Codebook construction, encoding and decoding operations}.
This will, in turns, lead to the compact representation of the achievable region \ref{sec:The Achievable Rate Region of the CGRAS}.

\medskip

Some more definitions that will be required for the proof are presented next.
These definitions are purely graph theoretic and are necessary to clearly present two lemmas, Lem. \ref{lem:Markov-equivalent conditions} and Lem. \ref{lem:only immoralities}.
These two lemmas establish conditions under which chain graphs are Markov equivalent.
By further elaborating on these two results, we come to prove the desired result.

The \emph{skeleton} of a graph is the underlying undirected graph of a graph $\Gcal(\Vv,\Ev)$ obtained by substituting all the directed edges with undirected edges,
that is the graph $\Gcal(\Vv,\Ev^u)$ with
\ea{
\Ev^u = \lcb (\al,\be), \ (\al,\be) \in \Ev \ \rm{or} \ (\be,\al) \in \Ev \rcb.
}
In a chain graph $\Gcal(\Vv,\Ev)$, the chain components $\tau \in \Tcal$ are the set of all the connected components obtained from $\Gcal$ when removing all the directed edges.
$\cl(\tau)$ denotes the closure of the chain component $\tau \in \Tcal(G)$, which is obtained by adding the boundaries of the chain component, i.e.
$\cl(\tau)=\bd(\tau) \cup \tau$.
$(G_{\cl(\tau)})^m$ denotes the moral graph of $\cl(\tau)$: this is the graph obtained by making the boundary of the set $\cl(\tau)$ complete,
 that is, by connecting all the nodes that have a common child.
The triple $(\al,B,\be)$ a \emph{complex} in $\Gcal$ if $B$ is a connected subset of a chain component $\tau \in \Tcal(\Gcal)$ and $\al$ and $\be$ are
two non-adjacent nodes in $\bd(\tau) \cap \bd(B)$.
Further, $(\al,B,\be)$ a \emph{minimal complex} if $B=B'$ whenever $(\al,B',\be)$   is a complex whenever and $B' \subseteq B$.
A minimal complex is an \emph{immorality} if $B$  contains only one node.
We will further require the following two lemmas:

\begin{lem}{\bf \cite[Th. 3.1]{andersson1997markov}}
\label{lem:Markov-equivalent conditions}
Two chain graphs $\Gcal_1$ and $\Gcal_2$ are Markov-equivalent if and only they have the same skeleton and the same minimal complexes.
\end{lem}

\begin{lem}{\bf \cite[Lemma 4.1]{andersson1997markov}}
\label{lem:only immoralities}
Let $\Gcal$ be a chain graph such that  $(\Gcal_{\cl(\tau)})^m$ is decomposable for every chain component $\tau \in \Tcal(\Gcal)$.
Then $\Gcal$ has no minimal complexes other than immoralities.
\end{lem}

A sketch of the proof is as follows: we want to show that there exists a DAG that is Markov-equivalent to the chain graph.
This equivalent DAG is obtained by orienting the edges of the chain graph so that no cycle is formed.
When the existence of this DAG has been established, we will provide an algorithm to orient the edges of the chain graph and successively show that any
variation of this orientation still yields a DAG which is Markov-equivalent to the original chain graph.

The first part of the proof is as follows: we want to show that the chain graph representing the CGRAS with is equivalent to some DAG using
Lem. \ref{lem:Markov-equivalent conditions}.
This DAG is obtained by orienting the undirected edges of the chain graph, so the two graphs have the same skeleton.
In a DAG all the connected components are composed of one vector, therefore all the minimal complexes are immoralities.
This implies that the equivalence holds only when the chain graph contains only immoralities.
A way to insure that all the minimal complexes are immoralities, is to have that the moral graph is decomposable.
Assumption \ref{ass:Joint binning forms cliques}, the CSJB-restriction, indeed guarantees that this is the case.
This is shown in the next theorem.

\begin{thm}
In a CGRAS for which assumption \ref{ass:Joint binning forms cliques}, the CSJB-restriction, holds all the minimal complexes of $\Gcal(\Vv,\Ev)$ are immoralities.
%
%
\end{thm}

\begin{IEEEproof}
Because of assumption \ref{ass:Joint binning forms cliques}, the CSJB-restriction, all the chain components are fully connected.
Additionally, since all the nodes have the same parents, a complex $(\al,B,\be)$, each node $\gamma \in B$  is an immorality $(\al, \gamma, \be)$.
\end{IEEEproof}

Next we show that there must exists a DAG which is Markov-equivalent to $\Gcal(\Vv,\Ev)$ obtained by orienting the edges of the connected components of $\Gcal(\Vv,\Ev)$.

\begin{thm}

Let be the graph $\Gcal(\Vv,\Ev)$ be associated to a CGRAS for which Assumption \ref{ass:Joint binning forms cliques}, the CSJB-restriction, holds, then any non-cyclic orientation of the edges produces a Markov-equivalent DAG.
Moreover at least one orientation must exists.
\end{thm}
\begin{IEEEproof}
Since each chain component is a complete graph, there exists an orientation which produces a DAG.
This orientation cannot create a cycle in the overall graph, since this would implies that a cycle existed in the original graph.
All the immoralities that are in $\Gcal(\Vv,\Ev)$ are also in the DAG produces through the non-cyclic orientation.
We now have to show that the orientation of the edges does not introduce new immoralities.
These immoralities $(\al,\gamma,\be)$ can be introduced in two ways: either both $\al$ and $\be$ are in the connected component, or only one of them is in the
connected component while the other one is in the boundary.
If $\al$ and $\be$ are in the connected component, then they must be adjacent, since the connected components are complete.
If $\al$ is in the connected component and $\be$ is in the boundary (or viceversa), then $\al$ and $\be$ must still be adjacent, since $\al$ has the same parent nodes than $\gamma$.
\end{IEEEproof}

Now we show that any non-cyclic orientation of the binning edges in the Markov-equivalent DAG produces yet another Markov-equivalent DAG.
For this part of the proof the next criterium to establish Markov equivalence among DAG is necessary.

\begin{thm}{\bf Markov equivalence of DAGs \cite{neapolitan2004learning} \\}
\label{th:Markov equivalence of DAGs}
Two DAGs $\Gcal_1$ and $\Gcal_1$ are Markov-equivalent if and only if they have
the same links (edges without regard for direction) and the same set of uncoupled
head-to-head meetings.
\end{thm}

A head-to-head meeting is defined as a set of edges $X \rightarrow W \leftarrow Y$ and an uncoupled head-to-head meeting is a head-to-head meeting for which $X$ and $Y$ are adjacent.

When Assumption  \ref{ass:Binning is transitive} (the TB-restriction) holds, there can be no uncoupled head-to-head meeting among nodes  that are not source nodes in $\Gcal(\Vv,\Bv)$, since transitivity implies that the nodes causing a head-to-head meeting must be adjacent.

We next show that changing the orientation of edges connecting non-sink nodes in $\Gcal(\Vv,\Bv)$ produces another Markov-equivalent DAG.

\begin{thm}
Changing the orientation of edges connecting non-sink nodes in $\Gcal(\Vv,\Bv)$ produces another Markov-equivalent DAG.
\end{thm}
\begin{IEEEproof}
The proof is carried out by induction, showing that changing the direction of one edge does not introduce a head-to-head meeting.
Iterating this procedure for each desired direction change, proves the desired result, under the assumption that the change of direction does not introduce a cycle.

Under Assumption  \ref{ass:Binning is transitive}, the TB-restriction , there can be no uncoupled head-to-head meeting among non-sink nodes.
Hence we only need to show that changing the direction of one edge does not introduce any uncoupled head-to-head meeting.
Th. \ref{th:Markov equivalence of DAGs} guarantees then that the two DAG are Markov equivalent.
The uncoupled head-to-head meeting can be of two types: a binning edge meeting a superposition coding edge and binning edge meeting another binning edge.
We start from this latter case: select the binning edge whose direction we wish to change and let us denote such edge $U_{\iv \sgoes \jv} \bin U_{\vv \sgoes \tv}$.
Since the departing node $U_{\iv \sgoes \jv}$ cannot be a source nodes, there must be at least an incoming edge $U_{\lv \sgoes \mv} \bin U_{\iv \sgoes \jv}$.
A head-to-head meeting $U_{\vv \sgoes \tv} \bin U_{\iv \sgoes \jv}$,$U_{\lv \sgoes \mv} \bin U_{\iv \sgoes \jv}$ is then created when reversing the direction of the selected edge and all the incoming edges.
But since transitivity must hold, we have that $U_{\lv \sgoes \mv} \bin  U_{\vv \sgoes \tv}$ and thus the newly created head-to-head meeting is not uncoupled.

Consider now the case in which the uncoupled head-to-head meeting occurs between a binning and a superposition coding edge.
More precisely let $U_{\iv \sgoes \jv} \bin U_{\vv \sgoes \tv}$ be again the edge whose direction we wish to change and let
$U_{\lv \sgoes \mv} \spc U_{\iv \sgoes \jv}$.
Reversing the direction of the edge, we obtain the head-to-head meeting  $U_{\vv \sgoes \tv} \bin U_{\iv \sgoes \jv}$, $U_{\lv \sgoes \mv} \spc U_{\iv \sgoes \jv}$.
Given the condition under which binning can occur, we must also have $U_{\lv \sgoes \mv} \spc U_{\vv \sgoes \tv}$, so the head-to-head meeting is not uncoupled.
%
%
%
\end{IEEEproof}

\subsection{Proof of Theorem \ref{th:Decoding errors only supeperposition} }
\label{app:Decoding errors only supeperposition}

In the following we simplify the notation when indicating specific codewords as in \eqref{eq:codeword indexing} to
\ea{
U_{\iv \sgoes \jv}^N \lb w_{\iv \sgoes \jv} \rb = U_{\iv \sgoes \jv}^N \lb w_{\iv \sgoes \jv} , \{ l_{\lv \sgoes \mv}, U_{\lv \sgoes \mv}  \in \pa_{\Sv}(U_{\iv \sgoes \jv}) \} \rb,
}
with the implicit understanding that the missing indices are determined by the bottom codebooks on top of which $U_{\iv \sgoes \jv}^N $  is superimposed.
Also, without loss of generality, we assume that the message set $\{w_{\iv \sgoes \jv}'=1, \ \forall \ (\iv,\jv) \in \Vv \}$ is to be transmitted.
Since the codewords are generated in an i.i.d. fashion, the probability of error is the same for any given message set.

\medskip

A decoding error is committed at decoder $z$ whenever  $\wh_{\iv \goes \jv}^{z} \neq 1$ for any $z$ and any $(\iv,\jv) \in \Vv$.
In general, any possible combination of errors can occur: let's assume that the set $\Fv$ corresponds to the set of incorrectly decoded codewords and
$\Fvo=\Vv^z \setminus \Fv$   to the set of correctly decoded codewords.

Since the decoder is a typicality decoder, an error implies that
\ea{
& \lb Y_z, \lcb U_{\iv \sgoes \jv}(\wt_{\iv \sgoes \jv}, \ (\iv,\jv) \in \Fv^{\Bv}\rcb ,
\lcb U_{\iv \sgoes \jv}(1), \ (\iv,\jv) \in \Fvo^{\Bv}\rcb
\rb \nonumber \\
& \quad \quad \in \Tcal_{\ep}^N \lb Y_z , \{ U_{\iv \sgoes \jv}, \  (\iv, \jv) \in \Vv^z \} \rb,
\label{eq:decoding errror event typicality}
}
for some $\wt_{\iv \sgoes \jv} \neq 1$.

The probability of a decoding error can be bounded using the union of events bound as the sum over all the possible subsets of $\Fv$
\ea{
& \Pr [ \ {\rm decoding\,NOT\,successful \, at \, Dec. } \   z \ ] \leq \sum_{ \Fv \subseteq \Vv^z  } \Pr   \lsb D_{\Fv}  \rsb,
\label{eq:factorization covering lemma decoding}
}
where $D_{\Fv}$  is the event defined as
\ea{
    D_{\Fv} = & \lb Y_z, \lcb U_{\iv \sgoes \jv}(\wt_{\iv \sgoes \jv}), \ (\iv,\jv) \in \Fv^{\Bv}\rcb ,
\lcb U_{\iv \sgoes \jv}(1 ), \ (\iv,\jv) \in \Fvo^{\Bv}\rcb \rb \nonumber \\
& \quad \quad \in \Tcal_{\ep}^N \lb Y_z , \{ U_{\iv \sgoes \jv}, \  (\iv, \jv) \in \Vv^z \} \rb.
\label{eq:decoding errror event typicality}
}

The number of error events to be considered can be reduced by noticing that the probability of incorrectly decoding a codeword when its parent nodes have
been incorrectly decoded goes to one as $N$ goes to infinity.
This is because codewords are created conditionally dependently from the parent codewords and the incorrect decoding of a parent node implies that the decoder looks for the transmitted message in an independent set of codewords.
In this set of codewords, the probability of correctly identifying the transmitted message is goes to zero as the block length increases.
%
This means that we need to consider only the sets $\Fv$ for which \eqref{eq:condition decoding errors superposition} holds, that is parents of correctly decoded codewords are also correctly decoded.

We now bound each term in the RHS of \eqref{eq:factorization covering lemma decoding} as:
\eas{
& \Pr \lsb D_{\Fv} \rsb  \leq
\Pr \lsb  \bigcup_{ \Fv }
 ( Y_z^N,
\{ U_{\iv \sgoes \jv} ( \wt_{\iv \sgoes \jv}), \ (\iv,\jv) \in \Fv^{\Bv}\}
\cup  \rnone\nonumber \\
&\quad \quad  \lnone \{ U_{\iv \sgoes \jv} ( 1 ), \ (\iv,\jv) \in \Fvo^{\Bv}\}
) \in
 \Tcal_{\ep}^N  ( P_{Y_z, \{ U_{\iv \sgoes \jv}, \ (\iv,\jv) \in \Vv^z  \} }^{\rm encoding} ) \rsb \\
 &=2^{N (\sum_{(\iv,\jv) \ \in \ \Fv }  R_{\iv \sgoes \jv}} \\
& \quad \quad  \Pr [ \{
Y_z^N,
\{ U_{\iv \sgoes \jv} ( \wt_{\iv \sgoes \jv}), \ (\iv,\jv) \in \Fv^{\Bv}\}
\nonumber \\
&\quad \quad  \quad \quad   \cup  \
\{ U_{\iv \sgoes \jv} ( 1 ), \ (\iv,\jv) \in \Fvo^{\Bv}\}
\}\in
 \Tcal_{\ep}^N  ( P_{Y_z, \{ U_{\iv \sgoes \jv}, \ (\iv,\jv) \in \Vv^z  \} }^{\rm encoding} )
   ],
 \label{bound probability II joint decoding last}
}{\label{bound probability II joint decoding}}
where \eqref{bound probability II joint decoding last} follows from the fact that codewords are obtained from i.i.d. draws. Using the  packing lemma \cite{kramerBook,el2011network},we can bound the probability term in \eqref{bound probability II joint decoding last} as:
\eas{
& \Pr [
\{ Y_z^N,
\{ U_{\iv \sgoes \jv} ( \wt_{\iv \sgoes \jv}), \ (\iv,\jv) \in \Fv^{\Bv}\}
\cup  \nonumber
\\ & \quad \quad   \{ U_{\iv \sgoes \jv} ( 1 ), \ (\iv,\jv) \in \Fvo^{\Bv}\}
\}
\in
 \Tcal_{\ep}^N  (  P_{Y_z, \{ U_{\iv \sgoes \jv}, \ (\iv,\jv) \in \Vv^z  \} }^{\rm encoding} )]  \\
& \leq  \sum_{  \Tcal_{\ep}^N  \lb  P_{Y_z, \{ U_{\iv \sgoes \jv}, \ (\iv,\jv) \in \Vv^z  \} }^{\rm encoding} \rb }
 \Pr [
Y_z^N \cup \nonumber \\
& \quad \quad  \{ U_{\iv \sgoes \jv} ( \wt_{\iv \sgoes \jv}), \ (\iv,\jv) \in \Fv^{\Bv}\}
\cup
\{ U_{\iv \sgoes \jv} ( 1 ), \ (\iv,\jv) \in \Fvo^{\Bv}\}
]  \\
& \leq  \labs \Tcal_{\ep}^N  ( P_{Y_z, \{ U_{\iv \sgoes \jv}, \ (\iv,\jv) \in \Vv^z  \} }^{\rm encoding}) \rabs
\Pr [ Y_z^N | U_{\iv \sgoes \jv}, \ (\iv,\jv) \in \Fv^{\Bv}] \nonumber \\
& \quad \quad  \Pr [ U_{\iv \sgoes \jv} ( \wt_{\iv \sgoes \jv} ), \ (\iv,\jv) \in \Fv , \ U_{\iv \sgoes \jv} ( 1 ), \ (\iv,\jv) \in \Fvo^{\Bv}] \\
& \leq 2^{+N H(Y_z, \{ U_{\iv \sgoes \jv}, \ (\iv,\jv) \in \Vv^z  \} )} 2^{- N H(Y_z | U_{\iv \sgoes \jv}, \ (\iv,\jv) \in \Fv )} \\
& \quad \quad  2^{-N H (U_{\iv \sgoes \jv}, \ (\iv,\jv) \in \Vv^z )}\\
& = 2^{ - N I(Y_z;  U_{\iv \sgoes \jv}, \ (\iv,\jv) \in \Fvo^{\Bv}| U_{\iv \sgoes \jv} ( 1 ), \ (\iv,\jv) \in \Fv , Q)}.
\label{eq:calculation I joint decoding last}
}{\label{eq:calculation I joint decoding}}

Plugging the bound in \eqref{eq:calculation I joint decoding last} in \eqref{bound probability II joint decoding last} we have
\ea{
\Pr \lsb D_{\Fv} \rsb  \leq 2^{N ( (\sum_{(\iv,\jv) \ \in \ \Fv }  R_{\iv \sgoes \jv} - I(Y_z;  U_{\iv \sgoes \jv} ( 1 ), \ (\iv,\jv) \in \Fvo^{\Bv}| U_{\iv \sgoes \jv} ( 1 ), \ (\iv,\jv) \in \Fv , Q))},
}
and therefore we conclude that the probability of error goes to zero when \eqref{eq:Decoding errors only supeperposition rate bound} holds for all the sets $\Fv$.

\subsection{Proof of Theorem \ref{th:Decoding errors only supeperposition and one-way binning}}
\label{app:Decoding errors only supeperposition and one-way binning}
Similarly to the proof of Th. \ref{th:Decoding errors only supeperposition}, we simplify the indexing of the codewords as
\ea{
U_{\iv \sgoes \jv}^N \lb w_{\iv \sgoes \jv} , b_{\iv \sgoes \jv} \rb = U_{\iv \sgoes \jv}^N \lb w_{\iv \sgoes \jv} , b_{\iv \sgoes \jv}  , \{ l_{\lv \sgoes \mv}, U_{\lv \sgoes \mv}  \in \pa_{\Sv}(U_{\iv \sgoes \jv}) \} \rb,
}
with the implicit understanding that the missing indices are determined by the bottom codebooks on top of which $U_{\iv \sgoes \jv}^N $  is superimposed or binned.
%


\medskip
\noindent
\emph{ENCODING ERROR ANALYSIS}
\smallskip

We start the proof by applying the Markov inequality to the encoding error probability as follows:
\eas{
& \Pr[{\rm encoding\,NOT\,successful}] \\
&= \Pr \lsb \not \exists \ \bv \  \lcb U^N_{\iv \sgoes \jv}(w_{\iv \sgoes \jv}, b_{\iv \sgoes \jv}), \ b_{\iv \sgoes \jv} \in \bv    \rcb\in \Tcal_\epsilon^N  \lb P^{\rm encoding}  \rb \rsb
\\
&=\Pr [ \{  \bigcap_{\bv} \lcb U^N_{\iv \sgoes \jv}(w_{\iv \sgoes \jv}, b_{\iv \sgoes \jv}), \ b_{\iv \sgoes \jv} \in \bv    \rcb   \in \Tcal_\epsilon^N  \lb P^{\rm encoding}  \rb \} = \emptyset ]
\\
&= \Pr[\Kv=0] \leq \f{\var[\Kv]}{\Ebb^2[\Kv]},
\label{eq:first bound mutual covering last}
}{\label{eq:first bound mutual covering}}
where
\eas{
& \bv  = \lcb b_{\iv \sgoes \jv}, \ (\iv,\jv) \in  \Vv \rcb  \ \ \ \ \forall \  b_{\iv \sgoes \jv} \in \lsb 1 \ldots 2^{N R'_{\iv \sgoes \jv}} \rsb, \\
& \Kv  = \sum_{\bv }  K_{\bv }, \quad K_{\bv}  = 1_E(\bv), \\
& E  = \lcb  U^N_{\iv , \jv}(w_{\iv \sgoes \jv},b_{\iv \sgoes \jv}), \ b_{\iv \sgoes \jv } \in \bv \rcb  \in \Tcal_\epsilon^N \lb P^{\rm encoding} \rb.
\label{eq: typicality condition markov lemma}
}{\label{eq:markov lemma indicato function definintion}}

The probability of encoding error is equivalent to the probability that no choice of binning indices $\bv$ produces the desired typicality among the codewords.
We then associate the indicator  function $\Kv_{\bv}$ to the event that a sequence of binning indices $\bv$ produces a successful encoding and the indicator function $\Kv$ to the event that encoding is successful for some sequence $\bv$.
By applying the Markov lemma to the indicator function $\Kv$, we can obtain bounds on the binning rates by estimating the mean and variance of $\Kv$.

We start by evaluating the term $\Ebb \lsb \Kv \rsb$ :
\eas{
& \Ebb    [\Kv]  =   \sum_{ \bv }  \Pr [K_{\bv} =1]
\label{eq:bound mutual covering lemma 1, a} \\
&=2^{N \sum_{(\iv,\jv) \in \Vv} R'_{\iv \sgoes \jv}} \ \Pr [K_{\bv} =1]
\label{eq:bound mutual covering lemma 1, b} \\
&=2^{N \sum_{(\iv,\jv) \in \Vv } R'_{\iv \sgoes \jv}}
\label{eq:bound mutual covering lemma 1, c} \\
& \quad \quad  \Pr \lsb \lcb  U_{\iv \sgoes \jv}^N(w_{\iv \sgoes \jv},b_{\iv \sgoes \jv}),   \ b_{\iv \sgoes \jv } \in \bv \rcb  \in \Tcal_\epsilon^N \lb P^{\rm encoding} \rb \rsb,
\label{eq:bound mutual covering lemma 1 last}
}{\label{eq:bound mutual covering lemma 1} }
where \eqref{eq:bound mutual covering lemma 1, a}  follows from the fact that the codewords in each bin are i.i.d.,
\eqref{eq:bound mutual covering lemma 1, b} from the fact that the total number of sequences $\bv$ is $2^{N \sum_{(\iv,\jv) \in \Vv }}$. 
We now bound the probability term in \eqref{eq:bound mutual covering lemma 1, c} as:
\eas{
& \Pr \lsb \lcb  U_{\iv \sgoes \jv}^N(w_{\iv \sgoes \jv},b_{\iv \sgoes \jv}),   \ b_{\iv \sgoes \jv } \in \bv \rcb  \in \Tcal_\epsilon^N \lb P^{\rm encoding} \rb \rsb \\
&\leq  \sum_{U_{\iv \sgoes \jv}^N \in \Tcal_\epsilon^N \lb P^{\rm encoding} \rb }
P^{N \ \rm codebook}
\label{eq:bound mutual covering lemma 2, a} \\
& = \labs \Tcal_\epsilon^N \lb P^{\rm encoding} \rb \rabs  2^{ - N H(P^{\rm codebook})} \\
& \leq 2^{ N H(P^{\rm encoding})} 2^{ - N H(P^{\rm codebook})},
}{\label{eq:bound mutual covering lemma 2}}
%
where
\eas{
H(P^{\rm encoding})
&= H \lb \prod_{(\iv,\jv) \in \Ev }P_{U_{\iv \sgoes \jv} | \pa_{\Ev}(U_{\iv \sgoes \jv})} \rb  \\
&= \sum_{(\iv,\jv) \in \Ev} H(U_{\iv \sgoes \jv} | \pa_{\Ev}(U_{\iv \sgoes \jv})),
}
and
\eas{
H(P^{\rm codebook})
&= H \lb \prod_{(\iv,\jv) \in \Sv } P_{U_{\iv \sgoes \jv} | \pa_{\Sv}(U_{\iv \sgoes \jv})} \rb \\
&= \sum_{(\iv,\jv) \in \Ev} H(U_{\iv \sgoes \jv} | \pa_{\Sv}(U_{\iv \sgoes \jv})),
}
so that we can write
\eas{
&\Pr \lsb \lcb  U_{\iv \sgoes \jv}^N(w_{\iv \sgoes \jv},b_{\iv \sgoes \jv}),   \ b_{\iv \sgoes \jv } \in \bv \rcb  \in \Tcal_\epsilon^N \lb P^{\rm encoding} \rb \rsb \\
& \quad \quad \leq 2^{N \sum_{(\iv,\jv) \in \Vv} I(U_{\iv \sgoes \jv} ; \pa_{\Ev}(U_{\iv \sgoes \jv}) |  \pa_{\Sv}(U_{\iv \sgoes \jv}) ) } \\
& \quad \quad = 2^{N \sum_{(\iv,\jv) \in \Vv} I(U_{\iv \sgoes \jv} ; \pa_{\Bv}(U_{\iv \sgoes \jv}) |  \pa_{\Sv}(U_{\iv \sgoes \jv}) ) },
\label{eq:E K I}
}

Combining \eqref{eq:E K I} and \eqref{eq:bound mutual covering lemma 1 last} we obtain
\ea{
\Ebb    [\Kv]  \leq 2^{N \sum_{(\iv,\jv) \in \Vv } (R'_{\iv \sgoes \jv} - I(U_{\iv \sgoes \jv} ; \pa_{\Bv}(U_{\iv \sgoes \jv}) |  \pa_{\Sv}(U_{\iv \sgoes \jv}) )) }.
\label{eq: bound E K}
}

To evaluate $\var[\Kv]$ we write:
\eas{
& \var[\Kv]=  \Ebb[\Kv \tilde{\Kv} ] - \Ebb[\Kv]^2 \\
&=  \Ebb [ ( \sum_{\bv }  K_{\bv } ) ( \sum_{\bvt }  K_{\bvt } ) ]
  - \Ebb [ \sum_{\bv }  K_{\bv } ]^2 \\
&= \sum_{\bv} \sum_{ \bvt }  (  \Pr[K_{\bv }=1,K_{\bvt }=1]\\
& \quad  \quad -\Pr[K_{\bv}=1]\Pr[K_{\bvt}=1] ).
\label{eq:independent bins mutual covering}
}{\label{eq:bound mutual covering lemma 3}}
%
It is possible to remove the terms in \eqref{eq:independent bins mutual covering}  for which
\ea{
\Pr[K_{\bv}=1,K_{\bvt}=1]& =  \Pr[K_{\bv}=1] \Pr[K_{\bvt}=1],
}
which correspond to the event that
\ea{
\lcb \{ U^N_{\iv \sgoes \jv}(w_{\iv \sgoes \jv}, b_{\iv \sgoes \jv}) ,  \ b_{\iv \sgoes \jv} \in \bv  \} \in \Tcal_\epsilon^N  \rcb
}
is independent from
\ea{
\{ \{ U^N_{\iv \sgoes \jv}(w_{\iv \sgoes \jv},\bt_{\iv \sgoes \jv}), \ \bt_{\iv \sgoes \jv} \in \bvt \}  \in \Tcal_\epsilon^N  \}.
}
The independence among these two set of codewords happens in two cases:

\begin{itemize}
  \item all the indices $b_{\iv \sgoes \jv}$ and $\bt_{\iv \sgoes \jv}$ are different.

  In this case the codewords selected by the two vectors are all different. Since codewords are produced in an i.i.d. fashion, there is no relation between different codewords in the same bin.

  \item some of the indices $b_{\iv \sgoes \jv}$ and $\bt_{\iv \sgoes \jv}$ are the same,  but when this is the case, the codewords belong to different codebooks.
Multiple codebooks are generated only by superposition coding, so this can be expressed as
 \ea{
 \forall \ (\iv,\jv) \ b_{\iv \sgoes \jv}=\bt_{\iv \sgoes \jv} \implies     \exists \ b_{\lv \sgoes \mv}, \  b_{\lv \sgoes \mv} \neq \bt_{\lv \sgoes \mv},  \ U_{\lv \sgoes \mv} \spc U_{\iv \sgoes \jv}.
 }
 In this case, since codewords in different codebooks are created independently, the fact that two binning indices match does not imply that the codewords are correlated.
\end{itemize}

Given these considerations, one only needs to consider the sequences for which
\ea{
& \bvt (\bv)\ \ST
\label{eq:conditionally independent codewords 2} \\
& \ \ \ \exists \ \bt_{\iv \sgoes \jv}=b_{\iv \sgoes \jv} \implies  \ b_{\lv \sgoes \mv} = \bt_{\lv \sgoes \mv} \ \forall  \ (\lv,\mv), \ U_{\lv \sgoes \mv} \spc U_{\iv \sgoes \jv} \ {\rm or} \ \pa_{\Sv}(U_{\iv \sgoes \jv})=\emptyset.
\nonumber
}

In the following we restrict our attention to the sequences $\bvt$ such that this condition, \eqref{eq:conditionally independent codewords 2}, holds.
%
We now continue bounding the variance as:
\eas{
& \var[\Kv]  =  \sum_{\bv} \sum_{\bvt  , \ \eqref{eq:conditionally independent codewords 2} } \nonumber \\
& \quad  \quad \lb   \Pr[K_{\bv}=1,K_{\bvt}=1]
-\Pr[K_{\bv}=1]\Pr[K_{\bvt}=1]\rb
\label{eq:bound variance a} \\
& \leq     \sum_{ \bv}  \sum_{ \bvt, \ \eqref{eq:conditionally independent codewords 2} } \nonumber  \\
& \quad \quad \quad   \Pr \lsb K_{\bv }=1,K_{\bvt}=1 \rsb
\label{eq:bound variance b} \\
& = \sum_{ \bv}  \Ebb \lsb K_{\bv} \rsb \nonumber \\
& \quad \quad \quad   \sum_{ \bvt, \ \eqref{eq:conditionally independent codewords 2}} \Pr  \lsb K_{\bvt}=1 | K_{\bv }=1  \rsb,
\label{eq:bound variance last}
}{\label{eq:bound variance}}
where \eqref{eq:bound variance b} follows from dropping the negative defined term in the RHS of \eqref{eq:bound variance a}.
Also \eqref{eq:bound variance last} follows from the fact that $K_{\bv}$ is independent of $K_{\bvt}$.

Consider now the probability term of \eqref{eq:bound variance last}:
the condition in \eqref{eq:conditionally independent codewords 2}  does not rule out the case in which $\bt_{\iv \sgoes \jv}\neq b_{\iv \sgoes \jv}$ but
there exist an $U_{\lv \sgoes \mv}$ such that $U_{\iv \sgoes} \spc U_{\lv \sgoes \mv}$  with $\bt_{\lv \sgoes \mv}= b_{\lv \sgoes \mv}$.
When this is the case, the fact that $\bt_{\lv \sgoes \mv}= b_{\lv \sgoes \mv}$ does not influence the conditional typicality of $\{ U^N_{\iv \sgoes \jv}(w_{\iv \sgoes \jv},\bt_{\iv \sgoes \jv}), \ \bt_{\iv \sgoes \jv} \in \bvt \}$ since $U_{\lv \sgoes \mv}$ is drawn from two different codebooks.

By extension, we have that
\ea{
\Pr  \lsb K_{\bvt}=1 | K_{\bv }=1  \rsb = \Pr  \lsb K_{\bv'}=1 | K_{\bv }=1  \rsb,
}
where
\ea{
\bv' = \lcb \p{
 \bt_{\iv \sgoes \jv }      &  \bt_{\iv \sgoes \jv} \neq  b_{\iv \sgoes \jv} \\
 b_{\iv \sgoes \jv}         &  \bt_{\iv \sgoes \jv} = b_{\iv \sgoes \jv}, \
  \bt_{\lv \sgoes \mv} = b_{\lv \sgoes \mv} , \ \forall \ (\lv,\mv) \ U_{\lv \sgoes \mv} \spc U_{\iv \sgoes \jv}  \\
 b'_{\iv \sgoes \jv }       &   {\rm otherwise},
&
}\rnone
\label{eq:definition b prime}
}
for some $b'_{\iv \sgoes \jv } \neq b_{\iv \sgoes \jv }$.
%
%
%
%
%
By the definition  of $\bvt'$, we can now say that $b_{\iv \sgoes \jv}=b_{\iv \sgoes \jv}'$ if and only if
 $U_{\iv\sgoes \jv}^N(w_{\iv\sgoes \jv},b_{\iv\sgoes \jv})=U_{\iv\sgoes \jv}^N(w_{\iv\sgoes \jv},b_{\iv\sgoes \jv}')$.
Since the codewords are generated in an i.i.d. fashion, this term does not depend on the specific values of $\bv$ and $\bv'$, but solely on whether $b_{\iv \sgoes \jv}=b'_{\iv \sgoes \jv}$.
With this consideration, we can rewrite the summation not in terms of the sequence of binning indices $\bv$ and $\bv'$ as a sum over all the patterns in which the indices in the two sequences can match each other.
That is, take two sequences $\bv'$  and $\bv''$ and a set $\Fv$ together with its complement $\Fvo=\Vv\setminus \Fv$, then
\eas{
{\rm If}   \quad & \ \forall \ (\iv,\jv) \in \Fv, \ b_{\iv \sgoes \jv}'= b_{\iv \sgoes \jv}''=b_{\iv \sgoes \jv},  \\
{\rm and}  \quad & \ \forall \ (\iv,\jv) \in \Fvo, \  b_{\iv \sgoes \jv}' \neq b_{\iv \sgoes \jv}, \  b_{\iv \sgoes \jv}'' \neq b_{\iv \sgoes \jv}, \\
{\rm then} \quad & \Pr  \lsb K_{\bv''}=1 | K_{\bv }=1  \rsb =\Pr  \lsb K_{\bv'}=1 | K_{\bv }=1  \rsb,
\label{condition binning vector}
}
while the specific value of the indices for which both the sequences $\bv'$ and $\bv''$ differ from $\bv$ have no influence.
%

For a given subset of indices $b_{\iv \sgoes \jv}$, $\Fv$, for which \eqref{condition binning vector}  hold,
the number of sequences having the same probability is
\ea{
\prod_{(\iv,\jv) \in \Fv}  (2^{N R'_{\iv \sgoes \jv}} -1)\leq 2^{N \sum_{(\iv,\jv) \in \Fv } R'_{\iv \sgoes \jv}}.
\label{eq:all sequences}
}
Next we need to translate the definition of $\bv'$ in \eqref{eq:definition b prime} into a condition on $\Fv$.
Two indices are the same across $\bv$ and $\bv'$ only if all the parent nodes in the superposition coding graph are the same or if the nodes have no parent nodes. : this is indeed the condition in \eqref{eq:condition enocding errors one way binning} which includes the condition in \eqref{eq:conditionally independent codewords 2} as a subcase.
%
%
%

Using this notation we have
\eas{
& \sum_{ \bv'} \Pr  \lsb K_{\bv'}=1 | K_{\bv }=1  \rsb \\
& \leq  \sum_{\Fv, \ \eqref{eq:condition enocding errors one way binning} } 2^{\sum_{(\iv,\jv) \in \Fvo} N R'_{\iv \sgoes \jv}} \Pr  \lsb K_{\bvt(\Fv)}=1 | K_{\bv }=1  \rsb,
\label{eq:bound mutual covering lemma 4 last}
}{\label{eq:bound mutual covering lemma 4}}
where \eqref{eq:bound mutual covering lemma 4 last}  follows from the fact that each ``pattern'' $\Fv$ appears
$\prod_{(\iv,\jv) \in \Fvo}(2^{R'_{\iv \sgoes \jv}}-1) \leq 2^{\sum_{(\iv,\jv) \in \Fvo} R'_{\iv \sgoes \jv}}$
times and where $\bvt(\Fv)$ is any fixed sequence that follows the pattern $\Fv$.
%
Finally we bound each probability term in \eqref{eq:bound mutual covering lemma 4 last} as
\eas{
& \Pr  [ K_{\bvt(\Fv)}=1 | K_{\bv }=1  ] \\
&=\Pr [ \{  U_{\iv \sgoes \jv}^N(w_{\iv \sgoes \jv} ,b_{\iv \sgoes \jv}), \ (\iv,\jv) \in \Fv \} \ \cup \nonumber \\
& \quad \quad \ \{ U_{\iv \sgoes \jv}^N(w_{\iv \sgoes \jv} ,\bt_{\iv \sgoes \jv}),  \ (\iv,\jv) \in \Fvo^{\Bv}\}  \in \Tcal_\epsilon^N \lb P^{\rm encoding} \rb  \nonumber \\
& \quad \quad  \quad \quad  |  \{ U_{\iv \sgoes \jv}^N(w_{\iv \sgoes \jv} ,b_{\iv \sgoes \jv}),   \ (\iv, \jv) \in \Vv \}  \in \Tcal_\epsilon^N \lb P^{\rm encoding} \rb  ]\\
&=\Pr [ \{ U_{\iv \sgoes \jv}^N(w_{\iv \sgoes \jv} ,\bt_{\iv \sgoes \jv}),  \ (\iv,\jv) \in \Fvo^{\Bv}\}  \in \Tcal_\epsilon^N \lb P^{\rm encoding}|  U_{\iv \sgoes \jv}^N, \ (\iv,\jv) \in \Fv^{\Bv}\rb  \nonumber \\
& \quad \quad  \quad \quad  |  \{U_{\iv \sgoes \jv}^N(w_{\iv \sgoes \jv} ,b_{\iv \sgoes \jv}),   \ (\iv, \jv) \in \Vv \}  \in \Tcal_\epsilon^N \lb P^{\rm encoding} \rb,
}
where
\ea{
\Tcal_\epsilon^N \lb P^{\rm encoding}| U_{\iv \sgoes \jv}^N, \ (\iv,\jv) \in \Fv^{\Bv}\rb
}
indicates the conditional typical set of $P^{\rm encoding}$ given the set
$\{ U_{\iv \sgoes \jv}^N, \ (\iv,\jv) \in \Fv^{\Bv}\}$

Next we write
\eas{
&=\Pr \lsb \{ U_{\iv \sgoes \jv}^N(w_{\iv \sgoes \jv} ,\bt_{\iv \sgoes \jv}),  \ (\iv,\jv) \in \Fvo^{\Bv}\}  \in \Tcal_\epsilon^N \lb P^{\rm encoding}|  U_{\iv \sgoes \jv}^N, \ (\iv,\jv) \in \Fv^{\Bv}\rb  \rnone \nonumber \\
& \quad \quad  \quad \quad \lnone  |  \{U_{\iv \sgoes \jv}^N(w_{\iv \sgoes \jv} ,b_{\iv \sgoes \jv}),   \ (\iv, \jv) \in \Vv \}  \in \Tcal_\epsilon^N \lb P^{\rm encoding} \rb \rsb \\
&=  \Pr \lsb \{ U_{\iv \sgoes \jv}^N(w_{\iv \sgoes \jv} ,\bt_{\iv \sgoes \jv}),  \ (\iv,\jv) \in \Fvo^{\Bv}\}  \in \Tcal_\epsilon^N \lb P^{\rm encoding}|  U_{\iv \sgoes \jv}^N, \ (\iv,\jv) \in \Fv^{\Bv}\rb \rsb,
}{\label{eq: conditional probability in typical set}}
since the codewords
\ea{
\{U_{\iv \sgoes \jv}^N(w_{\iv \sgoes \jv} ,\bt_{\iv \sgoes \jv}),   \ (\iv, \jv) \in \Fvo^{\Bv}\},
}
are conditionally independent on the codewords
\ea{
\{U_{\iv \sgoes \jv}^N(w_{\iv \sgoes \jv} ,b_{\iv \sgoes \jv}),   \ (\iv, \jv) \in \Fvo^{\Bv}\},
}
given the codewords in $\{U_{\iv \sgoes \jv}^N(w_{\iv \sgoes \jv} ,b_{\iv \sgoes \jv}),   \ (\iv, \jv) \in \Fv \}$.
This is because each codeword is created conditionally independently from the others given the parent codewords.

We next write
\eas{
& \Pr \lsb \{ U_{\iv \sgoes \jv}^N(w_{\iv \sgoes \jv} ,\bt_{\iv \sgoes \jv}),  \ (\iv,\jv) \in \Fvo^{\Bv}\}  \in \Tcal_\epsilon^N \lb P^{\rm encoding}|  U_{\iv \sgoes \jv}, \ (\iv,\jv) \in \Fv^{\Bv}\rb \rsb \\
& \leq \sum_{ \lcb U_{\iv \sgoes \jv}^N, \ (\iv,\jv) \in \Fvo^{\Bv}\rcb \in \Tcal_\epsilon^N \lb P^{\rm encoding}| U_{\iv \sgoes \jv}^N,  \ (\iv ,\jv) \in \Fv \rb } \nonumber \\
& \quad \quad \quad \quad  P^{\rm codebook} ( U_{\iv \sgoes \jv}^N,  \ (\iv,\jv) \in \Fvo^{\Bv}| U_{\iv \sgoes \jv}^N,  \ (\iv ,\jv) \in \Fv)  \\
& \leq 2^{N \lb H(P^{\rm encoding}| U_{\iv \sgoes \jv},  \ (\iv ,\jv) \in \Fv) - H(P^{\rm codebook}| U_{\iv \sgoes \jv},  \ (\iv ,\jv) \in \Fv)  \rb }, \nonumber
\label{eq:bound covering 2 last}
}{\label{eq:bound covering 2}}

where
\ean{
& H(P^{\rm encoding}| U_{\iv \sgoes \jv}  \ (\iv ,\jv) \in \Fv)\\
& = H(P^{\rm encoding}) - H \lb P^{\rm encoding} (U_{\iv \sgoes \jv}  \ (\iv ,\jv) \in \Fv)\rb.
}
Note now that the evaluation of the marginal distribution $P^{\rm encoding} (U_{\iv \sgoes \jv}  \ (\iv ,\jv) \in \Fv)$  cannot be easily determined, since some of the parents in the binning graph of $\Fv$ can be in $\Fvo$.
At this point we make use of Assumption \ref{ass:Binning is transitive}, the TB-restriction, to argue the following: if an edge in $\Bv$ crosses from $\Fvo$ to $\Fv$, then
$U_{\iv \sgoes \jv} \in \Fvo$ cannot be a source node in $\Gcal(\Vv,\Bv)$ since $(\iv,\jv) \in \Vv_{\Bv}$.
Since $U_{\iv \sgoes \jv}$ is not a source node, we can change the orientation of the edge crossing from $\Fvo$ to $\Fv$ and still obtain a Markov equivalent DAG.
With this consideration we can conveniently write
\ean{
H(P^{\rm encoding}| U_{\iv \sgoes \jv}  \ (\iv ,\jv) \in \Fv)
& \quad \quad - \sum_{(\iv,\jv) \in \Fv} H(U_{\iv \sgoes \jv} | \pa_{\Ev} (U_{\iv \sgoes \jv}))\\
& =\sum_{(\iv ,\jv) \in \Fvo} H(U_{\iv \sgoes \jv}| A_{\iv \sgoes \jv}(\Fv)),
}
where $A_{\iv \sgoes \jv}(\Fv)$ is defined as in \eqref{eq:Aij definition binnig}
and is the set of all the parent nodes of $U_{\iv \sgoes \jv}, \ (\iv ,\jv) \in \Fvo$ in the Markov-equivalent DAG where the edges of the graph are oriented so that the parents of nodes in $\Fv$ are also in $\Fv$.

First we make sure that the change of direction of the edges does not introduce a cycle.
\begin{lem}{\bf Flipping edges does not introduce cycles\\}
\label{lem:Flipping edges does not introduce cycles binning}
Consider a CGRAS graph $\Gcal(\Vv,\Ev)$ that does not employ joint binning, a subsets $\Fv$ with $\pa_{\Sv}(\Fv)\subset \Fv$ and
its complement $\Fvo=\Vv \setminus \Fv$.
Moreover, let $\Bv'$ be the set of edges obtained by reversing the direction of the edges in $\Bv$ that cross from $\Fv$ to $\Fvo$. Then $\Gcal(\Vv,\Bv'\cup \Sv)$ is  Markov-equivalent to $\Gcal(\Vv,\Ev)$.
\end{lem}
\begin{IEEEproof}
For a cycle to be formed by changing the direction of an edge from $\Fvo$ to $\Fv$, there must be an edge exiting from $\Fvo$.
The direction of all such edges is changed, so no cycle can be introduced by changing the direction of all the outgoing edges from $\Fvo$.
\end{IEEEproof}

The evaluation of $P^{\rm codebook}| U_{\iv \sgoes \jv}  \ (\iv ,\jv) \in \Fv$ is, instead, easier since it's determined only by superposition coding and $\pa_{\Sv}(\Fv)\subseteq \Fv$:
\eas{
& H(P^{\rm codebook}| U_{\iv \sgoes \jv}  \ (\iv ,\jv) \in \Fv)\\
& = H \lb \prod_{(\iv ,\jv) \in \Fvo^{\Bv}} P_{U_{\iv,\jv} | \pa_{\Sv}(U_{\iv,\jv})} \rb \\
& =\sum_{(\iv ,\jv) \in \Fvo} H(U_{\iv \sgoes \jv}| \pa_{\Sv}(U_{\iv,\jv})),
\label{eq:entropy codebook conditional last}
}{\label{eq:entropy codebook conditional}}
so that
\eas{
& 2^{N \lb H(P^{\rm encoding}| U_{\iv \sgoes \jv},  \ (\iv ,\jv) \in \Fv) - H(P^{\rm codebook}| U_{\iv \sgoes \jv},  \ (\iv ,\jv) \in \Fv)  \rb }   \\
& = 2^{- N  \sum_{(\iv ,\jv) \in \Fvo} I(U_{\iv \sgoes \jv} ; \pa_{\Ev}(U_{\iv,\jv}) | \pa_{\Sv}(U_{\iv,\jv})  )} \\
& = 2^{- N  \sum_{(\iv ,\jv) \in \Fvo} I(U_{\iv \sgoes \jv} ; \pa_{\Bv}(U_{\iv,\jv}) | \pa_{\Sv}(U_{\iv,\jv}))}.
}{\label{eq:encoding error probability bound}}
Combining \eqref{eq:bound mutual covering lemma 4} with \eqref{eq:encoding error probability bound} we have
\ea{
\sum_{ \bvt, \ \eqref{eq:conditionally independent codewords 2}} \Pr  \lsb K_{\bvt}=1 | K_{\bv }=1  \rsb
\leq 2^{N \lb \sum_{(\iv,\jv) \in \Fvo^{\Bv}} (R'_{\iv \sgoes \jv} - I(U_{\iv \sgoes \jv} ; \pa_{\Bv}(U_{\iv,\jv}) | \pa_{\Sv}(U_{\iv,\jv})) )\rb}.
\label{eq: conditional probability final bound}
}
We can now return to \eqref{eq:first bound mutual covering last}:
\ea{
& \Pr[\Kv=0]  \leq  \f {
 \sum_{\bv} \sum_{  \Fv, \ \eqref{eq:condition enocding errors one way binning}   } \Ebb \lsb K_{\bv } \rsb
   2^{N \lb \sum_{(\iv,\jv) \in \Fvo^{\Bv}} (R'_{\iv \sgoes \jv} - I(U_{\iv \sgoes \jv} ; \pa_{\Bv}(U_{\iv,\jv}) | \pa_{\Sv}(U_{\iv,\jv})) )\rb}
   }
   {\Ebb \lsb \Kv \rsb^2} \nonumber \\
& =  \f {
 \Ebb \lsb \Kv \rsb \sum_{\Fv, \ \eqref{eq:condition enocding errors one way binning}  }
   2^{N \lb \sum_{(\iv,\jv) \in \Fvo^{\Bv}} (R'_{\iv \sgoes \jv} - I(U_{\iv \sgoes \jv} ; \pa_{\Bv}(U_{\iv,\jv}) | \pa_{\Sv}(U_{\iv,\jv})) )\rb}
   }
   {\Ebb \lsb \Kv \rsb^2} \nonumber \\
&= \Ebb \lsb \Kv \rsb^{-1} \sum_{ \Fv, \ \eqref{eq:condition enocding errors one way binning} }
   2^{N \lb \sum_{(\iv,\jv) \in \Fvo^{\Bv}} (R'_{\iv \sgoes \jv} - I(U_{\iv \sgoes \jv} ; \pa_{\Bv}(U_{\iv,\jv}) | \pa_{\Sv}(U_{\iv,\jv})) )\rb}.
}{\label{eq:mutual covering last expression}}
%
Using the bound on  $\Ebb \lsb \Kv \rsb$ in \eqref{eq: bound E K} we have
\eas{
\Pr[\Kv=0] & \leq 2^{N \sum_{(\iv,\jv) \in \Vv } (R'_{\iv \sgoes \jv} - I(U_{\iv \sgoes \jv} ; \pa_{\Bv}(U_{\iv \sgoes \jv}) |  \pa_{\Sv}(U_{\iv \sgoes \jv})))}  2^{N \lb \sum_{(\iv,\jv) \in \Fvo^{\Bv}} (R'_{\iv \sgoes \jv} - I(U_{\iv \sgoes \jv} ; \pa_{\Bv}(U_{\iv,\jv}) | \pa_{\Sv}(U_{\iv,\jv})) )\rb} \\
& = 2^{N \lb
- \sum_{(\iv,\jv) \in \Fv} R' + \sum_{(\iv,\jv) \in \Vv} I(U_{\iv \sgoes \jv} ; \pa_{\Bv} \{ U_{\iv \sgoes \jv}\} | \pa_{\Sv} \{ U_{\iv \sgoes \jv}\} , Q) - I(U_{\iv \sgoes \jv} ; \pa_{\Bv}(U_{\iv,\jv}) | \pa_{\Sv}(U_{\iv,\jv})) \rb },
}
so that the probability of error goes to zero when \eqref{eq:binnig rates one way binning} holds.

\medskip
\noindent
\emph{DECODING ERROR ANALYSIS}
\smallskip

The decoding error probability can be performed in a similar way as in the proof of Th. \ref{th:Decoding errors only supeperposition} with two differences:
\begin{itemize}
  \item instead of only decoding $w_{\iv \sgoes \jv}$, we are now decoding both $w_{\iv \sgoes \jv}$ and $b_{\iv \sgoes \jv}$ and
  \item  the rate of $(w_{\iv \sgoes \jv},b_{\iv \sgoes \jv})$ is $L_{\iv \sgoes \jv}$.
\end{itemize}

For these reasons, the proof is analogous to the proof in App. \ref{app:Decoding errors only supeperposition}
up to the evaluation of the term in \eqref{eq:calculation I joint decoding}.
This term is, instead, can be bounded as:
\eas{
& \Pr [
\{ Y_z^N,
\{ U_{\iv \sgoes \jv} ( \wt_{\iv \sgoes \jv}, \bt_{\iv \sgoes \jv}), \ (\iv,\jv) \in \Fv^{\Bv}\}
\cup  \nonumber
\\ & \quad \quad   \{ U_{\iv \sgoes \jv} (1,b_{\iv \sgoes \jv} ), \ (\iv,\jv) \in \Fvo^{\Bv}\}
\}
\in
 \Tcal_{\ep}^N  (  P_{Y_z, \{ U_{\iv \sgoes \jv}, \ (\iv,\jv) \in \Vv^z  \} }^{\rm encoding} )]  \\
& \leq  \sum_{  \Tcal_{\ep}^N  \lb  P_{Y_z, \{ U_{\iv \sgoes \jv}, \ (\iv,\jv) \in \Vv^z  \} }^{\rm encoding} \rb }
 \Pr [
Y_z^N \cup \nonumber \\
& \quad \quad  \{ U_{\iv \sgoes \jv} ( \wt_{\iv \sgoes \jv}, \bt_{\iv \sgoes \jv}), \ (\iv,\jv) \in \Fv^{\Bv}\}
\cup
\{ U_{\iv \sgoes \jv} ( 1,b_{\iv \sgoes \jv} ), \ (\iv,\jv) \in \Fvo^{\Bv}\}
]  \\
& \leq  \labs \Tcal_{\ep}^N  ( P_{Y_z, \{ U_{\iv \sgoes \jv}, \ (\iv,\jv) \in \Vv^z  \} }^{\rm encoding}) \rabs
\Pr [ Y_z^N | U_{\iv \sgoes \jv}, \ (\iv,\jv) \in \Fv^{\Bv}] \nonumber \\
& \quad \quad  \Pr [ U_{\iv \sgoes \jv} ( \wt_{\iv \sgoes \jv},\bt_{\iv \sgoes \jv} ), \ (\iv,\jv) \in \Fv , \ U_{\iv \sgoes \jv} ( 1,b_{\iv \sgoes \jv} ), \ (\iv,\jv) \in \Fvo^{\Bv}] \\
& \leq 2^{-N I(Y_z|  U_{\iv \sgoes \jv}, \ (\iv,\jv) \in \Vv )} 2^{N H(P^{\rm encoding})} 2^{- N H(Y_z | U_{\iv \sgoes \jv}, \ (\iv,\jv) \in \Fv )} \nonumber \\
& \quad \quad  2^{-N H (P^{\rm codebook} | U_{\iv \sgoes \jv}, \ (\iv,\jv) \in \Fv )} 2^{N H(P^{\rm encoding} (U_{\iv \sgoes \jv}, \ (\iv,\jv) \in \Fv)) }\\
& = 2^{N \lb -I(Y_z ; U_{\iv \sgoes \jv}, \ (\iv,\jv) \in \Fvo^{\Bv}|U_{\iv \sgoes \jv}, \ (\iv,\jv) \in \Fv, Q)+ I(P^{\rm codebook}; P^{\rm encoding}| U_{\iv \sgoes \jv}, \ (\iv,\jv) \in \Fv) \rb}
}{\label{eq:calculation I joint decoding with binning}}
where
\eas{
H(P^{\rm encoding} (U_{\iv \sgoes \jv}, \ (\iv,\jv) \in \Fv))
& =H \lb \prod_{(\iv,\jv) \in \Fv} P_{U_{\iv \sgoes \jv} | \pa_{\Ev\cap (\Fv \times \Fv)}(U_{\iv \sgoes \jv})} \rb\\
& =\sum_{(\iv,\jv) \in \Fv} H \lb U_{\iv \sgoes \jv} | \pa_{\Ev\cap (\Fv \times \Fv)}(U_{\iv \sgoes \jv}) \rb
}
for $H (P^{\rm codebook} | U_{\iv \sgoes \jv}, \ (\iv,\jv) \in \Fv )$ defined in \eqref{eq:entropy codebook conditional}
so that the term  $H(P^{\rm codebook}| P^{\rm encoding}| U_{\iv \sgoes \jv}, \ (\iv,\jv) \in \Fv)$ is obtained as
\eas{
& I(P^{\rm codebook}; P^{\rm encoding}| U_{\iv \sgoes \jv}, \ (\iv,\jv) \in \Fv) \\
& = \sum_{(\iv,\jv) \in \Fv} \lb +H(U_{\iv \sgoes \jv}| \pa_{\Ev}(U_{\iv \sgoes \jv})) -  H(U_{\iv \sgoes \jv}| \pa_{\Ev \cap(\Fv \times \Fv)}(U_{\iv \sgoes \jv})) \rb \nonumber \\
& \quad \quad +\sum_{(\iv,\jv) \in \Fvo}  \lb +H( U_{\iv \sgoes \jv} | \pa_{\Ev} (U_{\iv \sgoes \jv}) - H( U_{\iv \sgoes \jv} | \pa_{\Sv} (U_{\iv \sgoes \jv}))  \rb \\
& = -\sum_{(\iv,\jv) \in \Fv} I(U_{\iv \sgoes \jv};  \pa_{\Ev\cap(\Fvo^{\Bv}\times \Fvo)}(U_{\iv \sgoes \jv})| \pa_{\Ev \cap(\Fv \times \Fv)}(U_{\iv \sgoes \jv})) \nonumber \\
& \quad \quad -\sum_{(\iv,\jv) \in \Fvo}  I( U_{\iv \sgoes \jv} ; \pa_{\Bv} (U_{\iv \sgoes \jv})| \pa_{\Sv} (U_{\iv \sgoes \jv})).
}{\label{eq: decoding boost bound}}
Note now that  $\pa_{\Sv}(\Fv) \subseteq \Fv$, so that the only edges in $\Ev\cap(\Fv \times \Fvo)$ are edges in $\Bv$.

\subsection{Proof of Theorem \ref{th:Decoding errors only supeperposition and joint binning}}
\label{app:Decoding errors only supeperposition and joint binning}

The proof follows along the same lines as the proof of Th. \ref{th:Decoding errors only supeperposition and one-way binning} in App. \ref{app:Decoding errors only supeperposition and one-way binning} and differs mainly in the way in which the encoding and decoding distributions factorize.

\medskip
\noindent
\emph{ENCODING ERROR ANALYSIS}
\smallskip
The encoding error analysis is analogous to the encoding error  analysis in App. \ref{app:Decoding errors only supeperposition and one-way binning}, although we need to re-evaluate typicality bounds.
In particular, the term  $\Ebb[\Kv]$ in \eqref{eq: bound E K} can be evaluated as
\ea{
\Ebb    [\Kv]  \leq 2^{N \sum_{(\iv,\jv) \in \Vv } (R'_{\iv \sgoes \jv} - I(U_{\iv \sgoes \jv} ; \pa_{\Bvt}(U_{\iv \sgoes \jv}) |  \pa_{\Sv}(U_{\iv \sgoes \jv}) )) }
\label{eq: bound E K 2}
}
since the encoding distribution is now described by $\Gcal(\Vv,\Evt)$ and the codebook distribution by $\Gcal(\Vv,\Vv)$.

Next, we wish to evaluate the term $\var[\Kv]$ which can be done by bounding the term
\ea{
\Pr \lsb \{ U_{\iv \sgoes \jv}^N(w_{\iv \sgoes \jv} ,\bt_{\iv \sgoes \jv}),  \ (\iv,\jv) \in \Fvo^{\Bv}\}  \in \Tcal_\epsilon^N \lb P^{\rm encoding}|  U_{\iv \sgoes \jv}^N, \ (\iv,\jv) \in \Fv^{\Bv}\rb \rsb
} in \eqref{eq: conditional probability in typical set}.
To evaluate this term we need to determine the conditional probabilities
$P^{\rm encoding}| U_{\iv \sgoes \jv}^N,  \ (\iv ,\jv) \in \Fv$ and $P^{\rm codebook}| U_{\iv \sgoes \jv}^N,  \ (\iv ,\jv) \in \Fv$ which are the encoding and codebook conditionally probabilities given the set of RVs $\{U_{\iv \sgoes \jv}^N,  \ (\iv ,\jv) \in \Fv\}$.
A convenient factorization for this distribution is available only when the direction of the undirected edges is chosen so that to go from $\Fv$ to $\Fvo$:
For this equivalent DAG, the conditional distribution of the random variables in $\Fvo$ can be expressed as in \eqref{eq:general conditional form}.

The following lemma grants us that such an orientation always exists.

\begin{lem}{\bf Flipping edges does not introduce cycles \\}
\label{lem:Flipping edges does not introduce cycles}
Consider given a CGRAS graph $\Gcal(\Vv,\Ev)$, its Markov-equivalent, a subsets $\Fv$ with $\pa_{\Sv}(\Fv)\subset \Fv$ and  its complement $\Fvo=\Vv \setminus \Fv$.
Moreover let $\Bv'$ be the set of edges obtained by reversing the direction of the edges in $\Bvt$ that cross from $\Fv$ to $\Fvo$.
Then $\Gcal(\Vv,\Bv')$ is also Markov-equivalent to $\Gcal(\Vv,\Ev)$.
\end{lem}

\begin{IEEEproof}
In Th. \ref{thm: DAG equivalence of graph representation} we have established that any non-cyclic orientation of the undirected edge in $\Gcal(\Vv,\Ev)$ produces a Markov-equivalent DAG.
Since $\pa_{\Sv}(\Fv) \subset \Fv$,  changing the direction of the edges can result in a cycle only when the outgoing edge from $\Fvo$ into $\Fv$ is a binning edge and there exists an undirected edge between  $\Fvo$ and $\Fv$.
More specifically, if there exists the path $[a_0...a_N]$ is in $\Fv$, another path $[a_{N+1}...a_M]$ is in $\Fvo$, moreover $a_N$  to $a_{N+1}$ are connected by a binning edge while $[a_0]$ and $a_M$ are connected by an undirected edge.
This scenario, though, implies that there exists a directed cycle in $\Gcal(\Vv,\Ev)$ which cannot occur because of Assumption \ref{ass:Joint binning forms cliques}, the CSJB-restriction.
%
%
%
\end{IEEEproof}

Lem. \ref{lem:Flipping edges does not introduce cycles} implies that we can choose the orientation of the undirected edges to be incoming to $\Fv$ from $\Fvo$ $\Bv'$ and obtain the factorization
\ea{
P^{\rm encoding}| U_{\iv \sgoes \jv}^N,  \ (\iv ,\jv) \in \Fv = \prod_{(\iv,\jv) \in \Fvo} P_{U_{\iv \sgoes \jv} | \pa_{\Bv'\cup \Sv}(U_{\iv \sgoes \jv})}
}
The factorization of $P^{\rm codebook}| U_{\iv \sgoes \jv}  \ (\iv ,\jv) \in \Fv$ is instead unchanged from \eqref{eq:entropy codebook conditional}: the distribution of these codewords is unaffected by binning since they are independent from the set of codewords which possesses the correct typicality properties.
We can now evaluate the analog of term \eqref{eq: conditional probability final bound} for this more general case as
\ea{
\sum_{ \bvt, \ \eqref{eq:conditionally independent codewords 2}} \Pr  \lsb K_{\bvt}=1 | K_{\bv }=1  \rsb
\leq 2^{N \lb \sum_{(\iv,\jv) \in \Fvo^{\Bv}} (R'_{\iv \sgoes \jv} - I(U_{\iv \sgoes \jv} ; \pa_{\Bv'}(U_{\iv,\jv}) | \pa_{\Sv}(U_{\iv,\jv})) )\rb}.
}
where the of the of a node in  the graph $\Gcal(\Vv,\Bv')$ for each $\Fv$ are obtained as
\ea{
\pa_{\Bv'}(U_{\iv,\jv}) = \pa_{\Bvt}(U_{\iv,\jv}) \cup \pa_{\Bv \cap (\Fv \times \Fvo)}(U_{\iv,\jv})
\label{eq:conditions joint binning}
}
that is, the parents of $U_{\iv,\jv}$ in $\Bv$ which go from $\Fv$ to $\Fvo$.
This corresponds to the definition of $A_{\iv \sgoes \jv}(\Fv)$ in \eqref{eq:A ij}.

\medskip
\noindent
\emph{DECODING ERROR ANALYSIS}
\smallskip

As for the encoding error analysis, the decoding error analysis follows the same steps as the decoding error analysis in App. \ref{app:Decoding errors only supeperposition and one-way binning} but the typicality bounds must be re-evaluated to account for the effects of joint binning.
These effect must be accounted for in evaluating the term  $I(P^{\rm codebook}; P^{\rm encoding}| U_{\iv \sgoes \jv}, \ (\iv,\jv) \in \Fv)$  in \eqref{eq: decoding boost bound}.
As for the encoding error analysis, a convenient factorization of $P^{\rm encoding}| U_{\iv \sgoes \jv}, \ (\iv,\jv) \in \Fv$ is available only when the
undirected edges in $\Gcal(\Vv,\Ev)$ are oriented in $\Gcal(\Vv,\Evt)$ to cross from $\Fv$ to $\Fvo$.
As established in Lem. \ref{lem:Flipping edges does not introduce cycles}, this is always possible in the graph $\Gcal(\Vv,\Ev)$.
The only difference in this case lies in the fact that decoder $z$ does not observe the entire graph $\Vv$ but only the sub-graph $\Vv^z$.
This does not substantially change the derivation but only the conditions in \eqref{eq:conditions joint binning} which are now restricted from $\Vv$ to $\Vv^z$. 
Other than this restriction, the proof follows a similar set of steps as Th.  \ref{th:Decoding errors only supeperposition and one-way binning}.